# The Hopkins Ultraviolet Telescope – A Historical and Personal Retrospective

**William P. Blair**
**Research Professor**
**The Johns Hopkins University**
**William H. Miller III Department of Physics & Astronomy**

**August 2026**

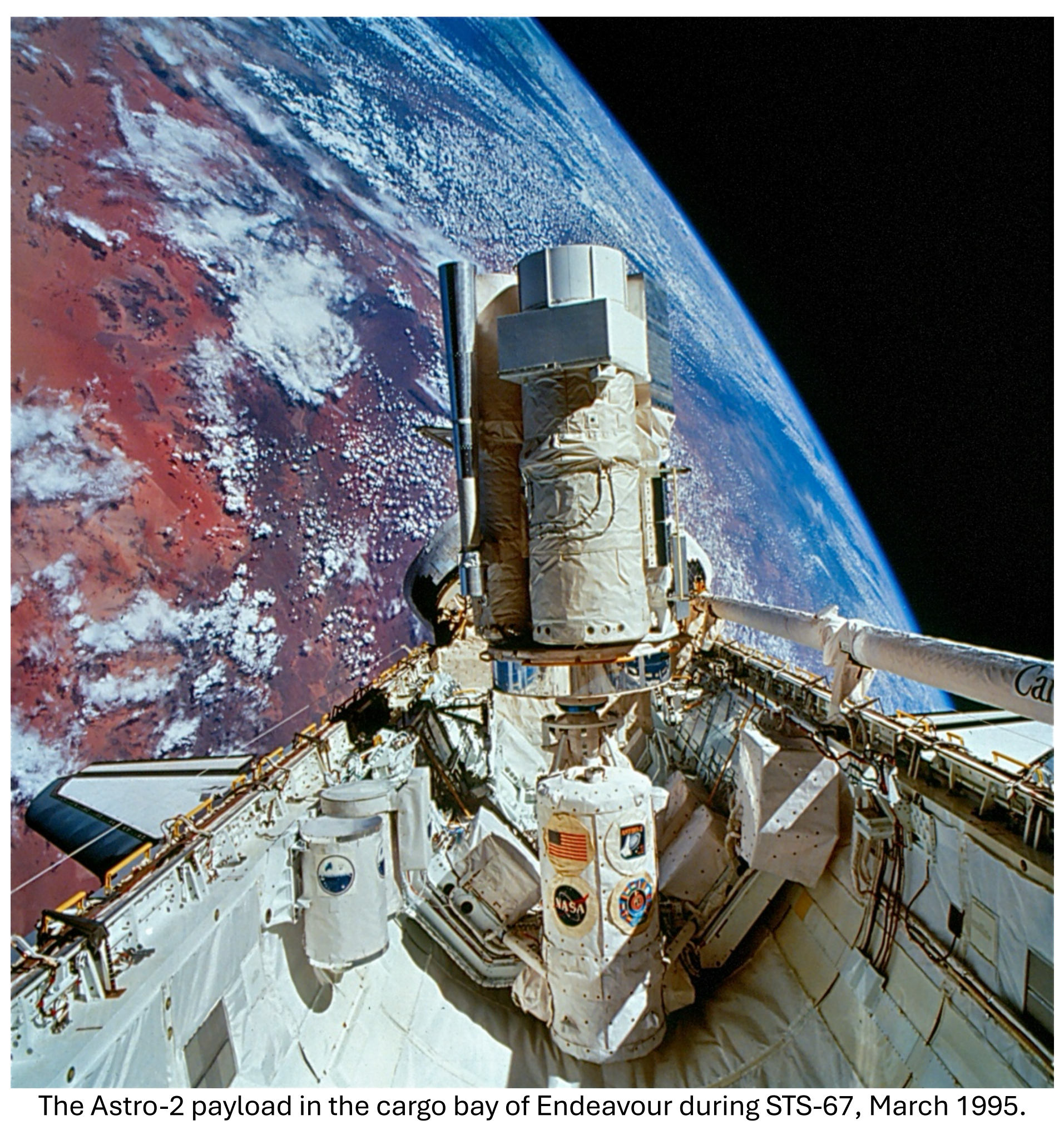

The Astro-2 payload in the cargo bay of Endeavour during STS-67, March 1995.

## Forward

It has always been a goal of mine to look back at various archived materials pertaining to the Hopkins Ultraviolet Telescope (HUT) project and put together a written history. Now, in my semi-retirement, the time seems right and this document is my attempt to perform this task. In the process, I have scanned and organized a record of documents, photos, and a selection of the most relevant press clippings and articles about the project to form an historical archive for future reference by whomever might be interested.

This project has been greatly aided by an extensive chronological collection of press clippings that were organized by Sharon (Busching) Doyle, Professor Arthur Davidsen's administrative assistant during much of the HUT project period. The materials I have scanned and organized are but a small fraction of all that is available in that collection, although I have tried to include the most representative and relevant articles and examples. These resources are called out with blue reference indicators in the text that follows, pointing to either the Documents section, the PR Articles section, or occasionally other sections of the archive. These are not links; rather, the indicators point to articles or documents in PDF format, that are archived in separate directories in the online HUT history archive, with the file names indicated in the reference listings provided in Appendices B and C of this document. (Not all scanned materials in the archive are referenced in this document.) Access to the PDF archive itself, photos, and other materials has recently become available as part of the HUT materials in the Mikulski Archive for Space Telescopes; see https://archive.stsci.edu/hut/hut-historical-archive/ (The file **contents-README** in that directory provides a more detailed description of the historical archive.)

I have also drawn on my own personal collection of writings, notes, annotated weekly calendars, and official HUT documents to find or confirm various dates and happenings over the life of the project. In support of this activity, I have put together a detailed timeline that includes notations for many details about travel or background activities that are not necessarily covered in this text. This timeline contains numerous annotations pointing to additional documents or articles in the archive where the reader can go for more information. I include the timeline here as Appendix A.

I do not regurgitate standard technical information about the HUT hardware or its science goals, as this sort of information is easily located in some of the archived materials or online. My goal is rather to try and tell the story of the project, adding context as I can and relating the importance of the project to the growth of astrophysics within the Department of Physics and Astronomy at Johns Hopkins University, especially during the 1970s and 1980s.

I make no claim to completeness or total accuracy, as some of this comes from my (faulty) memory or incomplete notes. Also, some comments may come from my own biased perspective of events. I make no apologies for that. It is my hope, as others familiar with HUT and the Astro missions read over these materials, that corrections and/or additional memories can be included in a future update as may be appropriate.

William Blair
August 2026

## Table of Contents

# Chapter 1: The Early Years

The Hopkins Ultraviolet Telescope project (HUT) was the brainchild of Prof. Arthur F. Davidsen, who was the principal investigator (PI) of the project and was the person responsible for putting together the team of people who ultimately built and flew the telescope on two space shuttle missions. Davidsen joined the Hopkins Physics department in 1975 as the newest member of the small astrophysics contingent within what was then the Physics department. It would not be until almost a decade later that the growth of the astrophysics side of the department resulted in it becoming the Department of Physics and Astronomy (as it is known today).

Davidsen became active in the extant sounding rocket program at JHU, under the tutelage of Prof. William (Bill) Fastie. In April 1977, Davidsen and (then) graduate student George Hartig launched a Black Brant rocket from White Sands New Mexico, carrying an 0.4 m Dall-Kirkham-type telescope known as the FOT (Faint Object Telescope) with a far-ultraviolet spectrograph. This flight successfully obtained the first ultraviolet spectrum of a quasar, 3C 273. An article was published in *Nature* in September 1977, which ultimately led to the awarding of the prestigious AAS Warner prize to Arthur in 1979. This was a big deal. *Sky & Telescope* magazine carried an article reporting the flight in its January 1978 issue, written by none other than Dennis Overbye of future NY Times fame (see P1).

George Hartig has reported that a real HUT precursor telescope, with a 0.9 m primary and a prime focus spectrograph with an open window microchannel plate detector, was also flown a few years later on April 27, 1981. Unfortunately, this flight was not a scientific success; the spectrograph was pushed out of focus by snubber screws that rattled loose during launch and the ion trap was not very effective, leading to high background noise. But important technical lessons were learned and applied as the real HUT instrument was developed. George left JHU for a postdoc at Cerro Tololo Interamerican Observatory in Chile before returning to Baltimore and STScI as a key player in Hubble Space Telescope development activities and later servicing missions.

The *International Ultraviolet Explorer* satellite (IUE) had launched in January 1978, and JHU professors Paul Feldman and Warren Moos were early users, observing Jupiter and the Io torus in the ultraviolet. Thus, by the late 1970's significant notoriety was brought to the ultraviolet astronomy efforts at JHU and to ultraviolet spectroscopy in particular.

For perspective, the mid-to-late 1970's was in the interim period between the conclusion of the Apollo moon program and the first flight of the space shuttle, which did not occur until April 1981. The space shuttle program, of course, was well behind schedule and way over budget, but the stated expectations were that the shuttle was going to provide cheap and frequent access to low earth orbit, so much so that "weekly" (or at least very frequent) flights were being discussed.*[1] In 1978, NASA was looking for things to fly on the shuttle and put out a "Request for Proposals" for exactly that, dated June 16, 1978.

*[1] Of course, in retrospect, this was "crazy talk," but it is a prime example of how differently things can be perceived looking forward versus looking backward.

Frustrated by the short observation times available on sounding rockets, Davidsen et al. proposed "A Far Ultraviolet Telescope/Spectrometer for Spacelab," in response to NASA AO-OSS-2-78, with an estimated total cost: $2,059,000. (See D1 and D2.) Original co-investigators include JHU professors Bill Fastie, Paul Feldman, Warren Moos, and Richard Henry. The proposal was submitted on Nov. 15, 1978. The original text discusses a broad range of science, mostly focused on AGN and QSO studies, but does not mention the He II Gunn-Peterson test specifically.*[2] While the hardware described is clearly what became HUT, it mentions several configurations that were never developed, including a higher resolution Cassegrain focus spectrometer in addition to the low resolution prime focus spectrograph. The proposal does state, however, "We are not proposing for a single flight aboard Spacelab, but are proposing a continuing program of investigations to be carried out over many flights." Depending on your definition of "many," it is clear that this was never realized, although there were three flights planned at the time of *Challenger* disaster in 1986 and an extended set of another three flights were under initial discussion at that time (see, for instance, discussion in D15).

*[2] I mention this because in later years, it was always said that the He II G-P test to detect the intergalactic medium was the main scientific driver for HUT, which doesn't seem to be accurate. I haven't located when it was first raised as a primary science goal, however.

The number of proposals received is reported differently in different publications, all somewhat after the fact. A good reference is toward the end of an article in the Chronicle of Higher Education is 1985 (P18), where it indicates about 200 proposals were received and initially down-selected to about 40. A mailgram in the archives dated 1979-08-30 announces the tentative acceptance of the initial proposal. A second RFP for what would now be called the Phase A/B study period, RFP-5-15117-278, was released in late 1980 or early 1981 but I have not located the actual proposal submitted for this. The three Astro UV telescopes, including the Ultraviolet Imaging Telescope (UIT) and Wisconsin Ultraviolet Photo-Polarimeter Experiment (WUPPE), were all apparently accepted again in this round and were combined by NASA to become the Astro Observatory. The resulting HUT contract can be viewed in document D6.

On the surface, and certainly from NASA's perspective, combining three UV telescopes onto a co-aligned pointing system made perfect sense. But at a deeper level, the capabilities and science priorities of the three telescopes were not a great match. With an 0.9 m mirror, HUT was the largest of the telescopes, geared toward modest resolution FUV 900-1850 Angstrom spectroscopy of faint point (or slightly extended) sources through small fixed apertures (with EUV 450-900 Angstrom capability as well on Astro-1). WUPPE performed spectropolarimetry in the FUV/NUV, requiring bright stellar targets and high signal-to-noise observations. UIT was a wide field (40 arcmin) imaging telescope with bands in the FUV and NUV, with extended galaxies and star clusters prominent in their priorities. It wasn't that there was no overlap in targets or science goals, but it was certainly not a match made in heaven, and these differences caused a number of complexities in planning joint timelines (of which many were planned).

It is an interesting side note that during this same time frame, when HUT had just been accepted, the NASA RFP looking for a host institution for the Space Telescope Science Institute for Hubble was released (1979), and an AURA/JHU proposal was written to host it on the JHU campus, led in large part by none other than Arthur Davidsen. This proposal was ultimately successful, resulting in a further huge boost to astrophysics at JHU. Thus, in many ways, Arthur was responsible for this success, partially leveraged from his recent success in getting the HUT proposal accepted (even though it had yet to be developed and flown). Also sometime in this period, Arthur was added onto the science instrument team for the Hubble first generation instrument, the Faint Object Spectrograph (or FOS; Richard Harms from UCSD, PI).

With the successful launch (finally!) of the space shuttle *Columbia* (STS-1) on April 12, 1981, the idea of three Astro mission flights on roughly 8-month centers starting in 1984 seemed like a real possibility. But the actual reality was that the complexity and difficulty of preparing, flying, and refurbishing the shuttle for the anticipated flight schedule was not yet fully understood, and ultimately was never achieved.

A memo from project manager Glen Fountain to JHU/APL personnel, dated Oct. 12, 1981, notes that "The HUT development program is now underway." A JHU Gazette article dated Dec. 10, 1981, announced the acceptance and start of a 4.5 year $5.9 million contract, presumably not only for the design study but also the construction phase. A later "HUT Investigation Development Plan," dated July 1983, provided the real design details for the hardware and construction, and was no doubt the result of this study period (see D8).

In parallel with the hardware build up, Arthur also began building the HUT team. Knox Long and Sam Durrance both came to JHU in 1982, with Knox being designated project scientist. Randy Kimble arrived in August 1983, and Hal Weaver, who had left JHU at the end of 1982 with a freshly minted Ph.D., returned to work on HUT in 1984. Bill Blair was hired in July 1984, mainly to focus on coordinating the science plan. JHU grad students Chuck Bowers and Harry Ferguson were also early and integral team members. Jerry Kriss joined JHU/HUT in August 1985.

Randy Kimble recalls "The first project monthly that I attended, down at APL (so August or September 1983), the Headquarters folks absolutely REAMED the HUT team for how far behind schedule they were. Arthur therefore announced that we were officially working 6-day weeks until we got back on track. So our hands-on hardware folks were in every Saturday for quite a while. (It wasn't that you could NEVER miss, but the baseline was working Saturdays.) Arthur and Knox came in too and worked in their offices to share the pain."

Interestingly, an APL memo by Glen Fountain et al. from Feb. 25, 1986 (less than a month after the *Challenger* disaster) gives a good retrospective summary of the timing of the main steps of the HUT hardware development (see D15); the reader is encouraged to look at that memo for details if desired. However, a couple of dates of interest include when actual construction started in the lab in JHU Rowland Hall (May 1984), when it left the Homewood campus for APL (September 1984), and when it was shipped from APL to NASA KSC (March 1985).

On the NASA side, Ted Gull from NASA/GSFC was appointed as NASA Mission Scientist for Astro on June 15, 1982, to coordinate activities between all three UV instrument teams and NASA. There was apparently a bit of a power tussle within NASA between GSFC and MSFC, and ultimately Marshall was designated as the primary NASA center responsible for Astro, but Ted remained Mission Scientist all the way through the actual Astro-1 mission that finally flew in late 1990.

A back story during this period was the selection of three Payload Specialists by NASA, to fly with Astro. Any two of the three would fly on a given flight, with the other working the console during each mission as Alternate Payload Specialist. An initial round of interviews was done in August 1983 at NASA/GSFC. An article from the Baltimore Sun, dated Oct. 31, 1983, highlights that Arthur Davidsen, Knox Long, and Sam Durrance were three of eight finalists, and that the NASA vetting process would be in a few weeks (P4). A follow-up article in the JHU Newsletter (P5) confirms that NASA candidate PS vetting occurred in mid-November 1983 at JSC and that a selection would be made in spring of 1984.*[3] The selection of Sam Durrance (HUT), Ron Parise (UIT), and Ken Nordsieck (WUPPE) was announced in a NASA/JHU press release June 20, 1984, with Sam and Ron assigned to Astro-1.

*[3] There is some information about this competition and the vetting process in a long retrospective *JHU magazine* article that was published after *Challenger*; see P24.

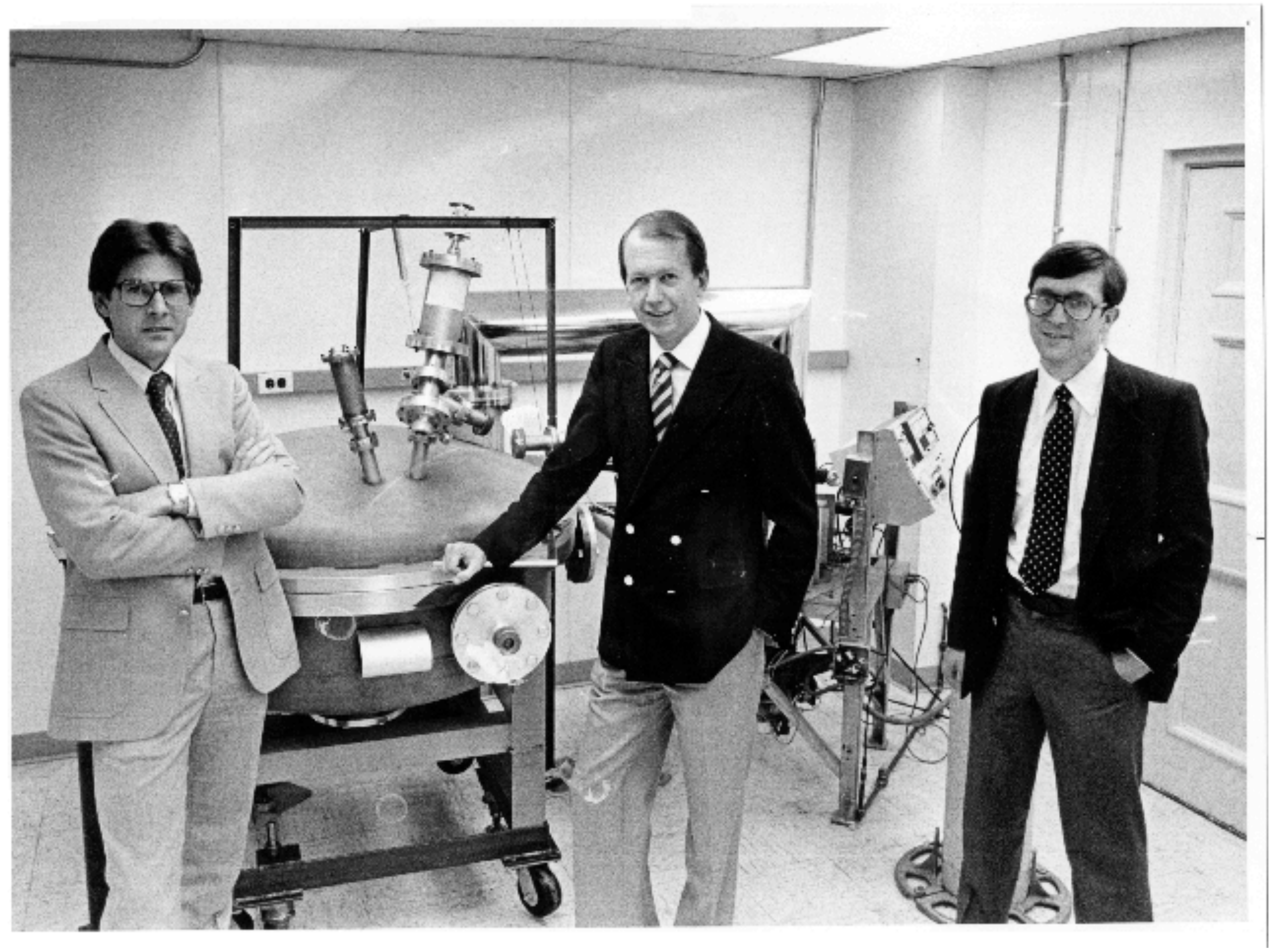

Sam Durrance, Arthur Davidsen, and Knox Long, HUT finalists for Astro Payload Specialist.

## Comet Halley

As many will recall, a major focus and driver for the original Astro-1 mission (which I will now refer to as "Astro-Halley"), especially from NASA's perspective, was to observe comet Halley. Because of cost and budget, NASA had punted on more expensive flyby options, and observations with the Astro UV telescopes became one of the main things NASA was doing for Halley's visit. In addition to the UV observations, a "Wide Field Camera" was also added to the payload to record optical imagery. Early on (P3), a three-flight sequence all concentrating on Halley was apparently under discussion, but at the time I joined the project in mid-1984, Astro-1 was slotted into Nov. 1985 to observe the comet at closest approach to earth and when it was inward bound. For schedule reasons that have been lost to me, the launch date slipped to March 6, 1986, shortly thereafter, and observing the comet as it was outward bound was the one shot NASA had to do something significant to contribute to the visit of the comet. As important as the comet observations were (especially to Paul Feldman, Hal Weaver, and to NASA), plans were of course to observe many different classes of objects in the UV window for everyone on the HUT team, even as part of this first mission.

## Jabba the HUT?

In 1983, the third movie in the *Star Wars* franchise, *Return of the Jedi*, was released to critical acclaim, including the infamous space villain Jabba the Hutt. Always the entrepreneur, and with a name so close to the HUT acronym, it was irresistible to Arthur to think of using Jabba as the HUT mascot. He wrote to George Lucas in late February 1984 asking permission. About a month later, a reply was received from Lucas' executive assistant granting permission. And thus the famous graphic of Jabba sitting in the space shuttle cargo bay that graced many T-shirts and mugs came to be. The response letter contained an apology for the delayed reply because they had been, "...extremely busy with the production of INDIANA JONES AND THE TEMPLE OF DOOM." (See D9.)

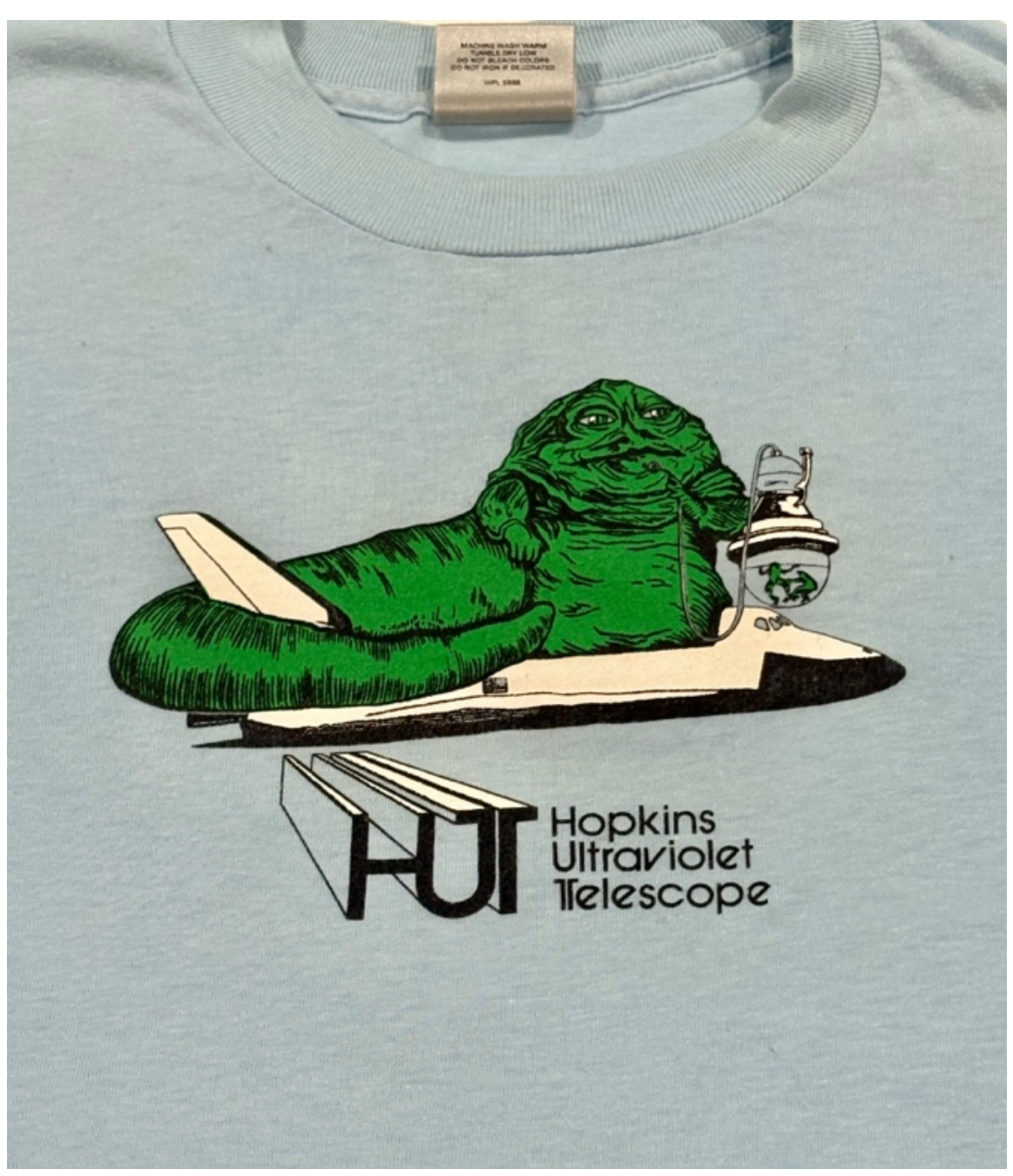

Jabba the "HUT"!

## The Light Bulb Incident

The HUT hardware dodged a number of near disasters over its time of development. One of the more serious events occurred after HUT was delivered to KSC, and after HUT was integrated with the other telescopes into the Astro payload on the so-called cruciform structure that would mount it to the Spacelab Instrument Pointing System (IPS). The accident was described in some detail in a JHU magazine article published shortly after the *Challenger* disaster (see P24).

It was May 1, 1985, and an alignment test was in progress at KSC. The payload was oriented vertically, and the HUT doors were wide open, exposing the mirror surface to light for the alignment. NASA engineers had built scaffolding up around the payload. A 1000 watt halogen light was swung into position above the telescope when the unthinkable happened: the glass light bulb burst, showering glass shards down into the telescope. What was even worse, the bulb's tungsten filament melted and splattered down onto the primary mirror. The most significant damage was a solidified "blob" of what had been molten metal about the size of a 50-cent piece, directly on the iridium mirror coating. (Iridium is what provided the best reflective performance for FUV/EUV light that HUT was to observe.) While there was a backup mirror for HUT back at JHU, it was not clear that the schedule would allow the time needed to disassemble the payload and HUT itself and replace the mirror. (Remember the pressure to observe comet Halley.)

Over the ensuing month or more, HUT team members (primarily Sam, Hal Weaver, and engineer Steve Conard*[4] ) were able to access the telescope and mirror assembly, remove the glass shards as best they could, and clean and assess the mirror reflectivity. Less than 1% of the mirror's surface was badly damaged, and reflectivity measurements indicated no significant degradations resulted. While it was hard to have full confidence that every last shard of glass had been found and removed, it was decided that reassembly and testing could move forward. A memo giving an "all clear" on the sensitivity assessment was sent to NASA in early June 1985 (see 1985-06-05_KSL Mirror_assess.pdf ). Randy Kimble remembers the final realignment activities stretched into Labor Day weekend, to the delight of KSC alignment techs who got double-time on Sunday and triple-time on the holiday!

*[4] Hal Weaver or Steve Conard could in principle provide further details here since they lived through this event!

## On to…launch?

With the hardware built and delivered, the emphasis for much of 1985 shifted to what we called "mission planning." This was a complicated set of sub-processes that included such things as generating and prioritizing science programs and specific targets and observation times per target (for each of the three science teams), defining a system to combine and manage those inputs, interfacing the science teams to NASA Marshall mission managers and planners, and overseeing the process of generating not only the timeline of expected observations and activities on orbit, but also producing numerous "products" that would be used by the astronaut-observers during the mission. Combine this with the requirements levied on the process by the shuttle program itself, where a launch date and time had to be assumed months in advance, and nearly all of the products were planned to that specific assumption. It was not clear to any of us, let alone the MSFC planning staff, how the timeline or these planning products could be updated in real time, given the unlikelihood of everything going exactly as planned. We joked that our MSFC mission manager, Jack Jones, used to say, “Sooner or later we are gonna launch on time.”

To say that planning a mission was a cumbersome and painful process is a gross underestimate of its difficulty. It fell to me to be the primary organizer of this effort for HUT, and to be the liaison to the other teams as well as NASA. I will spare the reader the details partly because it is too painful and because I have suppressed the worst parts into the portions of my memory that are thankfully no longer recoverable without hypnosis.

It is worth recounting something about the state of technology back in the early to mid 1980’s. Computer access was very limited. E-mail and the internet were not yet in play. Zoom and the like were not in general use. US mail, FedEx, fax machines, and telephones (telecons) were the coin of the realm. When I arrived at JHU in mid-1984, our HUT administrative assistant was the only one in our group with a crude word processor, and the printer was dot-matrix. We not only did not yet have workstations, we did not even have terminals on our desks or general access to central computing facilities. As I recall, the high energy physics group in our department was the first group to get a Vax 11-750, and “others” were only allowed to use it after 5 p.m.

While these shortcomings impacted many aspects of the mission and preparations, I will recount only one here in some detail as an example. Once the list of potential targets was available, accurate (and verified) coordinates and other supporting information needed to be compiled. Then what were essentially standardized “finder charts” had to be generated as part of the production of the “Target Book”, a hardcopy of which actually flew on the shuttle for reference by the astronauts. There were no digitized sky surveys at the time, but the vast majority of targets were visible on the Palomar Observatory Sky Survey (POSS) plates/prints, so that was the primary source material. Polaroid had one version of their famous film products that not only generated a positive (hard copy), but also produced a negative that could be chemically processed and saved for future use. Thus, hundreds of finder charts centered on targets located on the POSS prints, each including the edge of a millimeter ruler in the frame for scale, had to be produced and tracked. Then standardized 5x7 inch photographic prints of each target field, made to the specifications of the erstwhile Target Book, had to be produced. Many fun hours in the darkroom in old Rowland Hall were spent by myself, Chuck Bowers, and several undergrads to produce these charts.

Of course, to standardize this process, we had to do this job for the targets from ALL THREE teams, not just for HUT. This was a complication to two reasons. Many WUPPE targets were very bright stars, which were just large black blobs on the POSS prints, and thus did not show the presence of any nearby stars that might be visible in the HUT TV guider. Also, many UIT targets were large and bright nearby galaxies or star clusters that were also just burned in on the POSS prints. So alternative image materials had to be located and the scales of these materials had to be matched to the needs of the standardized finder charts. While different sources were used for these alternate materials, one of the more useful ones was the glass plate collection at the Harvard-Smithsonian Center for Astrophysics in Cambridge, MA. Chuck Bowers and I made a trip there for a full nine days in late September 1985 and were able to locate images that were not as deep as the POSS, and hence more suitable for our task. This trip happened at the time Boston was being walloped by hurricane Gloria, which roared up over Long Island and across the Boston region. Chuck and I remember taking a break from photography to go up on the roof of the observatory and watch the nearby trees swaying in the wind. It was rather surreal.

Each page of the Target Book contained much more information than just the finding charts. Encoded was detailed set up and configuration information for each of the three instruments on Astro, as well as annotations as to the target, the guide stars, and expected angle for the planned observation (how to rotate the finding chart to match what would be seen on the HUT TV guider*[5] at the time of the observation). Hence, each instrument team had to determine their setup details for each potential observation and track it in a database after the observation timeline was constructed. When possible, simulated spectra for HUT and WUPPE were generated and scaled to the expected count rates as well. Target book pages were then printed and the finder charts were affixed to each one by hand, sometimes with careful hand mark ups! The assembled pages were then delivered to MSFC for the actual production of the Target Book. (Whew!) Some examples of the final product for Astro-Halley are in A-Halley_Tbook_examples.pdf in the General directory of the archive. (Comparison to the more sophisticated digital charts produced for Astro-2 are in A-2_Tbook_examples.pdf in the same directory.)

*[5] The HUT TV camera was not just used by HUT. Rather, it functioned as the primary guider for all three UV instruments. Once HUT was locked on guide stars, WUPPE and UIT would know that we were in the correct position and (hopefully) pointing was stable. They could then start observing. (Recall, UIT was taking images on film, so stable pointing was important prior to exposing. An Image Motion Compensation System provided fine guidance information to UIT and WUPPE.)

Suffice it to say, with integration and testing of the hardware still going on and mission planning work, and with a few trips to Huntsville for mission simulations thrown in, the run up to the mission was a frenetic time for all.

## Goodbye Astro-Halley

There is really no need for me to recount the details of January 28, 1986, and the *Challenger* disaster, an event that I am sure is seared into the memories of all involved. Earlier in January, "our" shuttle, *Columbia*, had completed its first mission (STS-61C) after two years of refurbishment, and its next flight was STS-61E carrying the Astro payload into orbit, scheduled for March 6, 1986. Thus, we were less than six weeks from launch and were the next shuttle to go after *Challenger*. But it was not to be.

A brief JHU Newsletter article from Jan. 31, 1986 (P23), carries a photo of a number of HUT team members standing in the HUT conference room in Rowland Hall staring at a small B&W TV, watching coverage of the disaster. I have a curious memory of that time, whereby we thought there was still some chance that we might fly as scheduled, or at least with only a month or two delay. The article itself quotes Arthur as saying, "chances are slim" that we will go on schedule "but there has been no official change" yet. In the rearview mirror, it was more than obvious that we would not fly in March, but at the time we were so invested in the mission that we couldn't imagine not flying and somehow were holding out hope. Needless to say, NASA's hopes of observing the great comet Halley were dashed by this terrible event.

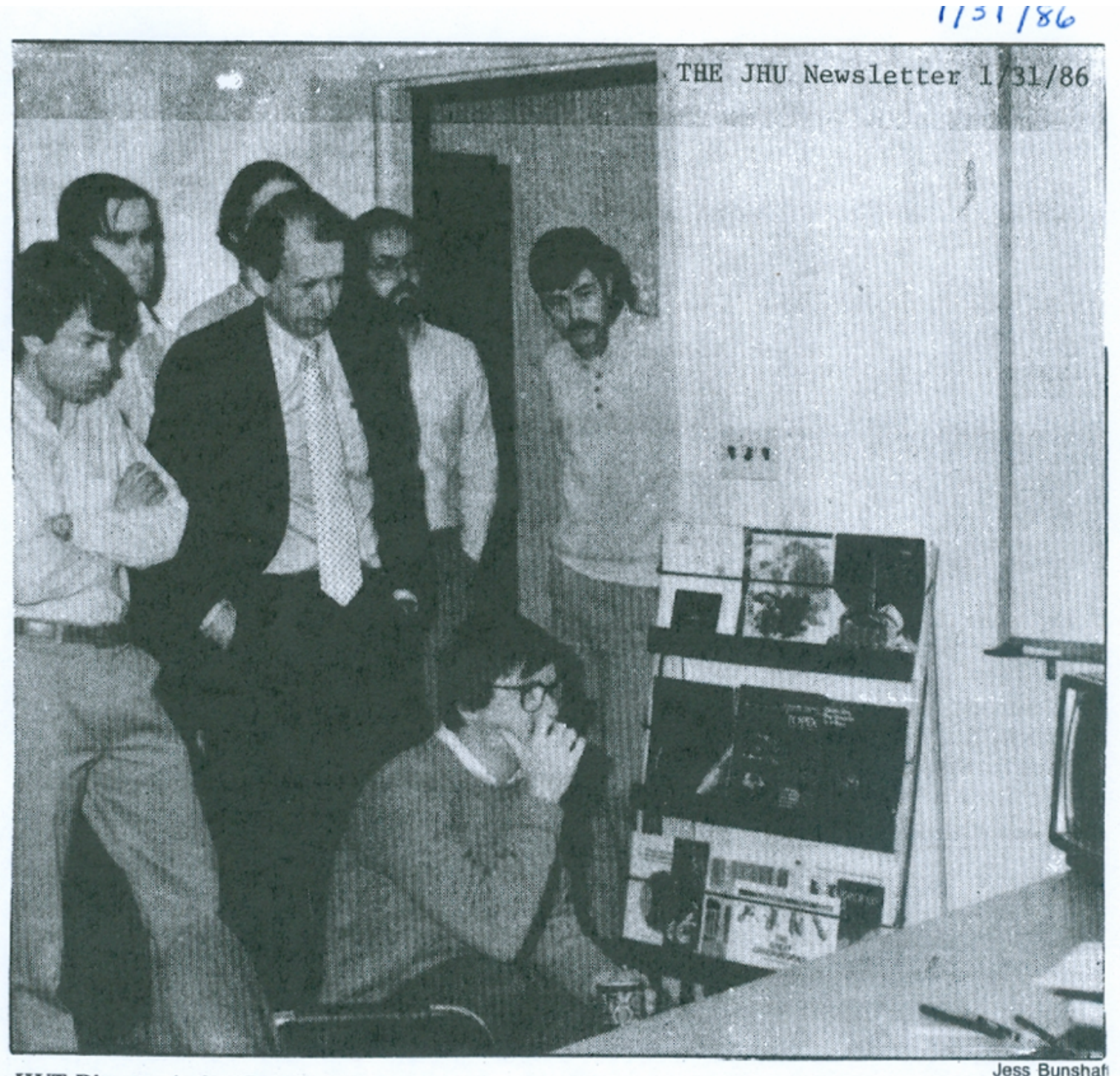

1/31/86

THE JHU Newsletter 1/31/86

Jess Bunshaft

HUT Director Arthur Davidsen and members of the Department of Physics and Astronomy who watched the fate of the Space Shuttle Challanger during its tragic launch on Tuesday. Davidsen's colleague Samuel Durrance, also from Hopkins, was scheduled to blast-off with the HUT* on the next space shuttle mission, although the scheduled launch will more than likely be delayed because of Tuesday's disaster.

**Shuttle Tragedy to Postpone March Blast-Off of HUT**

As a side note of interest, the Hubble Space Telescope was in its final preparation stages and prior to *Challenger* was planned for launch in the summer or early fall of 1986. Of course, this mission was also delayed, ultimately launching April 24, 1990.

One anecdote that has been confirmed by Scott Vangen[*6] deals with the post-processing of the payload. Scott, Michael Haddad, John-David Bartoe, and others were directly involved with the post-*Challenger* assessment of the Astro payload in the O&C building at KSC that showed there was a lack of clearance between modified Instrument Pointing System (IPS) bumper rings (that apparently were added late in the integration process) and the IPS itself as it moved into the deployment position. If we had launched in March 1986, the IPS would likely have bound in a partially deployed position. An unscheduled spacewalk might have been able to free it or it is possible it could have stuck in the partially-deployed state, resulting in the payload being jettisoned and the shuttle returning to earth. This would have been an unmitigated disaster for Astro-Halley, and thus for NASA, had it flown as planned.

*[6] Scott was an integration engineer at KSC for many years, and ultimately became an Alternate Payload Specialist for the Astro-2 mission.

# Chapter 2:  The Road to Astro-1

It was quite some time before there was any clarity about when the shuttle might fly again and what the fate of the Astro observatory might be.  In late May 1986, Arthur and his wife Anita invited the HUT team and spouses out to their house in Deep Creek Maryland for the weekend.  Included was a white-water rafting trip for those interested.  It was quite a team-building experience. (Note: there had been an earlier HUT team trip to Deep Creek back in the period when the Payload Specialist selection was still pending.)  Not one to let the moss grow under his feet, Arthur rallied the team to submit a MidEx proposal for SpEx, the Spectroscopic Explorer, which was submitted at the end of July 1986.  (It was highly rated but not selected.)

A Presidential Commission (The Rogers Commission) was formed and turned in their final report in June 1986 (see D16 which shows the first few pages including the Table of Contents for reference).*[7]    An Aviation Week article dated Sept. 1, 1986, announced that NASA "will cancel 15-18 Spacelab shuttle missions that were planned to fly in the next five years…", although later in the article it notes that the only Spacelab candidates left for the remainder of the decade include "an Astro ultraviolet telescope pallet and a life sciences module in 1989."  A mention in passing is made that a second Astro flight in the 1992-93 timeframe is "likely."  This upbeat assessment did not hold up for very long (see P30).

*[7] For an intense presentation of the backstory leading up to the disaster, the book *Challenger* by Adam Higginbotham is recommended reading.

In September 1986, Van Dixon began his graduate career at JHU and became a key team member throughout the remainder of the Astro-1 and Astro-2 missions, ultimately receiving his Ph.D. in 1996.  Van worked the replanning desk during both missions, and was a key player in programming not only planning software, but ultimately the data processing software as well.

A Baltimore Sun article in late October 1986 announced that shuttle flights would restart in 1988 and that Astro was manifested in a January 1989 slot (P31).  Hubble was slotted into November 1988.  Of course, both of those dates slid further into the future, but it showed significant priority on NASA's part to eventually fly HUT and the Astro Observatory.  However, the idea of any additional flights of the payload was up in the air, and certainly we felt the writing was on the wall for the idea a multiple flight sequence.  It is to NASA's credit that they continued to support the instrument teams during this interim period.

While awaiting clarification of status, many of the HUT contingent found meaningful alternative activities. JHU hooked up with the Carnegie Institution in the early stages of planning for a large optical telescope to be known as Magellan, and many of us took advantage of the connection by getting observing time at Carnegie's  Las Campanas Observatory in Chile a number of times in the 1987-1989 timeframe.  Grad student Harry

Ferguson worked with visiting astronomer Allan Sandage for his Ph.D. thesis during this time. Sam Durrance returned to Baltimore and got heavily involved in the early stages of the development of the technique known as adaptive optics (with David Golimowski and others). The APL crowd had other missions to work on before eventually coming back to work on HUT.

In parallel with this was a major proposal effort for a HUT-like free flier as a Small Explorer payload, no doubt in response to the fear that Astro/HUT would be a one-time opportunity. This proposal for SUSI, *the Small Ultraviolet Spectroscopic Instrument*, was submitted to NASA in September 1988. Ultimately, like so many of these major proposal efforts, the proposal was not selected despite being well received by the review panels.

Arthur Davidsen was also a co-investigator on the Faint Object Spectrograph (FOS), one of the original Hubble Space Telescope instruments, and Jerry Kriss and I worked to support the testing and calibration for that as well. Even though Hubble was nominally getting close to launch back in 1986, there turned out to be plenty of activity still needed to get it ready for launch even with an additional several years of delay.

Memorial day in 1987 found the HUT team once more invited to Deep Creek for a second round of team building and whitewater rafting. This time we went on the Youghiogheny river, which was quite a thrill! I was in a raft with Sam, who went overboard just about the time the "photo" of our raft was taken; I can be seen reaching over the side to grab him.

On October 5, 1987, there was a ground breaking ceremony on the JHU campus hill across the street from the Space Telescope Science Institute, at the location of what would become the new Bloomberg Center for Physics and Astronomy. This was a major strategic step, positioning the Department of Physics & Astronomy in close proximity to STScI and providing significantly more space for expansion of the department. Bloomberg included a high bay area and adjacent clean rooms, with an eye toward future potential instrument development activities.

At some point, Arthur had convinced the International Astronomical Union to hold their triennial General Assembly at the Baltimore Convention Center downtown in August 1988, no doubt thinking that HUT would have flown and Hubble would be launched by then. It was a major event with nearly 3000 participants from all over the world, and plenty of JHU and department resources went into coordinating it, once again raising the visibility of astrophysics at JHU. Arthur hired Karen Weinstock during this period to help coordinate this major event, and HUT business manager Harold Screen had a major role as well. Mary Romelfanger was hired about this time to oversee computer systems. Mary ended up staying through both Astro-1 and Astro-2 missions and was a key player in our data systems and data processing. Shortly thereafter, Jeff Kruk was hired in to work on HUT and quickly became a key team member.

## Back on track, sort of…

After a 32-month hiatus, the first shuttle launch occurred on Sept. 29, 1988, with STS-26*[8] on *Discovery*. By that time, there had been several official launch manifests with various dates, but with *Discovery*'s successful launch, the manifest took on more meaning, and hope for a flight of the Astro payload rose significantly.

*[8] In the interim, NASA simplified its mission numbering scheme so that missions were designated by STS and a number. But as planned missions moved around in the schedule, the missions were not monotonically increasing numbers, Astro-1 was designated as STS-35 and assigned once again to the oldest shuttle in the fleet, *Columbia*.

The unexpected appearance of supernova SN 1987A in the LMC in February 1987 provided a new target of high importance and was likely a strong driver for the addition of the Broad Band X-ray Telescope (BBXRT) to the reconstituted Astro-1 mission. BBXRT was run out of GSFC (as was UIT), but BBXRT was controlled on a separate pointing system from the UV telescopes and could choose to either point in the same direction as the UV payload or look at an offset sky position up to 37 degrees away. BBXRT and its pointing system would ultimately sit in the back of the cargo bay, behind the pallet containing the IPS and Astro UV telescopes package.

From an operations standpoint, the addition of BBXRT did not affect the astronauts very much because it was actually controlled by a team of scientists and engineers at NASA/GSFC. But it had a significant impact on the science planning in several ways. For one thing, the limited observing time (and control of the shuttle pointing direction) now had to be split four ways instead of three. The timeline planning now had to accommodate two interleaved observing sequences. If there was any good news out of this arrangement for HUT, it was that the sorts of targets BBXRT scientists were interested in (e.g., quasars and active galaxies) were typically of more interest to the HUT science team than many of the WUPPE and UIT targets were! Once the BBXRT primary pointings and targets were selected, the UV teams could decide whether to co-point or offset the IPS (UV telescopes) pointing to a different nearby position or target. Once the UV telescope primary pointings were defined, BBXRT could do the same thing. It was just another level of coordination that was needed in the planning process. Otherwise, the mission planning process was pretty much the same cumbersome system we had used previously, with minor improvements and somewhat better computers, software, and support.

The purported launch date of course slipped several times in 1989, from Jan. 19 to June 29 and then to Nov. 16. There continued to be some "hopscotch" in the schedule of Astro versus the Hubble launch where various manifests had one or the other payload launching first. This was partly due to other flights being delayed or rearranged for various reasons, which could affect your mission's launch date indirectly (because of the time it takes to turn a given shuttle around for its next flight – we were on *Columbia*). With DoD and planetary missions, some of which had various other launch timing constraints, some shuffling of the schedule was unavoidable. As missions went off and our turn got closer, a launch date of Mar. 1, 1990, was set, and we thought we finally had a believable launch date to plan for. (How naïve we were.)

Harry Ferguson completed his Ph.D. in October 1989 and left for a post-doc in Cambridge, England. Harry had been a key player with HUT, and was able to stay involved in mission preparations as well as coming back to work the actual Astro-1 and 2 missions in person.

There were numerous meetings and trips in 1989, including several Astro Investigator Working Group (IWG) meetings, a HUT functional test at the end of May, a Joint Ops test at KSC in early August, and various training and mission simulation trips during the fall. (See the HUT timeline file in Appendix A for details.) Of course, mission planning activities had been intense throughout the period, and the launch had slipped to an Apr. 21, 1990, planning assumption, with a notation that the “final” JSCIPLAN (Joint Science Plan, including the integrated IPS and BBXRT pointing schedule) was delivered to MSFC on Oct. 20, 1989. The Target Book was not delivered until early April 1990. Timelines and mission support products were still dependent on specific launch date and time, with little ability to gracefully adjust to changes in launch dates, many of which befell the mission throughout 1990.

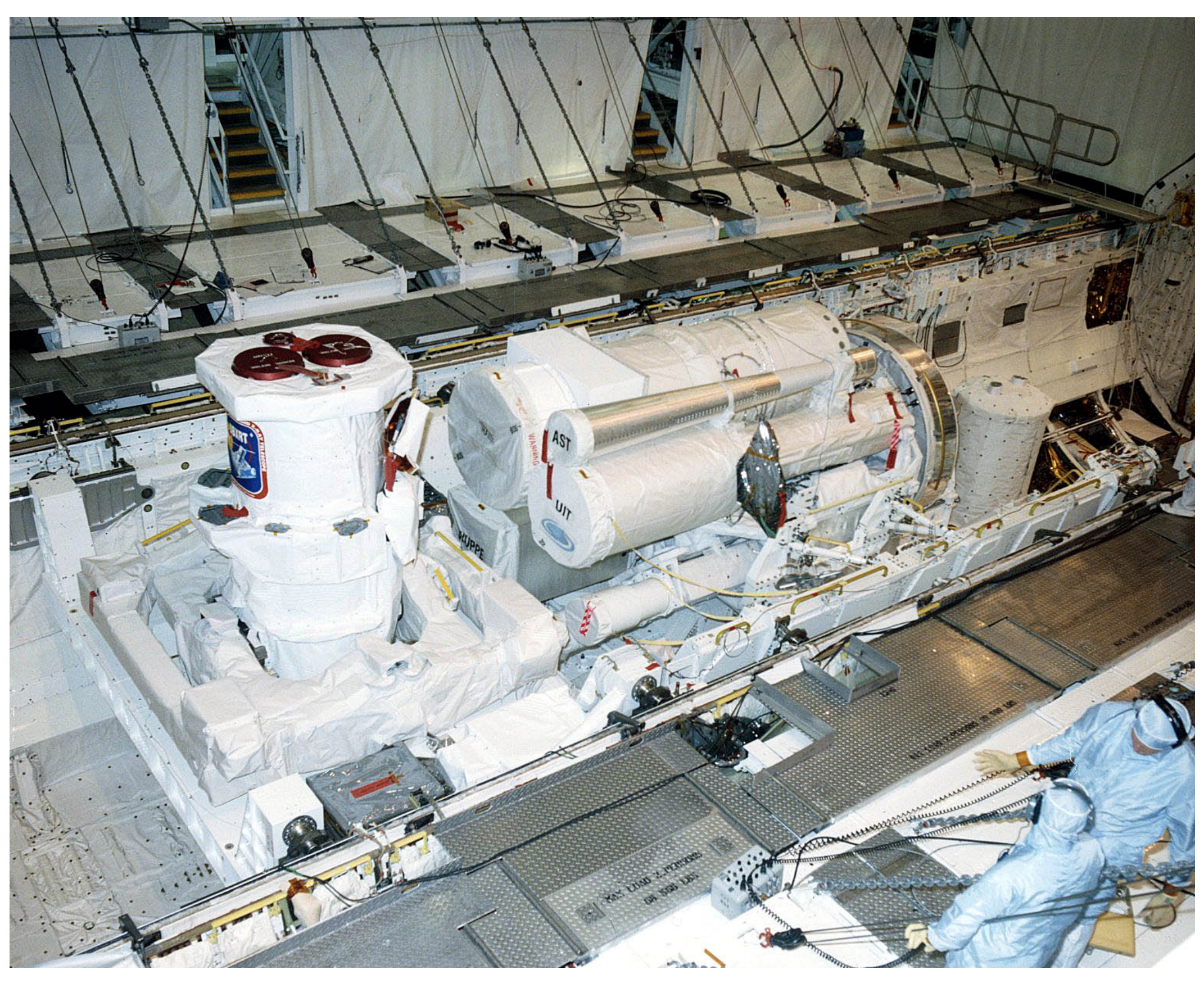

The integrated Astro-1 payload at KSC, ready for installation in *Columbia*.

By late 1989, NASA announced the decision that Astro-1 would be the only flight of the Astro payload. Arthur Davidsen wrote a strong, 3-page letter to NASA administrator Lennard Fisk expressing his disappointment and laying out the justification for additional flights, but to no avail. (See D18.) NASA funding constraints in combination with the greatly reduced launch schedule for the shuttle fleet in the post-*Challenger* period were the motivations for the decision. The old "cheap and easy access to space/multiple flight" model simply was no longer viable.*[9]

*[9] If one simply takes the annual budget for shuttle operations at that time and divides by the number of actual shuttle flights in a given year during this period, one finds that each shuttle flight cost approximately half a billion dollars, certainly not cheap! In retrospect, flying what had grown to be a roughly $100M payload where the "shipping costs" are $500M made for a tough sell.

The April 1990 JHU Magazine (no doubt written a month or two prior to publication) contained a long article talking with both Hubble staff as well as HUT staff about the upcoming launches (P33 covers the HUT sections of this article, quoting Sam, Arthur, and Steve Conard at length). Some interesting insights from this article include the following:
a) The March 1 date had slipped first to April 21, 1990, and then to May 9, 1990, making room for the Hubble launch scheduled for Apr. 12. The Hubble launch was delayed again and finally happened on STS-125, Apr. 24, 1990, on *Atlantis*);*[10] This then pushed Astro-1 to May 16, to leave enough time between launches (a minimum 21 days was mandated).
b) Arthur mentions ongoing lobbying activities with NASA and congress, trying save the idea of additional flights, and
c) Arthur mentions (without calling it by name) the importance of the observations of a high redshift quasar (HS1700+64), which is of course the He II Gunn-Peterson test to search for the intergalactic medium. To quote Arthur in the article, "If one experiment made me think HUT was worth devoting years to, that would be it."

*[10] HST was nearly launched on April 10,1990; the Hubble FOS team had scheduled a team meeting for KSC so we could watch the launch, but it was scrubbed a mere four minutes before launch due to a bad Auxiliary Power Unit. This was the closest I ever got to seeing a shuttle launch in person.

Early 1990 included a final mission planning coordination trip to MSFC in mid-January, a mission simulation exercise in mid-February, and the first full-up Joint Integrated Sim (JIS) March 20-22. A second JIS planned for the April timeframe was cancelled. I guess we were all SIMed out and just wanted to get on with it!

Ted Gull and I had written an article for *Sky & Telescope* magazine about Astro, and it was published in the June 1990 issue (although it came out in mid-April, timed for the "launch" – ha!). It is still a pretty good place to go to for a summary of the instrument characteristics (including BBXRT) and the planned science of the mission (see P35).

Toward the end of April when the May 16 launch assumption was still in play, there was an ominous medical issue that threatened to scrub Sam from the mission. He apparently developed a kidney stone! After what I am sure must have been a stressful time (not only for him, but for backup Ken Nordsieck!), the situation was rectified and Sam was cleared for flight. (See Balt. Sun article from May 4, 1990, which also mentions the current launch date had slipped to May 16, 1990.) This was the start of the tortuous Great Shuttle Leak sequence that eventually ended with the launch of STS-35 Astro-1 on the Dec. 2, 1990.

Relatively recently, I rediscovered a pair of web-articles on the site *AmericaSpace*, written in 2013 by Ben Evans, a writer who has written a number of carefully researched historical articles about various NASA missions, often using NASA records directly. In a two-article series "In the Shadow of *Challenger,*" he describes first the demise of STS-61E Astro-Halley (see P96) and then the sequence of events and delays leading up to STS-35 Astro-1 (see P97). Both are worth reading. The Table below, created from information in the second article, provides a synopsis of the incredible sequence of launch delays we experienced and their causes, starting from our first brush with a real launch opportunity in late May 1990 to the eventual launch on Dec. 2, 1990. Of course, as noted above, there had already been several launch slips earlier in the year. Selected articles from various press sources during this period are included in the scanned article archive. A headline in *Time* magazine in September intoned, "See you next leak!"

**Table: Astro-1 Launch Dates from May through December 1990**

| **Launch Date** | **Outcome** |
|---|---|
| May 16, 1990 | Faulty valve in freon cooling loop. |
| **May 30, 1990** | H leak in *Columbia* aft, from external tank disconnect hardware; launch scrubbed at T-6 hours. |
| **June 7, 1990** | Tanking on 6/6/90 show leak still there; leak diagnosis required return to VAB. |
| Aug. 12, 1990 | Two tanking tests in July showed "no joy"; Aug. 8 announced leak caused by faulty seal in the drive mechanism used to close the valve in the [external tank] disconnect; launched reset to 9/1/90. |
| Sept. 1, 1990 | Two days prior, faulty avionics box failed and was replaced, causing delay to 9/6/90. |
| **Sept. 6, 1990** | During tanking on 9/5, H leak again detected in the aft engine compartment, at roughly 10x allowed level. A second (different) leak was also discovered; H circulation pumps and bad Teflon seal replaced. Launch reset to 9/18/90. |
| **Sept. 18, 1990** | Tanking on 9/17 showed the leak still present at close to the previous level. Indefinite postponement. After further troubleshooting, a tanking test on 10/30/1990 confirmed "no leaks." A DoD flight was already scheduled for mid-November, so Dec. 2, 1990, was set for Astro-1. |
| Dec. 2, 1990 | Actual STS-35 Astro-1 launch at 6:49 UT. |

With the near-brush with launch in late May, we were afforded an opportunity to take what amounted to a full "team picture." Since all personnel supporting the launch were in Huntsville, we all gathered in front of the Payload Ops building where there was a scale model space shuttle, and took the now famous HUT team group shot. I set my camera on a chair we dragged out from the POCC and set the timer, so I could get in the picture as well.

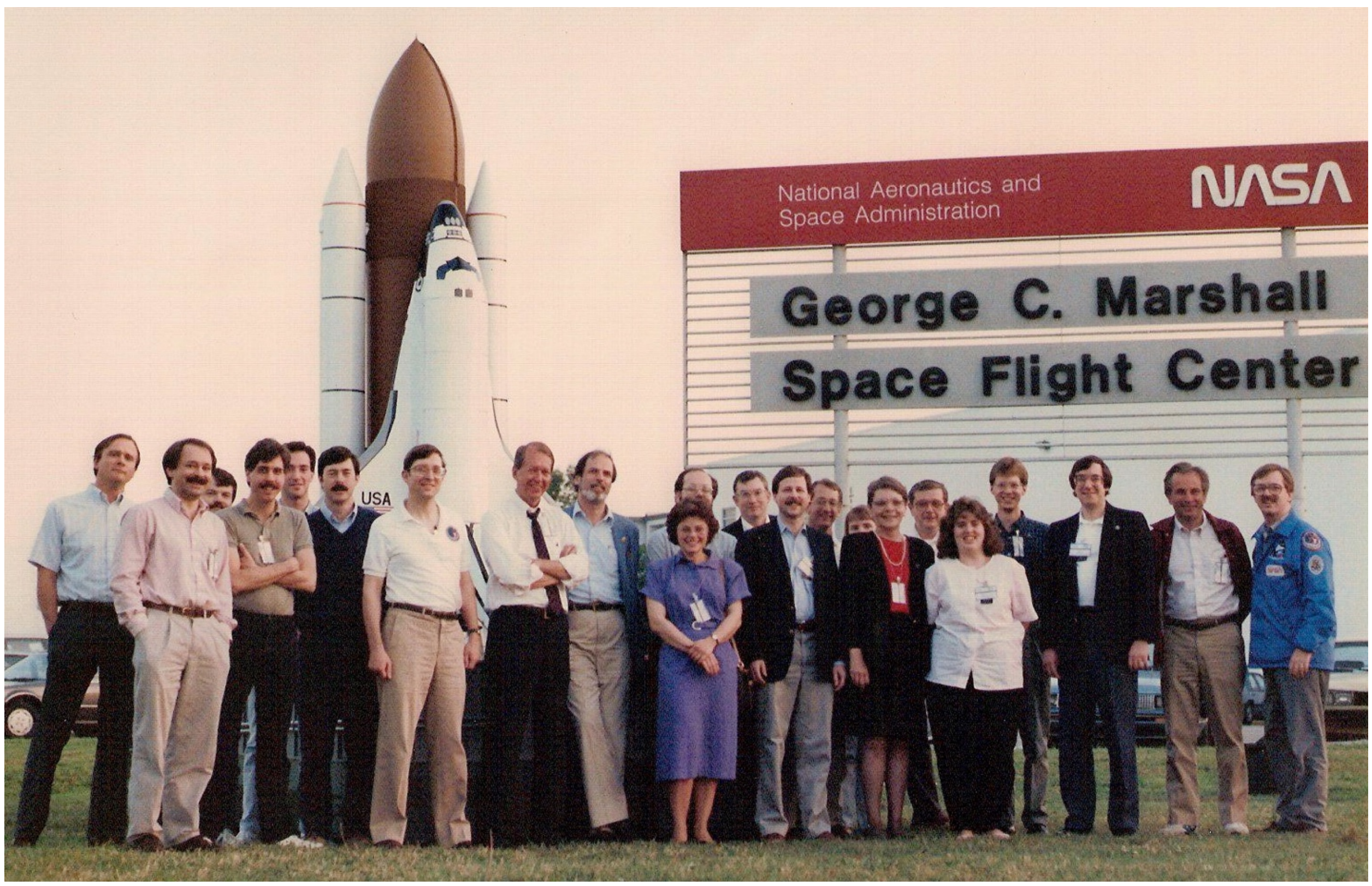

A nearly complete HUT team picture from May 30, 1990.

## Improvements to Planning

While it is entirely appropriate to have sympathy for the poor NASA engineers who were trying to diagnose the various shuttle problems, it is hard to describe how disheartening these launch delays were to those of us working mission planning, watching one carefully crafted observation timeline after another go down in flames. Various high priority observations came and went in various iterations of the observation timeline as the seasons literally changed and the targets available for observation came and went.

With so many "opportunities" to replan, we developed some ability to handle these frenetic situations with software, largely developed by Harry Ferguson. After crafting a timeline for a given future launch assumption, we found that we could slide that timeline forward or backward from the nominal time by days to weeks and make minor adjustments, keeping most of the observations in place. It wasn't a panacea—some things changed like the orbital day vs. night fractions of some observations or short observations scheduled near South Atlantic Anomaly passes, but by and large, it was a blessing for short term launch changes. The other thing the "launchslip" program allowed was studies ahead of time to assess what would happen to a given timeline for slips of the launch *time*, so we could know ahead of time what observations might be most adversely impacted. At the outside, we decided even a month slip was tolerable before the changes needed were too drastic. For slips larger than that, well, it was time to replan from scratch.

The summer of 1990 was a crazy time. With the slip from the May timeframe into August, we were again in replan mode. During a mid-June replanning trip to MSFC, we stayed an extra day in Huntsville to help populate the Science Operations Area (yes, SOA) in the POCC because President George H. W. Bush was visiting MSFC to give a speech about the moon-Mars program. He stopped briefly by the HUT area and shook hands with Arthur and chatted for a minute while the rest of us tried to look busy.

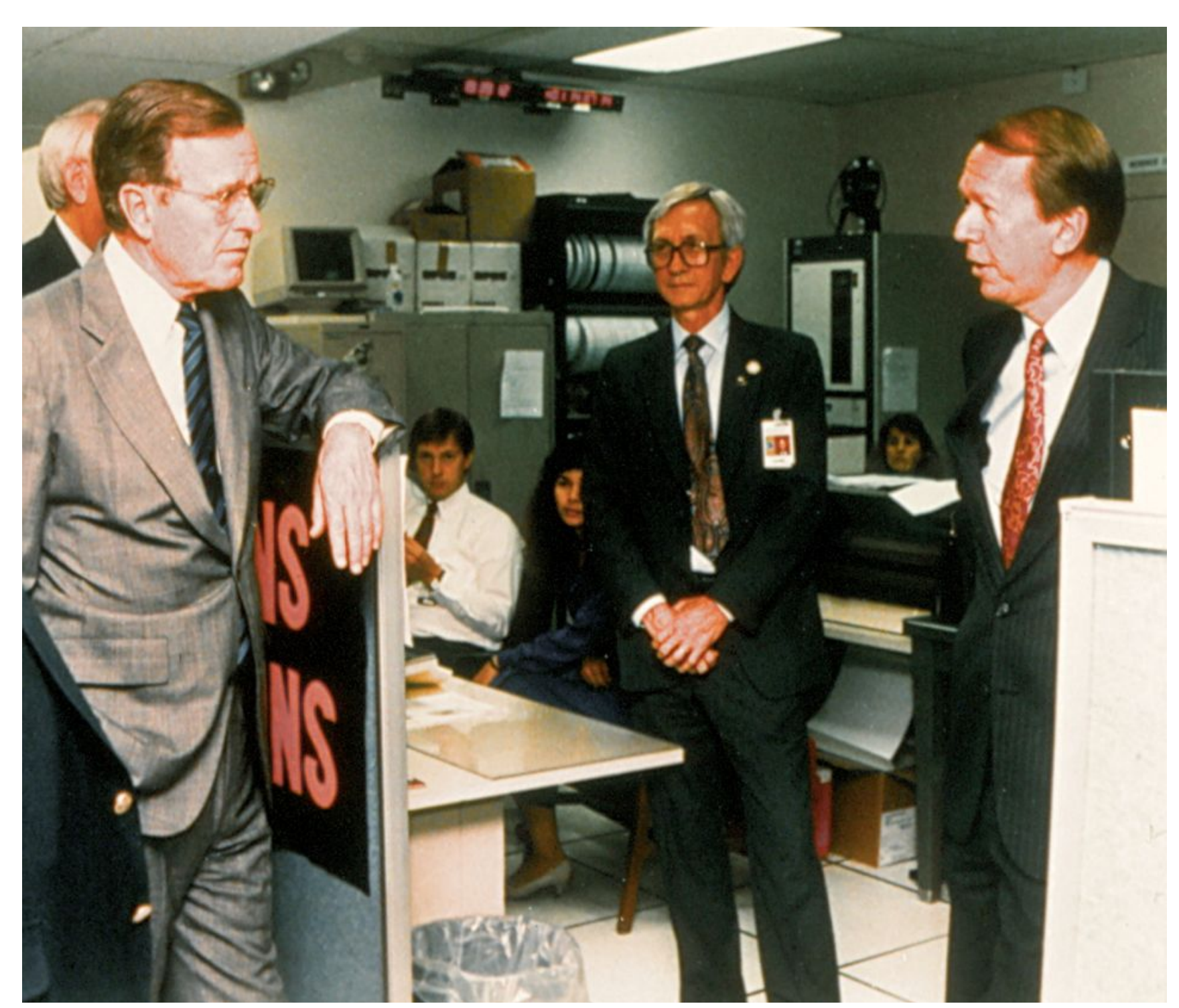

President George H. W. Bush (left) visits the HUT area of the POCC at NASA/MSFC. Mission manager Jack Jones is standing in center, and Arthur Davidsen at right. (Photo: NASA)

NASA's woes in trying to get our shuttle off the ground were only exacerbated by the official announcement of the famous Hubble primary mirror spherical aberration problem in late June 1990. Being involved with the FOS instrument on Hubble, we had heard rumblings about this earlier, but the official announcement was not made until June 27. The press had a field day bashing NASA, and there were many brutal political cartoons about both Hubble trouble and the shuttle leak problems over the ensuing weeks and months.

Finally, in the midst of all this turmoil, the new Bloomberg Center for Physics and Astronomy was nearing completion over the summer and our offices were moved from Rowland Hall to the new building at the end of July. I remember packing up my office in Rowland prior to a trip to Huntsville, and returning to unpack in the new building! The Bloomberg offices were beautiful, and very spacious compared with our previous offices, and with the change to distributed computing, almost everyone had their own Sun workstation on their office desk.

With a launch slip from the mid-August date to Sept. 1, a new Joint Integrated SIM was inserted into our schedule in mid-August. It had been a long summer, and "getting back into launch mode" and remembering how to operate in the MSFC environment was important. Of course, more launch woes lay ahead of us, but we didn't know it at the time.

A long (20 page!) article in the JHU Magazine, written by Robert Kanigel, was published in October 1991, well after Astro-1 finally flew (see P66). I mention this here because the first roughly half of that article covers the early days of HUT, focusing largely on Sam Durrance but recounting many aspects of the early days of the project. It is definitely worth a read if one wants more details of the backstory of this pre-mission period.

# Chapter 3:  The Run-up to the Astro-1 Mission

*(Author's note: My memoirs of the immediate period before the Astro-1 mission and the mission itself were first written (from copious notes and diary entries) shortly after the mission.  This chapter and the next assimilates the most important parts of those notes into this dialog along with additional comments and references added at this time.)*

"Where do I begin, to tell the story of the Astro-1 mission?"*[11]

*[11] To the tune of "The Theme from Love Story."

I begin in late September 1990, after our fourth launch attempt for the year had been scrubbed due to hydrogen leaks in the aft engine compartment of the shuttle *Columbia* (see P39).  Of course, these launch scrubs do not even tell the whole story of our woes, as many other launch date changes had occurred before we actually got into a launch configuration.  For instance, our official launch date of Apr. 26, 1990, had slipped to May 9, then May 16, and finally to May 30 before the shuttle leaks even came into the picture.  In Sept. 1990, Geoff Clayton from NASA HQ was hawking T-shirts that said, "Astro-1: Don't Worry Be Happy" on the back while on the front listing some 18 dates (dating back to 1983) on which the Astro-1 launch had been officially manifested by NASA.  I bought one.

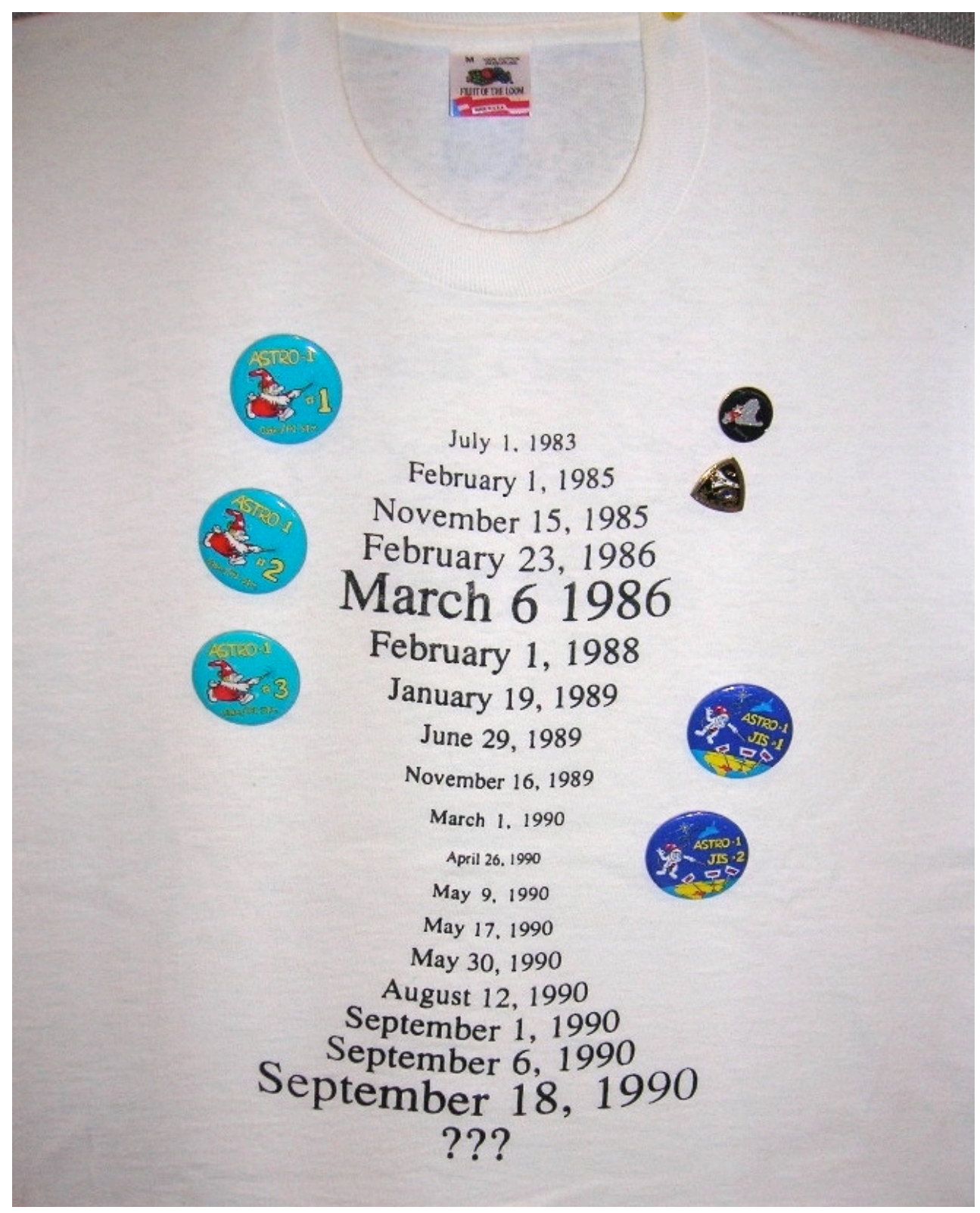

The launch date T-shirt

This latest launch scrub had left us without a viable timeline of observations. Timeline planning is a difficult task. First, it involves each of the four science instrument teams independently compiling a prioritized list of potential observations that they are interested in acquiring. This is a tedious and difficult task in its own right, since it depends on the detailed object visibilities, day/night breakdowns, target brightnesses, and science program assessments, all based on an assumed launch date. These inputs are all collated into one file and used as the basis for planning a "science timeline", which gives the order and start/stop times of each desired observation throughout an assumed (in this case) 10 day mission, using a certain launch date, time, and orbital configuration.

For Astro-1, this planning was further complicated by the addition of the Broad Band X-Ray Telescope (BBXRT) to the mission, since it was on a separate pointer, and hence could either choose to point at the same object as the UV telescopes, or at some other nearby object of more interest to their team. Hence, two interleaved timelines needed to be created in a single file called the joint science plan. The joint science timeline is then handed over to the NASA/Marshall planning team, who had to turn this into a shuttle maneuver timeline, checking thermal constraints on the shuttle, insuring that sufficient TDRS satellite coverage is available, and finding guide stars and roll angles for the Instrument Pointing System (IPS) to use at the time of each observation. It is also Marshall's responsibility to produce appropriate documentation of all this for the astronauts to carry on-board so they would know what is supposed to be happening (literally on a minute by minute basis) throughout the whole mission. This takes time. Even with all of our practice, this whole process took about two months of flat-out effort.

With our many slips of launch date, we had developed software to take a given science timeline and "slip" it by days, weeks, or (gulp) months to see what happened to the planned observations. It wasn't pretty. Because of changing visibilities of the targets, slips longer than about one month were in general totally unacceptable; the replanning became too complicated to deal with and it was better to plan again from scratch.

By the end of September 1990, we were once again in this situation. The timeline we had was planned in late June for the September 1 launch date. At the time, I swore that this was the last timeline that I was going to plan. Instead, it turned into "the best timeline we never flew."

To make matters worse, not only did we have to plan a new timeline, but we had no idea what launch date to plan it to! No one knew how long it would take to troubleshoot the shuttle hydrogen leak problem. However, because of the two month planning period required to produce a new timeline, we had to start planning right away. We chose to plan to January 1, 1991, the reasoning being that we could take this timeline and slide it forward or backward in time by a month, and cover the entire Dec. 1 to Jan. 31 time period. If we were delayed by more than that, we would have to replan yet again.

It was about this time that we started thinking about changing the words to Christmas songs to be appropriate for Astro. My previous attempt at rewriting the words for Paul Simon's "Fifty Ways to Leave your Lover" (I called it "Fifty Ways to Lose Your Pointings") had received critical acclaim from my co-workers. Here is an excerpt:

> *In Astro mission planning we all strive to not be crude.*
> *Furthermore we hope our changes don't get lost or misconstrued.*
> *But if we slip two hours on the pad we know that we're all screwed!*
> *There must be fifty ways to lose your pointings…*

My mood from the various launch scrubs is aptly recorded by text of "The Little Planner Boy" and "Working in the Shuttle Wonderland," both of which are included in the *Astro-1 Song Book* compiled after the mission. One of the better song possibilities that never got fleshed out was, "We'll be launched by Christmas…but only in our dreams!"

So it was time to replan once again. We spent the first week of October in lovely Huntsville, Alabama. The joint science plan was produced in record time--4.5 days--but only by spending some hellaciously long days in intense planning sessions at MSFC with representatives from all of the instrument teams. It's a record that will never appear in the Guinness Book of records, but it's one we worked awfully hard to set.

After handing the file over to the MSFC planners, our work concentrated on producing finding charts, choosing HUT guide stars, estimating count rates, and verifying coordinate accuracies for the roughly 230 observations that had been planned into the joint science plan. Plus, we had to analyze this timeline for the effects of potential slips -- which targets get too close to the sun for comfort or change visibility dramatically because they are close to the pole of the (assumed) orbit, and what objects should be planned in their place if needed?

Finally, by the end of October they put *Columbia* back together and performed a leak check (see P40), which seemed to indicate that the leaks were finally fixed and we were really flight-ready. On Nov. 19 the shuttle *Atlantis*, which had leap-frogged over us in the launch schedule, took off with a secret DoD (Department of Defense) payload. *Atlantis* had been delayed a week but because of problems with the payload, not the shuttle, so everything was looking up. We were once again next in line, and it was indeed "beginning to look a lot like Christmas" was when we were going to fly. Except for the downside of familial arrangements for the holidays, the closer we launched to January 1 the better because fewer changes would be needed to the timeline.

For reasons that were never explained to us, NASA decided to waive their ground rule that required a minimum of 21 days between flights and instead press ahead full steam toward an Astro-1 launch as soon as possible. Dates as early as November 30 were considered before finally settling in on late in the evening of Saturday, December 1 (technically early morning Dec. 2, 1:28 am EST). This left less than two weeks between the flights, and ironically also coincided almost exactly with the *full* moon*[12]!

*[12] As an aside, we had *always* lobbied for a launch near *new* moon to keep the moon out of the night part of the orbit. Our mission management always claimed they knew about this when we inquired about it, but NASA *always* ignored this request when setting launch dates for us! It wasn't a big deal, really, just one more detail that we had to worry about in the planning process. But it had become somewhat of a joke on "our" side of the project, and at some level was a good example of a systematic problem that we had all along in our dealings with NASA--the problem of communications.

## The Launch Slip Flap

As the early December launch began to firm up, we began to look more seriously at the launch slip situation for that time. Even with all of our previous experience with launch slips, this was a new situation for us because we had never had a serious occasion to slip a timeline *earlier* than the nominal assumed launch. Our initial look at the situation left the impression that things were pretty grim. We first moved the January 1 timeline back to December 1 at the nominal launch time, and it was about as bad as we had come to expect for a one month slip. But then we looked at what happened with slips on the pad by one, two or three hours and found to our surprise that these slips did not seem to make things very much worse. Hmm...

This got us thinking about how we were doing the slips (that is, in two steps). The way our "slip" program worked was to take an input timeline and an assumed new launch and orbit configuration and truncate observations that violate known constraints at the new time of observation. Hence, by doing these slips in two steps, we could be unknowingly truncating some observations that would be allowed for a late launch on the pad. We reran the slips starting from the January 1 timeline in one step, and we found an amazing thing -- *a launch ~2 hours later than the nominal launch time on December 1 minimized the changes necessary to the January timeline!* It was a very significant effect. Except for a few targets that became too close to the sun for the December 1 time frame, almost everything else was available for the planned observation times!

Without going into the details, this change would have recovered about 100 ks of observing time (about 30% of our total) that would have otherwise been lost. In retrospect, some of us that had been mission planning for years kicked ourselves for not realizing this earlier. Slipping a launch date earlier in time and slipping the launch time later from an assumed time offset each other and nearly preserved the visibilities of the objects to be observed! It was a miracle in the making.

It didn't help much that we realized this on Friday of the week before Thanksgiving. We had to inform the other teams of what we had found, and try to convince the NASA bureaucracy that it was necessary to change the launch time, even though we were less than two weeks from launch! It was a battle we didn't think we could win and yet we had worked too hard for this mission to let such a significant potential improvement to the scientific return slide by without trying.

In many respects, the battle that ensued typified many of our dealings with NASA and MSFC throughout the program. In a telecon with our Mission Scientist, Ted Gull, and the other instrument teams on Monday, Nov. 19th, we went over the case for the "late launch" and easily convinced people of the scientific case for doing it. On Tuesday afternoon in the normal weekly "PI telecon" (which included representatives from all of the various factions of Astro, including instrument teams, MSFC, NASA HQ, etc.), we made our case to the mission manager and NASA folks. There was certainly some resistance to the idea during this telecon, but I thought we made the case effectively. We were left with the task of supplying our arguments in the form of a "position paper" to the mission manager and NASA HQ, which we did by Tuesday evening (under great duress).

With the Thanksgiving holiday upon us, things were left somewhat up in the air. Clearly we needed a quick decision one way or the other so that we knew how to plan. We heard through the rumor mill that a NASA telecon was being held on Friday Nov. 23rd after lunch and that some decision would be forthcoming. Interestingly, no science team representation was sought for this telecon, and we knew that the two most likely NASA reps to make the case for us were away for the holiday weekend. Hence, we were not too surprised to find out that they turned down the idea; the only possible people on the telecon that could have argued in our favor were the MSFC representatives who we knew were against the idea! It was also interesting that we never received an official notification of the decision. The only thing we received was a forwarded e-mail communication from someone at MSFC to someone at HQ discussing the probable results of the telecon.

The reasons for being against the late launch were never successfully communicated to us. A few reasons put forward were totally ludicrous, like "Our analysis shows that there would be a 6% decrease in the chance for launch because of the decreased launch window." The closest thing to a reason that made sense was that they did not have time to update the various documentation that was going onboard, and they were afraid of the confusion that this might cause. This would have been a valid point were it not for the fact that the documentation onboard was *already out of date*--it was for the January 1st launch assumption and was going to have to be updated in real time anyway! I'm not so much saying that they made the wrong decision as I'm saying that they never made a clear case for their decision to the science teams.

Although we realized at this point that the battle was probably lost, we did push it a little further. Basically, HUT Principal Investigator, Arthur Davidsen, used some of his contacts to get in touch with several people high in the NASA decision-making hierarchy to see exactly how effectively our case had been communicated. The answer was surprising even to us who have become wary of these things: The people making the decision had no concept of why the late launch was desirable or what the scientific tradeoffs were. They had been fed the party line and simply reacted to it. Arthur did his best to make the case over the phone, more for information than anything, and was left with the impression that they might even go so far as to review the decision.

Well, this certainly caused a stir. This approach was unorthodox and it caused more than a few feathers to be ruffled. I don't know the details of who said what to whom, but it was clear from various rumors and several phone calls with contacts at MSFC that the mission management folks were hopping mad at us. There were a number of misunderstandings and miscommunications up and down the ranks which were just accentuated by the fact that most of the information was being transmitted second or third hand at this point. We spent a rather large fraction of time the week before launch just trying to trace down some of these misunderstandings and patch things up, instead of concentrating on preparations for the mission. The really sad fact in retrospect is that none of this made a hoot of difference in the outcome of the mission. We were so naïve.

# Chapter 4: Onward and…Upward! The Astro-1 Mission

*(Author's Note: For the mission itself, I will relate my recollections as a day-by-day diary of sorts, with dates added for reference. Since I was working the "night" shift, the days I refer to are skewed. For example, when I say "Wed. Dec. 5," that is late Wednesday through Thursday morning when my "day" ended.)*

**Wednesday, November 28, 1990:**
Most of the HUT team arrived in Huntsville late in the afternoon. I had worked into the wee hours of the morning the night before trying to tie up all of the loose ends that were not only needed for the mission, but also were going to be needed very shortly after we returned. (I was realistic enough to realize that if we had a successful launch and mission I was not going to be in any kind of shape to do anything between when I got back from the mission and Christmas.) These tasks included finishing an IUE proposal (due Dec. 15), writing a revised science justification for my *Hubble Space Telescope* proposal (due Dec. 14), and writing some letters of recommendation for a grad student, as well as getting the appropriate materials together to take to Huntsville. I managed to drag home at 4 a.m. and got four hours of sleep. At least it got me partially adjusted around to a night schedule since I was working the night shift for the mission.

After I got up, I packed and then took wife Jean and toddler daughter Amy out for breakfast at Bob's Big Boy. It was a bittersweet time, the anticipation and excitement of a possible launch, the possibility of being so close to "the event" after such a long road, blended with the imminent two-week absence from each other and the extra heavy burden that I was leaving on Jean. Jean decided to drive me out to the airport, and we swung by the office about noon to pick up Jerry Kriss and Van Dixon. We took the (by now) familiar American Airlines flight through Nashville to Huntsville, arriving about 4 p.m. CST under overcast but warm skies. I was supposed to get a midsize rental car, but they must have been out because they gave me a huge gray Ford Crown Victoria! It had a tape deck, which got put to good use over the next two weeks; I had brought a collection of Christmas tapes along just in case this happy coincidence might occur.

After checking into the Amberley Suite Motel, several of us decided to have a nice dinner at the Steak and Ale restaurant. This was our traditional dinner spot *after* launch scrubs, and we decided to try and break the jinx by going there before we got any closer to launch. After dinner we took about an hour to get settled in at the motel, then headed out to the Payload Operations Control Center building (the POCC) about 11 p.m., stopping at security on the way in to get our badges and access cards. Amazingly, I didn't have much trouble staying awake that first night. There was enough to do to keep us all busy, unpacking boxes, testing software, and organizing the paperwork. We were comfortable with this part of the activity since we had done it four times before! We left the POCC about 7 a.m., and I had breakfast at the Amberley with some of the day shift people that were just getting ready to go to work.

I got to bed about 9:30 a.m. and slept straight through until 4:30 p.m. Not bad, 7 hours. Little did I know at the time that it was the best sleep I would get for the next two weeks!

**Thursday, November 29, 1990**

It was hard to believe, but we were two nights away from our nominal launch. A group of HUT folks had dinner at a middle eastern restaurant called Sahara's. We always joked about this place being a front or a tax write-off for some rich Arab because there were never any cars in the parking lot and there was almost never anyone there besides us when we went there. This night was no different. They also played this tape of Middle Eastern musac too loud over speakers which always seemed to be carefully placed right over wherever I was sitting.

We got out to the POCC about 8 p.m. to relieve the Blue shift. Many of the MSFC folks worked 8 hour shifts, which were called the red, white, and blue shifts. The astronauts and MSFC planning people worked 12 hour shifts, and called them the red and blue shifts. We nominally worked 12 hour shifts as well, but the handover times were offset from the astronauts schedule by about two hours. We called our shifts Black and Blue, in rough accordance with Hopkins school colors, but with an obvious second meaning. Van Dixon had spent most of the day changing one of our programs to allow us to more readily enter changes into the JSCIPLAN (Joint SCIence PLAN) file. This was the file that showed explicitly what each of the two telescope pointers were observing as a function of time throughout the mission. (Recall that the X-ray telescope was on a separate pointing device from the UV telescopes, and could choose to point at the same target, or a different target within 37 degrees.)

My job for the night was to take all of the changes that had been proposed to "fix" the Jan. 1 timeline for the Dec. 2 launch, and enter them into the JSCIPLAN with Van's program. This was a crucial step to verify that the proposed changes would really work, and to provide all of the instrument teams with a plot of what we thought we were really going to observe. There were many changes necessary to "repair" the timeline; recall that a month is about as far as we liked to slip a timeline, and we were right at that limit. There were close to 100 pre-mission RRs [replanning requests] in the system addressing changes needed to the official Jan. 1 timeline to make it appropriate for Dec. 2. This nominally should have been done before this time. The fact that it was not was entirely due to the flap about the launch time. Indeed, slipping the launch later would have obviated the need for many of these RRs in the first place! Now we were behind the 8-ball. I worked on this *all night*, literally only getting up to go to a status briefing at 1 a.m., and got about 70% of the changes made. Van's program worked, but in a fairly cumbersome way that took significant time and concentration to avoid making mistakes.

My other task for the night was supposed to be to write a position paper for Ted Gull (our Mission Scientist) on how we calculated visibilities of targets (relative to how MSFC does it). Part of the fallout from the "two hour launch delay" business was a misunderstanding that left the wrong impression there was an inconsistency in the way this was done. This was totally unfounded, but Ted was getting a lot of grief from the mission manager's office about it and he wanted to put it to bed. I had told Ted I would have it for him Friday morning, so I wrote a memo about target visibilities from 7:30 to 9:00 a.m., and finally left for the motel about 9:30. Whew! A 13.5 hour shift--what a way to get adjusted over to a night schedule!

**Friday, November 30, 1990:**
Unfortunately, I didn't sleep all that well during the day on Friday. I woke up about 2:30 p.m., and had phrases of what became "Twas the Night before Astro" floating through my head. I had gotten the idea of rewriting Clement Moore's famous poem as I was driving back to the motel in the morning, and as was often the case with these songs and poems I rewrote, my brain just wouldn't drop it! I finally got up and started writing some of the phrases down just to get them out of my head. We had dinner at TGIFridays, and people were fairly pensive. I actually finished writing the first draft while waiting for dinner to come, and ran it by Knox, Jerry, and Steve Conard (who were sitting at my table). They got a real kick out of it, so I decided it was a keeper.

Dinner went faster than expected, and we actually got out to the POCC early that night, about 7 p.m. It was one night to launch and counting (assuming everything went on schedule), so everybody was fairly pumped up. Even though we had been this close to launch multiple times, it seemed different this time having some confidence that the leak problem was fixed. At least if we were going to get delayed again it was because of the weather or some other problem!

Warren Moos (a Co-I on the HUT project and the other replanner on my shift) arrived that evening, so I split time between getting him up to speed on planning issues and trying to finish fixing up the JSCIPLAN for Dec. 2. On the day shift, Van had spent most of his time improving the program that edited the file to make it faster and more user friendly, but hadn't gotten around to finishing what I had started last night. Also, there were a number of small holes in the timeline that we might have been able to fit something into. Everyone was pretty much in place now, with the exception of PI Arthur Davidsen who was in Florida for the launch (see 1990-09-17 Pre-launch letter from Arthur Davidsen.pdf for context about that), and we were settling in to our patterns. I had some breakfast at the Kettle (a 24 hour restaurant adjacent to the motel) and tried to call Jean (she was out) before turning in about 9:45 a.m.

**Saturday, December 1, 1990:**
I awoke at 4:30 p.m. and called out to the POCC to see what the status was. The countdown was going smoothly, but the weather sounded touch and go, officially 30% chance at the beginning of the launch window, increasing to 50% by the close of the window a few hours later. (It would be fine with us if the weather delayed us a couple of hours!) I also got ahold of my wife Jean, and she said to call her if it got off, even though the launch was scheduled for 1:28 a.m. EST. Several of us had dinner at Chili's, and had a good time kidding around and releasing nervous tension. As we were finishing up dessert, Paul

Simon's "Fifty Ways to Leave Your Lover" came on over the musac speakers. We hoped it wasn't an omen!*[13]

*[13] Recall I had rewritten this to be "50 Ways to Lose Your Pointings."

We got out to the POCC about 7 p.m. again. I ran the ephemeris targets (objects whose observations were time critical) through our programs to determine what happened to them for various slips. The weather continued to sound marginal, but the mood was still upbeat. Most folks from the day shift straggled back in after they had dinner to watch the launch with us. No way they could go back and go to sleep! As the time of launch approached, they held at T - 9 minutes (the usual place) as they continued to monitor the weather. On the loops, there seemed to be some confusion as to whether the weather conditions were go or not. The Air Force said "no" because they weren't convinced the 8000 foot ceiling requirement was met. The launch director got them to dispatch a helicopter to go see. After holding for 21 minutes, the all clear was given and the countdown restarted! Launch occurred at 6:49:01 GMT = 12:49 CST/1:49 EST.

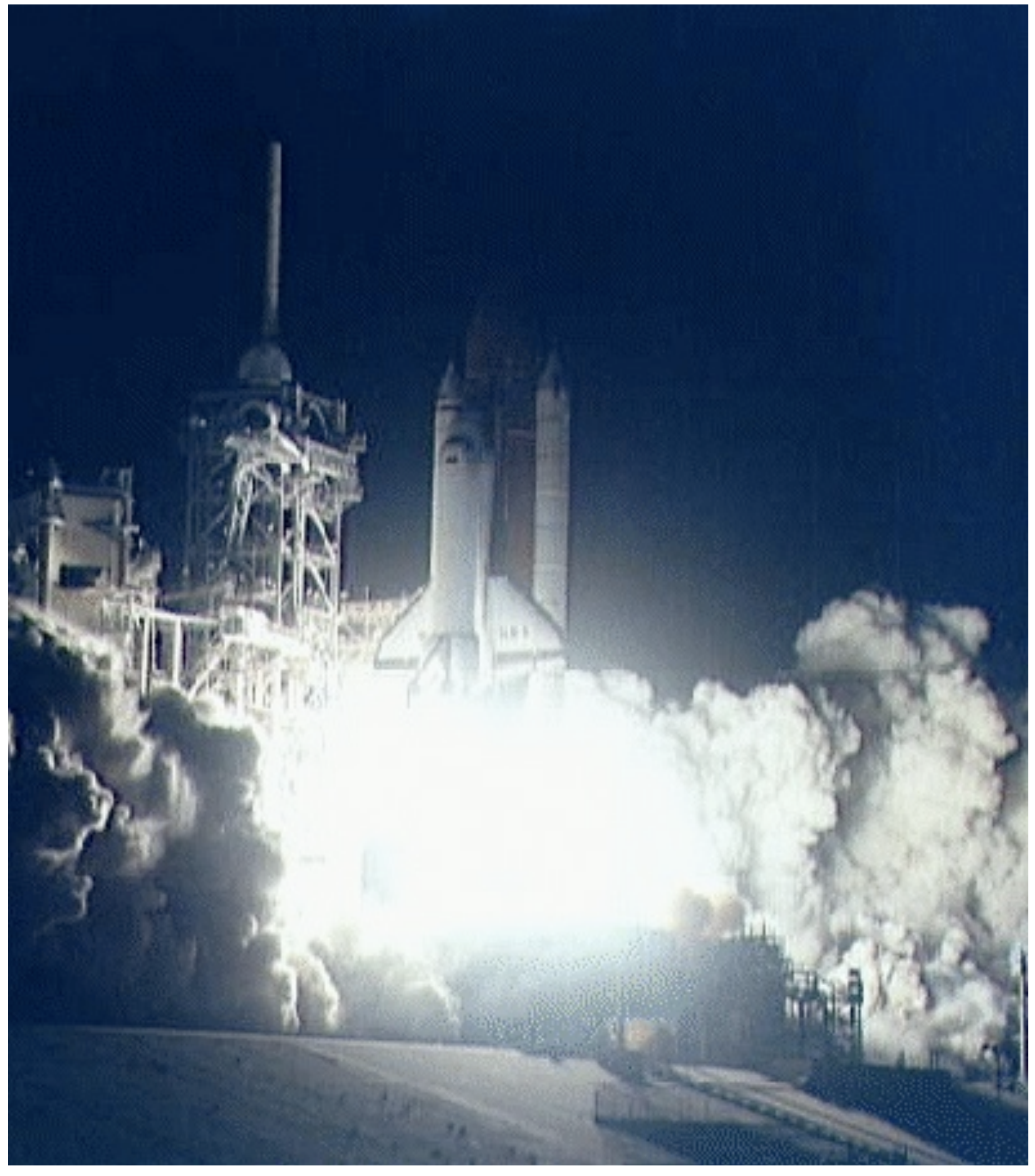

Astro-1 STS-35 Liftoff at 12:49 CST December 2, 1990.

## The Mission Itself…

The launch was picture perfect, absolutely stunning. Night launches are spectacular anyway, but this one was a sight to behold. (See 1990-12-03_BaltSun_article.pdf.) With all the delays and uncertainties we had faced (including the marginal weather), we could hardly believe it when it finally occurred! Even with the broken clouds the cameras were able to follow it all the way up, and we got a clear view of booster separation. The nominal orbit achieved was perfect – the initial numbers indicated we were within about 50 meters of the expected altitude. After things calmed down, the shift basically went well. We ran our launchslip program for the real orbital information and launch delay, and submitted the appropriate RRs for the current replan shift (which is always 12 hours ahead), based on our pre-mission analysis. There were some start up problems as the instruments were activated. For instance, the PS (payload specialist) could not see HUT TV images on their screen at first even though we could from the ground. This was resolved just before the morning shift change. I went to the morning planning meeting, then had breakfast with Warren and called my Mom before turning in at 10 a.m.

**Sunday, Dec. 2, 1990:**
I woke up at 4:30 p.m., having slept like a brick for the first time in several days. At dinner we found out that the activation activities were about three hours behind schedule, that the IPS (Instrument Pointing System) was having problems, and that WUPPE (the Wisconsin telescope) might be dead! In addition to that, at 10 hours into the mission one of the two DDUs (Data Display Units) on the shuttle had overheated and shut down.*[14] Wow, we were off to a roaring start! However, we had received first light! The slit wheel moved, the HUT detector was alive, and we could see emission from the earth's residual atmosphere. HUT was alive and well! (Now we just had to get onto some real targets!)

*[14] Normally one DDU was used by the Mission Specialist to control the IPS and the other was used by the Payload Specialist to set up the UV instruments and start the observations. With only one DDU, they had to waste a lot of time switching back and forth between IPS and instrument control modes as well as the unfortunate logistics of one guy being in the other guy's way all the time.

When we arrived at the POCC, the news was more encouraging about WUPPE, which was in the process of overcoming its problems. It turned out that one of WUPPE's Dedicated Experiment Processors [of which it had two] was fried when it was turned on, but the fault may have been an error in the turn-on sequence, not the fault of the instrument. Some Spacelab heaters were supposed to be turned on before WUPPE was activated, but they weren't. Luckily, they could operate off of the backup DEP-B.

However, the rest of the shift was pretty discouraging. Things were so behind the timeline that no one knew an hour ahead what was going on. It was clear listening on the communication loops that everybody was thrashing around trying to figure out what was going on. Since we had missed some crucial activation activities and because we had a number of high priority targets early in the nominal timeline, we had a busy shift looking for the best places to put these targets back into the near term timeline. However, it was clear that the IPS was continuing to have problems locking onto and holding guide stars. Until that got fixed, very little else mattered because we were continuing to miss every observation.

We had a long SOPG (Science Operations Planning Group) meeting in the morning to try and put together a strategy that would get us back on track. We were more than 30 hours into the mission and (nominally) should have been well into the science timeline, but had not accomplished most of the activation and early calibration activities. As I left the POCC that morning, it was gray, cold and raining hard. I got pretty wet on the way out to the car, and as I sat in the car getting ready to go, I had a passing thought about how the weather seemed to be tracking how things were going so far. It had gone from partly cloudy at launch to gray and overcast, and now to pouring rain. It fit my mood perfectly. I called Jean when I got back to my room and didn't get to bed until 10:15 a.m.

**Monday, December 3, 1990:**
On Monday afternoon, I got up at 4:15 p.m. and decided to turn on the TV and watch NASA Select (NASA's cable channel) that was carrying the mission live.*[14] Someone had figured out how to twiddle the cable controls on the motel TV sets to get NASA Select to come in, although it was a little fuzzy. As the TV came on, there was a HUT spectrum! They had acquired an active chromosphere star called HR 1099, and although it was bouncing in and out of the slit, it was bright enough that they were able to get a spectrum. All three UV instruments were in ` observe' mode for the first time, so maybe things were looking up.

*[14] As an aside, the SOA (science operations area) was inundated with NASA TV people, and cables were running all over the place. With such long shifts, we often had food available right in the HUT area. You had to watch out that you didn't have a TV camera pointed in your face before grabbing a donut and stuffing it in your mouth.

We got out to the POCC about 8 p.m. after another wonderful dining experience at Chili's, only to find out that one of the Spacelab computers had crashed. As the shift developed, it became clear that this was a significant impact. It basically put us back into a "useless" mode of operation where the IPS couldn't track or hold onto guide stars again. Some of the precious and painfully obtained IPS calibration data had been lost in the crash. We didn't know it until much later, but this was a JSC blunder -- they actually uplinked some stuff to the wrong computer! It was probably just as well that we didn't know this at the time; someone could have gotten hurt.

We worked very hard this shift to get Capella (our spectrograph focus target) scheduled into the end of the *current* shift so we could try to run our focus sequence again. (We normally replan the timeline for the period 12 - 24 hours after *the end* of the current shift!)

However, by the time of the observation, the IPS was still not performing properly. We did get some data on Capella, but not very much and not any data with sufficient stability to allow us to check the focus.

Even all of this craziness was not enough to stifle my creativity. It was about this time that I penned "Don't Rest Ye Mission Planners" (to the tune of "God Rest Ye Merry Gentlemen") and "We Planners from Baltimore Are" (to the tune of "We Three Kings..."), which aptly reflect the mood at the time of their writing. For instance:

*Don't rest, ye mission planners, there's too much to be done.*
*We're missing targets right and left. This isn't very fun.*
*Until the IPS can calibrate we'll miss them every one.*
*Oh tidings of discomfort and woe, woe for Astro,*
*Oh tidings of discomfort and woe...*

*We planners from Baltimore are,*
*Replanning and writing RRs.*
*IPS is shaking, nerves are breaking,*
*Searching for yonder star. Ohhhhhhh...*
*Star of Wonder, star of light.*
*Star with UV light so bright,*
*Slitward leaning, then careening*
*Off the slit and out of sight!*

I went to the morning SOPG meeting, then got drafted to stay for a special IPS meeting to get more information on where they stood. They had some more tests scheduled in the next shift that were supposed to get them back on track. Hmm...

I was hungry by the time I got back to the Amberley at 10:45 a.m. so I had breakfast at the Kettle. Greg Madjeski of the BBXRT team came in and sat with me. They had been having problems as well determining the alignment of their telescope with their pointing system (called TAPS, Two Axis Pointing System), although they had gotten some data. All in all, things had been real frustrating to this point. The good news was that it had finally stopped raining, so there was hope our situation would improve as well! I got to bed about 11:30 a.m.

**Tuesday, December 4, 1990:**
I awoke at 5 p.m. to a call from Randy Kimble out at the POCC. Things had been working at some level during the afternoon and they were coming up on the CYGLOOPA observation (one of my highest priority observations). We doubled checked a few things about the set up of the observation, and I wished him luck! I ate at O'Charlies with Glen Fountain and Steve Conard, two of the engineers on the project. It was refreshing to a) eat in a smaller group for a change, b) eat with some people who were interested in a different aspect of the project from myself, and c) eat at a different restaurant from where we had been eating!

I arrived at the POCC about 7:45 p.m. It was a beautiful, dark, clear evening. The moon, which was full the night we launched, had not risen yet and Orion (with Mars above) was hanging beautifully in the southeast. The stars of Orion are on the Astro-1 Mission patch, and with the "weather" connection I mentioned earlier, I had a real positive feeling that things were about to start looking up. The first report on the CYGLOOPA observation was good, but they were not in contact with the shuttle and we had to wait awhile for the data to be downlinked and "replayed"; it would be several hours before we knew what we had. In a way, I was happy because I wanted to be there when the data for which I had been waiting 6-1/2 years finally came in. From what they could tell in real time, though, the acquisition went well and Sam hand guided the observation, holding the target within about 2 arcsec for most of the observation. Hence, while the IPS was still not locking onto guide stars, they had at least been able to increase its stability to the point where hand guiding could keep the object in the spectrograph slit. Although a few targets were observed that afternoon, this was the first important HUT target to be (successfully?) observed (some 65 hours into the mission!), and everyone was quite excited. We also had another important target, the white dwarf star G191B2B, coming up a couple of hours into the shift.

About an hour into the shift, our data evaluator, Prof. Dick Henry, realized that there were some Cygnus Loop data files on our data analysis computer! Actually, about the first 600 seconds of Cygnus Loop data had been downlinked before the signal was lost, so we had some data sitting there on disk! I swallowed hard as he displayed the "first look" spectrum and…there it was! It included the acquisition period and was all "daytime" data (so the airglow contamination was bad), but I could see the salient features of the spectrum even with these drawbacks! It was quite a moment. Unfortunately, I barely had time to enjoy it at this point. I was in the middle of a replanning activity that had a deadline, and the real time G191B2B data were about to come in, which had to take precedence.

Well, the white dwarf spectrum came in, too! We really felt like we were on a roll. As luck would have it, though, there were no more high priority HUT targets on our shift, although we continued to obtain spectra at every target position. Then the mood was altered by additional "hardware" problems which occurred in close succession. First, the HDRR (High Data Rate Recorder) failed, which was the best way (in terms of speed and flexibility) of storing data for periods when no TDRS contact was available. Luckily, there was a backup system, the PFDR (Payload Flight Data Recorder), but it was much slower than the HDRR and would not hold as much data. Next, the TAGS (Text And Graphics System--basically the shuttle's FAX machine) failed. This was the primary way we uplinked new information to the crew, including new target book pages with revised observing information. Because of subsequent problems, this didn't cause as much grief as it could have, but it was still bad news at the time. Also during this shift, more IPS testing was done trying to get the data they needed on the ground to figure out what the problem(s) was (were). It was amazing that we were getting any data at all! We were getting real tired of giving up observation time to IPS tests when they didn't seem to be making any progress.

I talked with our press relations person, Lisa Hooker, in the morning and brought her up to speed on the evening's activities. When Arthur Davidsen arrived for the morning shift, he asked if I would stay after the SOPG and tape an interview with Byron Lichtenberg to be used that afternoon on the "Today in Space" program on NASA-Select, which I did. Also, Randy took the G191B2B spectrum up to the 9 a.m. press briefing, and did a great job. It was 10:30 a.m. before I got back to the motel.

**Wednesday, December 5, 1990:**
I got up about 5 p.m. and my message light was lit. The folks at the front desk said I had a package. It was a CARE package from home, including fudge and M&M cookies! I had "breakfast" at the Olive Garden (Italian) and it struck me that I had been having some of the most bizarre breakfasts that I have ever had since breakfast was at normal dinner time!

The report at the POCC was basically good. They had gotten some objects and missed some objects. For HUT, some data were obtained on 3C273 and Capella, but an important cataclysmic variable VW Hydri had been missed due to a shuttle thruster problem (subsequently fixed). They had also discovered the detrimental effects of having one of the astronauts walking the exercise treadmill during an acquisition or observation -- things bounce all over the place!

Although some IPS tests continued to be done each shift, the astronauts had settled into using the Contingency Target Acquisition (CTA) mode of operations. This mode of acquiring targets was somewhat slower than nominal IPS acquisitions were supposed be, but IPS operations were not nominal. CTA seemed to be working effectively in every case where it had been tried. The mode basically used the Image Motion Compensation System's star tracker as a "finder" telescope. The IPCS's primary purpose was to supply error signals to the UIT and WUPPE telescopes permitting them to remove pointing jitter. (UIT was taking pictures on film, after all.) The fact that they could also use this star tracker as a finder telescope was saving us from disaster at this point.

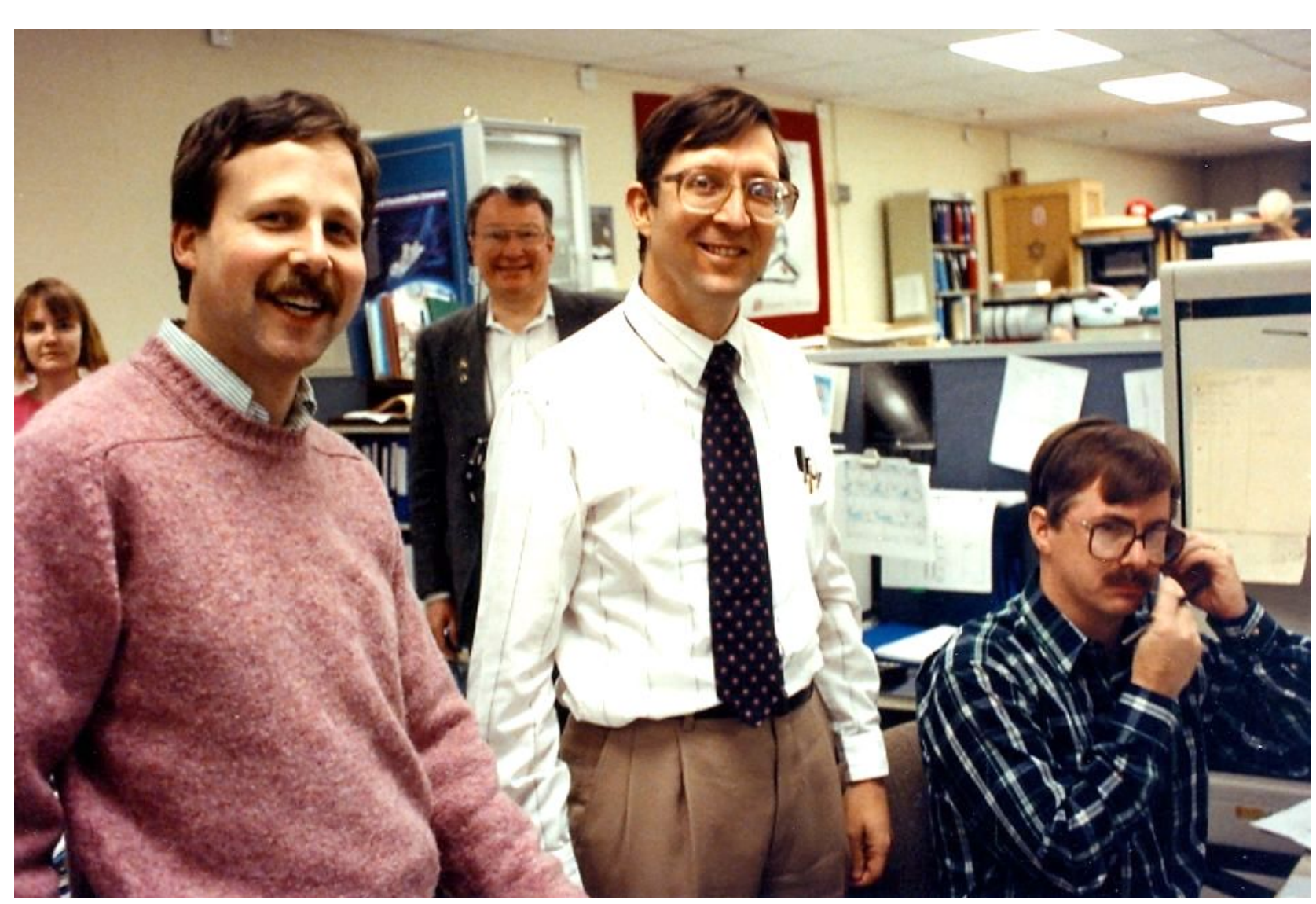

Harry Ferguson, Dick Henry, and Knox Long in the Science Operations Area of the POCC, with Bill Blair at the Replanner's desk.

The main excitement that evening was another Cygnus Loop observation early in the shift. This time we observed a bright optical radiative filament and got another good, solid observation. It was a big thrill for me. The O VI 1035 Angstrom emission was stronger than I had expected, meaning that the shock velocity at this position was high, at about 170 km/s. Most of the other targets this shift were prime for WUPPE or UIT, so progress on our overall science program had been fairly slow. (And there I sat with both of my Cygnus Loop observations in the bag!) We were about 100 hours into the mission now, or roughly half way. The other main excitement tonight was my CARE package from home, which was thoroughly enjoyed by everyone. "Mama Jean's" famous fudge was a big hit about 2 in the morning.

As part of its energy generation onboard, the shuttle produced waste water that occasionally had to be dumped through a vent pipe in the side. Of course, we had to close our doors and stop observing while this took place to avoid possible contamination of our optics. A nominal dump would take about an hour, with a 20 minute dissipation period, or about a full orbit. Nominally, there were three dumps scheduled during the mission at about 65 hour intervals. Well, even this didn't work quite as well as it was supposed to, so the NASA folks kept coming to us on very short notice (as far as our planning cycle was concerned) and saying "Which target do you want to wipe out for an emergency water dump?" This wreaked havoc on the timeline more than once and complicated the delicate negotiations between the various instrument teams as we replanned. One of these "fire drills" occurred during this shift.

But little did we know at that point that the worst lay ahead. About an hour before the end of the shift yet another disaster struck. The second (and only remaining) DDU on the shuttle overheated just like the first one had, and it had to be shut down. (See 1990-12-07_BaltSun_article.pdf.) This left us in a real bind. The astronauts had no way of pointing the IPS or controlling the instruments without the DDUs, which basically meant that we were dead in the water again! Although each of the UV instruments could be controlled from the ground *individually* by the instrument teams (basically for trouble shooting purposes), there was no plan in place that would allow the IPS to be controlled from Houston, and all three of the UV instrument teams to be simultaneously controlled from Huntsville! And even if there was a way, it would only work when we had TDRS coverage to allow communications, which was not all the time.

At the morning SOPG, we discussed the options for recovering from our problems. Basically, the UV instruments could not operate in the near term until they figured out whether ground commanding of the instruments would be feasible. The IPS team felt that they could have ground commanding of the pointing operational soon (to the extent that they could point it anyway!). We decided to stay on the timeline as planned for the next 12 hour shift, but really only BBXRT would be making observations. This would give BBXRT an unhindered chance to try and work out their continuing alignment problems. For the next 12 hour block, the planners were tasked with coming up with a repeating sequence of attitudes that was good thermally for the shuttle and would contain objects of interest to as many teams as possible so that whoever was operational by the time we got to that point in the timeline would have something to observe. We were having loads of fun now.

Since I was one of the only HUT folks who had any significant data at this point, they sent me up to the morning press briefing to talk about the Cygnus Loop. This would have been fun and interesting had I not been a walking zombie at this point. I managed to say a couple of things that got quoted in the press anyway, one about "If you don't like how the mission is going, just wait ten minutes" (paraphrasing the old statement about the weather), and the other something like "I don't know whether to cry or smile from ear to ear" since I had gotten my Cygnus Loop spectra but otherwise things were dead. (Notice the complete lack of any scientific information in what was quoted.) Anyway, it was my one chance in the spotlight of the press, and people said I did a good job. I just remember trying not to fall asleep while the other people were talking as fatigue was really setting in. I called Jean when I got back to the room, and didn't get to bed until 11:30 a.m.

**Thursday, December 6, 1990:**
I awoke at 5 p.m. and just staggered next door to "The Kettle" for dinner. I was really feeling dragged out at this point. News from the POCC was encouraging; they seemed to be on the verge of achieving control of the telescopes effectively *from the ground!* An excellent description of the miraculous recovery is recounted in the Robert Kanigel JHU Magazine article published in October 1991 after the mission (see pages 32-36 of P66). Basically, many of the comm protocols that were in place were dropped in favor of allowing each of the three instrument teams to control their own telescopes from the ground, with JSC controlling the IPS and the astronauts performing the final acquisition and guiding from the AFT flight deck. The heroic efforts of astronaut and Alternate Payload Specialist John-David Bartoe in spearheading this effort must be acknowledged here.

HUT was able to demonstrate control capability from the ground very shortly after I arrived for work at 8 p.m. The day planners had laid out a plan of repeated observations on "compromise" targets of interest to two or more teams for this shift. A continuing assumption of what we then called "block scheduling" had been adopted for the near term replanning, whereby each team was given a two orbit block (out of each 12 hour [=8 orbit] shift) to specify targets of interest.

This worked fine for the day shift scheduling, but we had unending problems planning the night shift this way. This was because the day shift had (very nearly) a clean slate of "good" orbits, while the night shift had to deal with many SAA passages,*[15] shuttle tests, and water dumps that forced an uneven number of useful orbits to be distributed amongst the four teams. I won't go into the details of the negotiation process, but this turned out to be a very tiring and time consuming part of my effort for the remainder of the mission. This shift was only the beginning!

*[15] The SAA (South Atlantic Anomaly) is a region above the south Atlantic (where else?) where the Van Allen belts dip down, causing more intense high-energy particle background rates. The instruments could not be used in the portions of the orbits where we passed through this region.

As the shift unfolded, it became clear that we really could work effectively controlling from the ground. We obtained three consecutive good observations of the giant elliptical galaxy NGC 1399, which was one of our highest extragalactic priorities! Not only were we working again, we were getting quality data! This was followed by two consecutive solid observations of an intermediate redshift quasar (Q1821+64), one of which even got 106% of its scheduled time (because the acquisition went smoother than expected)! That was a first. It was also on the first of the quasar observations that *full joint operations of the three UV instruments was re-established* for the first time since the second DDU died. It had only taken 22 hours to recover from a potentially devastating problem. Amazing. (See P45.)

Toward the end of the shift, as things were looking rosy, I began to push the idea of pitching the "block scheduling" and returning to the pre-planned timeline. The block scheduling was very difficult to negotiate on my shift, and we were approaching a portion of the nominal timeline that was very heavy in HUT targets, many of which were the objects that we would be trying to get into the timeline through block scheduling anyway. I was quite disappointed when others didn't pick up on this and push it; it would have made the rest of the mission much less difficult and quite possibly would have returned more HUT priority science targets. Having lost that battle, I retired to the Kettle (again!) for a quick bite of breakfast and sour grapes, and got to bed about 11 a.m.

**Friday, December 7, 1990:**
I got up at 5:15 p.m. and watched some of a press conference on NASA Select. Chris Anderson from the WUPPE group did a credible job presenting some of their quick look results. Also, astronauts Jeff Hoffman and Sam Durrance gave an "Astrocast" science presentation for use in classrooms. Several of us met at the Classic Cafe for dinner. This was basically a 50's diner atmosphere with 1990's glitz (mirrors, neon lights, and TV sets mounted in the corners of the dining room).*[16] The food was ok, and at least it wasn't the Kettle. The report from today was good; they had gotten a good shot at Jupiter and the Io torus, a high priority planetary nebula, two full observations of M31 (the Andromeda galaxy), and a smattering of other possibly interesting observations. It's almost fun when things go well.

*[16] Actually, they were playing a tape of an old Elvis concert on the TV. It was one of his last concerts where his face was all bloated and sweaty and he was wearing one of those white sequined suits. It was really awful.

The observations continued to go well into the evening and a few more targets of high priority to both HUT and the other teams were observed, including NGC 4151 and NGC 1068 (the brightest Seyfert 1 and 2 galaxies). The WUPPE team, who purported pre-mission that they had no interest in extragalactic targets, announced that they detected about 16% polarization from NGC 1068, which is extremely strong polarization. Most of the other targets this shift were of no particular interest to HUT, but just to have things running smoothly was exciting enough.

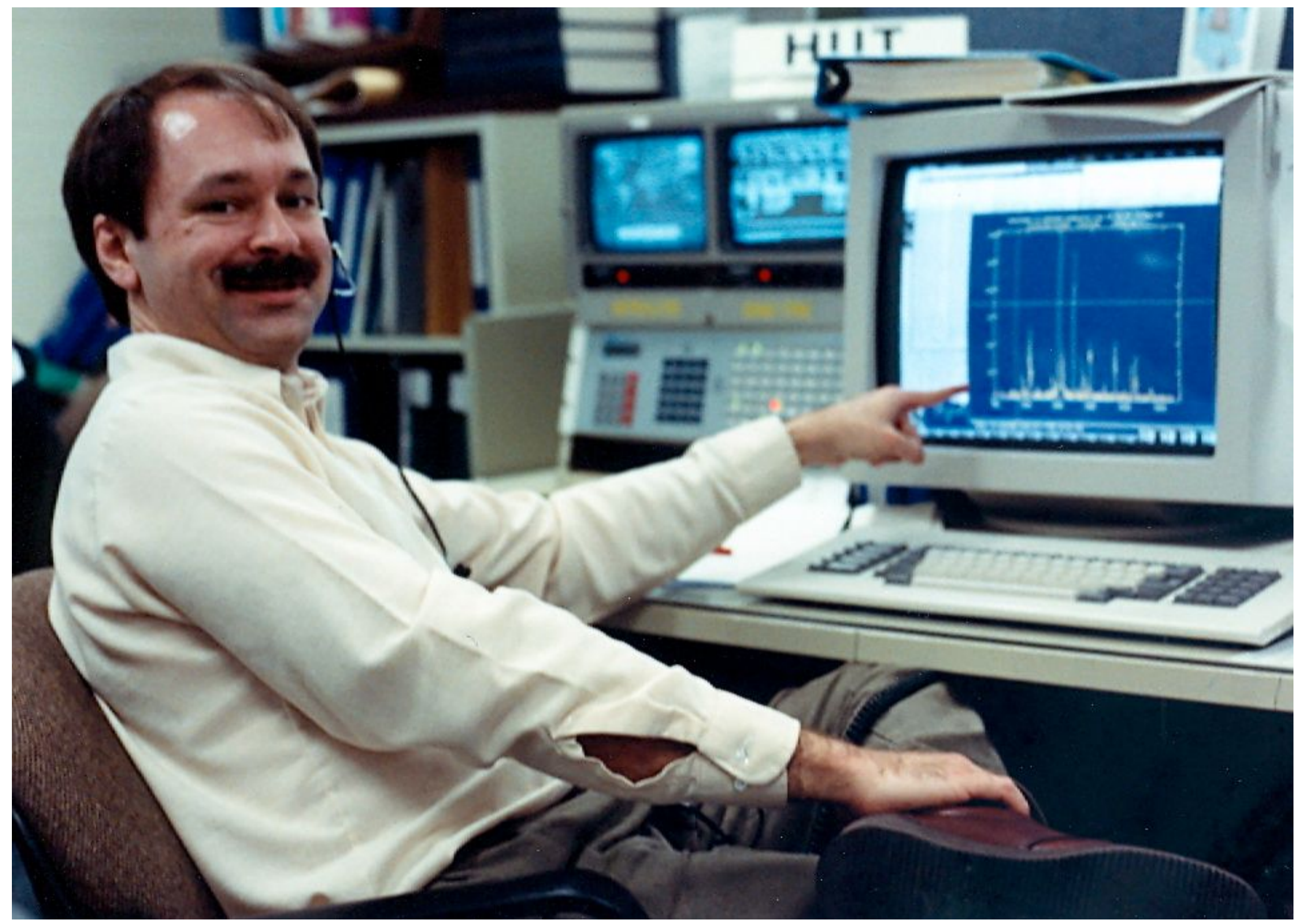

Data Evaluator Jeff Kruk showing off a nice HUT spectrum in near-real time.

Unfortunately, things on the planning front were not going that smoothly. There were only seven good night passes in the 12 hour shift we were planning, and one of those was lost to a planned water dump. Six nights divided by four teams doesn't work very well.*[17] Also, some technical problems during the day shift today caused WUPPE to lose their prime observations in that shift, so they were looking for extra dispensation to begin with.

*[17] Again, much of our rough accounting was based on usage of orbital night time, which is the darkest and most useful part of each orbit.

We just about had a compromise hammered out when the NASA planners dropped a bombshell on the whole deal. They started talking about moving a water dump and/or a shuttle test (called the RCS Hotfire test) back into the shift we were planning! This issue came up very shortly before the whole plan for the shift was supposed to be finalized and delivered, and it really threw things into turmoil. The morning planning meeting was long and heated. It was clear that everyone was getting tired and testy; the long hours had started to take their toll. I dropped a lot of unfinished business on the next shift, called Jean for the news from home, and got to bed about 11 a.m.

**Saturday, December 8, 1990:**
I got up about 5:15 p.m., and ate at the Kettle with Glen Fountain. While I was actually getting quite sick of eating at the Kettle, it was convenient, and it served "breakfast" 24 hours. Since I was eating breakfast at dinner time, it worked out. I got out to the POCC about 7:30 p.m. amidst rumblings that the ongoing problems getting rid of waste water might actually shorten the mission to 9 days (+2 days contingency). If that were really the case, then the current replan shift I would be working would be the last one that mattered. As the night progressed, it sounded like perhaps they had another trick up their sleeve that would let us stay up longer, but it was far from certain. We were asked by Ted Gull (mission scientist) to assemble a list of high priority targets we would observe if we got the extra day, so he could waive it in front of the NASA administrator if necessary.

A fun thing happened that evening: about 10:30 p.m. we had a chance for a representative from each instrument team to talk to the crew about how things were going. Knox Long gave the update from HUT, and after providing a science update, he actually read "Twas the Night Before Astro" to them! Here it is, from the actual PAO transcript of the event:

> *Knox: I should tell you that we have been kept in a good mood by Bill Blair who seems to be able to either write a poem or rewrite lyrics to various things every day. One of them is called, "Twas the Night Before Astro." If you have a moment, I'll read it to you.*
>
> *Sam: Yes, go ahead. Sounds great.*
>
> *Knox:*
> *Twas the night before Astro, and all through the POCC,*
> *Not a creature was stirring; we were all in our spots.*
> *The posters were hung on the walls and the door,*
> *In hopes that we'd soon hear the Astro launch roar.*
> *The Blue team was nestled all snug in their beds*
> *While visions of "real data" danced in their heads.*
> *The MUM in her kerchief and the PAP in his cap*
> *Had just settled their brains from some JSC flap.*
> *When out on the loops their arose such a clatter*
> *We put on our headsets to hear all the chatter.*
> *To VT-240's we flew like a flash.*
> *We signed on to OMIS, RRs to rehash.*
> *(The moon in the midst of the cluster Virgo*
> *Caused a lot of RRs in the system below.)*
> *When what to our wondering ears should we hear,*
> *But the shuttle was fixed! There's no leak in the rear!*
> *Like wild leaves before a wild hurricane fly,*

*When they meet with an obstacle, mount to the sky,*
*So out on the pad they were ready to sail,*
*Thanks to friend "Uncle Max" and his crew biting nails.*
*The spacesuits were hung on the crew with much care*
*In hopes that the shuttle would launch like a flare!*
*The little clean room at the top of the gantry*
*Was smaller than most people's kitchen or pantry.*
*The astronauts entered Columbia's port,*
*They strap themselves in – "All is well" they report.*
*In the POCC back in Huntsville we sat and we waited*
*(Although more than one veteran's mood was quite jaded.)*
*The countdown went smoothly; the APUs worked.*
*The SRBs fired, it lurched and it jerked.*
*It cleared the launch tower; we hooted and whistled,*
*And away they all flew on the back of that missile!*
*The smoke it encircled the pad like a wreath,*
*And we all breathed a big heavy sigh of relief.*
*And we heard KSC as they rose out of sight,*
*"Happy liftoff to all, and to all a good flight!"*

*Sam: That sounds great, Bill. You have really produced some great poetry there. Thanks a lot Knox.*

The uncertainty about the end time caused a lot of frustration and tension on the replanning front this shift, and the end of the night was very hectic. The RRs for the next shift were late, an important observation of cataclysmic variable star UX UMa was screwed up, and the SOPG meeting was a mess. At the very end of the shift, the cataclysmic variable star Z Cam was observed, which was a highlight, and the data looked great. Lisa Hooker asked me to stay and go on Frank Six's "Today in Space" program, but then it turned out they didn't have a spare camera available to do it. They decided instead to have Knox go on the 5 p.m. press briefing, so I wrote up a "cheatsheet" on Z Cam for him before leaving. I got back to the Amberley and to bed about 11 a.m.

**Sunday, December 9, 1990:**
I got up about 5:15 p.m. and at 5:45 p.m. I got a call from Marcia Dunn of the AP. She mostly wanted to talk about the "Night before Astro" poem and the songs I had written. I guess they needed a little human interest stuff for the paper. After talking with her for about 10 minutes, I left to meet folks at La Fiesta for dinner, but when I got there it was closed! Since I didn't know where people had gone, I went across the street to O'Charleys but didn't find anybody. So I ate by myself. After dinner, I called Jean, and then headed out to the POCC for one last night of frustrating replanning activities.

There was still a lot of uncertainty about the mission end time, both because of the waste water situation and because of projected weather conditions at the various potential landing sites. Either tonight was going to be "it", or we would have one or two more days on orbit. Because of the uncertainty, we pushed two OCRs (operations change requests) this shift. OCRs impact the *current* shift of operations, as opposed to the normal RRs that after the next replan shift (12 –24 hours out). One was to move up a target that had just been replanned last night to get it into the current shift, and the other was to insert a pointing at a new comet, Comet Levy, to try to get it in before the end of the mission. (This was important partially for historical reasons: back in 1986 we were supposed to observe Halley's comet before the big delay from the *Challenger* accident occurred.)

Well, the replanning was just horrendous. We did end up getting the Comet Levy OCR through the system, but we also had to do the normal RR planning for the following shift, and the WUPPE planning reps in particular were extremely uncooperative. Then as the RR deadline approached, the PAP (Payload Activity Planner) decreed a limit on the number of RRs that would be permitted in the upcoming shift. This was done only 5 minutes before the nominal deadline! In the SOPG meeting that followed a couple of hours later, we were in the process of negotiating our way through this situation when the word came down that this was it—we were coming home tonight because of weather concerns. (See 1990-12-10_HunstvilleTimes_article.pdf.) The need for planning activities was over. As I walked back into the SOA, we were just setting up for the last observation that would be allowed. It was the Comet Levy observation that I had worked for so hard earlier in the shift.

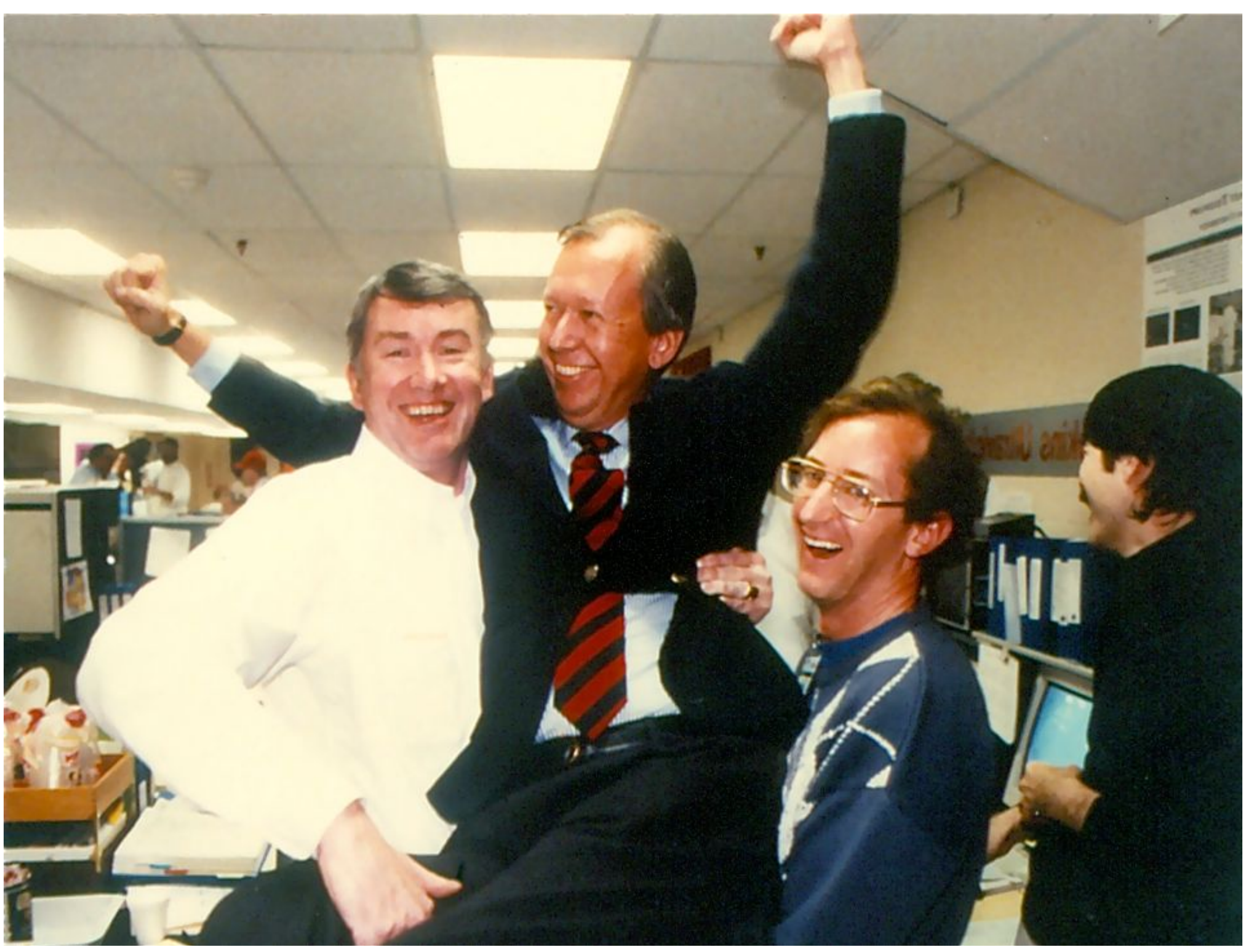

Victory from the jaws of defeat! Glen Fountain, Arthur Davidsen, and Ben Ballard celebrate the successful end of Astro-1 on-orbit operations.

Things quickly turned into a party atmosphere in the SOA. I was absolutely dead tired and could hardly even participate. Even adrenalin couldn't help me at this point. It struck me how very little time I had actually had to enjoy what was going on in real time, including the eventual success we had in obtaining data over the last few days. If I had really known this was the last shift, I could have just kicked back and enjoyed it a bit, instead of experiencing the frustrations described above, all for no good reason. C'est la vie.

Mid-morning, a call came in from CBS radio and they interviewed me about the mission, and eventually got around to asking me about the songs and poem stuff. I guess I read them the "Night before Astro" poem and maybe even did a bit of "50 ways to lose your pointings." I can't even remember. I was a zombie at this point. But it must have put me in the mood for one more song. After I got back to the Amberley about 12:30 p.m. I penned "Deck the Walls with Plots of Data," (see the 1991-12_A1_songbook.pdf file in the Diaries-Memories portion of the online archive) which was debuted at the post-mission party later that evening. I got to bed about 1:30 p.m.

**Monday, December 10, 1990:**

My Monday only started at 6:15 p.m. when I dragged out of bed. The whole HUT crew was assembling at the Fogcutter restaurant for the post-operations party. I guess they had a good time at the POCC after I left, handing out "Astro-1: More Bang for the Buck" T-shirts and generally whooping it up. While I was sad to miss the fun, I simply had no more gas in the tank after all the planning frustrations and the lack of sleep.

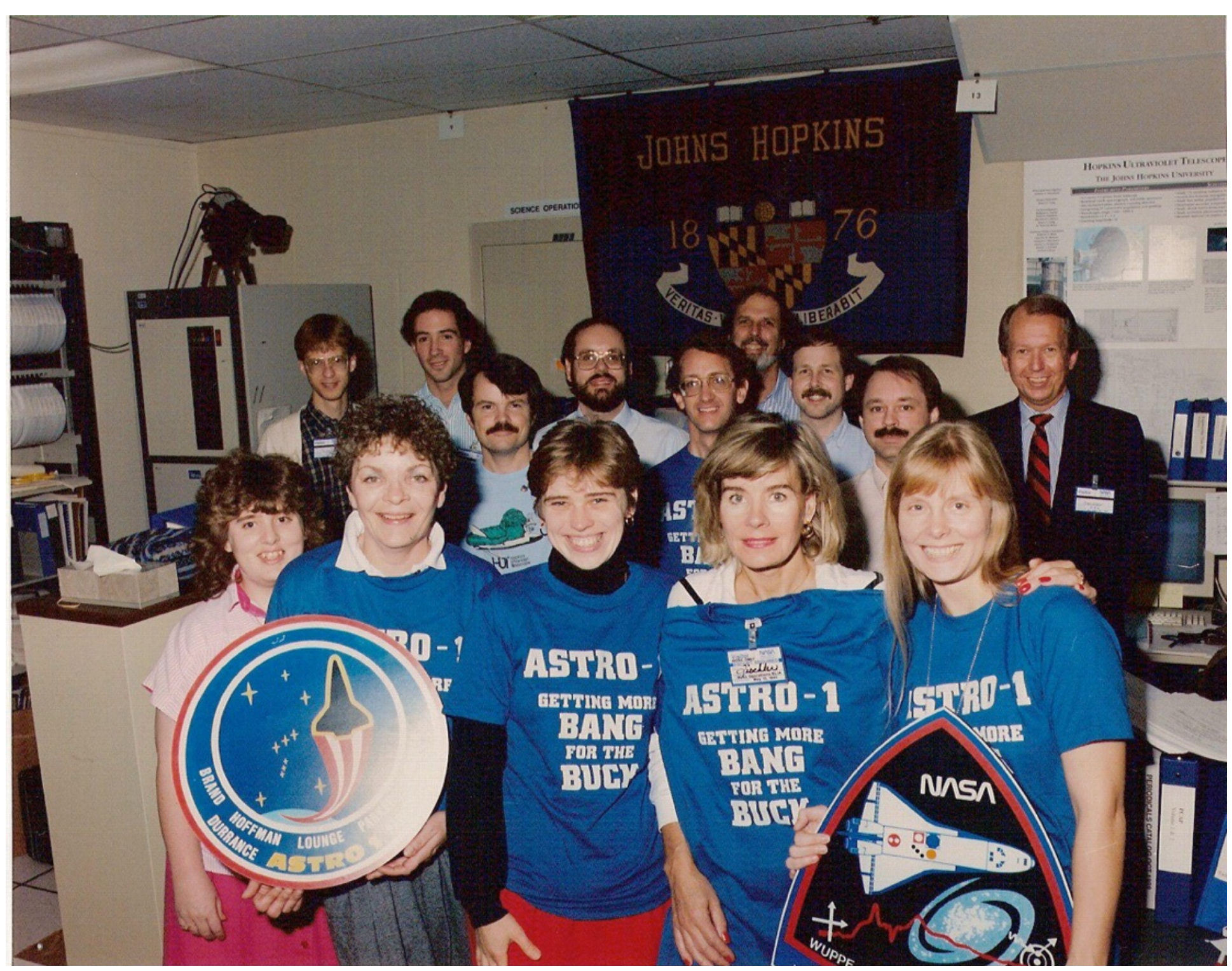

The day shift celebrates the end of operations with "Astro-1 Getting more BANG for the BUCK" T-shirts.

After less than five hours of sleep, I wasn't in much better shape even now. We had a really nice dinner and the wine flowed. After dinner, most of us were at least halfway in the tank and the singing began. They got me up there to do the "Little Planner Boy," and I coerced Van Dixon, Harold Screen, and Lisa Hooker to do the "par um pa pum pums" in the background. Then I and others started writing songs in real time, with Ben Ballard and Randy Kimble getting into the act. I was apparently inspired from somewhere to pen "I'm Dreaming of the Next Astro," which even talks about involving "GOs" (guest observers) in the next (then non-existent) mission. Quite remarkable to look back at this after the fact. Then I did "Deck the Walls with Plots of Data" before packing it in for the night. It was a great evening of camaraderie and thoroughly enjoyed by all.

We all met back in the lobby at the Amberley about 11:15 p.m. because the shuttle was supposed to land about midnight. (Yes, a night landing as well as a night launch!) And land it did. While waiting for the astronauts to disembark, Randy Kimble wrote "Oh Little Town of Huntsville" and we performed it to fill the time. The astronauts finally came out about 1:45 a.m. and walked around the shuttle. By 2 a.m. I was in bed.

**Tuesday, December 11, 1990:**
I got up about 9:30 a.m. (still not quite even 8 hours of sleep), partially shifting back toward a normal sleep cycle. Because we hadn't expected the mission to end so abruptly, a bunch of us had to get reservations changed. I got changed to a flight returning to Baltimore that evening. After packing and checking out of the Amberley Suites, I headed out to the POCC one last time. Many folks were already out there packing things up and getting ready to ship stuff back to Baltimore. Our home away from home was almost unrecognizable. It almost seemed like a dream at this point. The mission really was over.

## Astro-1 Epilog

At this point, my notes stopped, and so details become scarce. I obviously made it home to Baltimore and reunited with the family. And I remember a very strange thing about the time after we got home. Most of our colleagues back at JHU and STScI were almost reticent to talk to us about the mission. It turns out that the press coverage of the early part of the mission was so negative that most people thought it had been a complete failure, and thus people were almost embarrassed to bring it up! About the time we got things working reasonably well, the press more or less dropped the story and barely covered our amazing turn around. It could have been great story, but because of all of NASA's problems over the course of the previous year (shuttle leaks, and let's not forget that the Hubble "spherical aberration" problem was discovered in June 1990), all the press wanted to do was "bash NASA." No success stories were allowed in real time I guess.

We worked very hard (and very quickly) to turn this perception around. Within a week or so, more upbeat "mission summary" style articles in *Space News*, *Aviation Week*, and even the journal *Science* started to appear that were more balanced reports. (See articles from P50 to P56 in the article archive.). Locally, the JHU Gazette ran a follow-on article that was upbeat and mentioned some of the successes (see P55).

By just a few weeks later, at the January American Astronomical Society meeting in Philadelphia, we were already present in full force and presenting preliminary science results (see P58, with title "Astro's brilliant discoveries dimmed by news that it may not fly again"*[18]). Arthur and others started giving colloquia in various venues to make the early results more visible. A picture of Astro-1 on-orbit made the cover of Sky & Telescope's April issue (no doubt it was published in late January as the issues always come out ahead of the issue date; see P61). NASA press officer (and UIT team member) Steve Maran published a full article, "Astro: Science in the Fast Lane," in the June 1991 Sky & Telescope (but would have come out in late March or early April) that also provided wide visibility to the successes of the mission (see P65).

*[18] Recall, given the launch pressure on the shuttle schedule when missions started flying again after the *Challenger* disaster, NASA administrator Lennard Fisk had announced in December 1989 that Astro-1 would be the only Astro mission.

While the details may be lost to the annals of time, somewhere in the February or March 1991 timeframe, Arthur Davidsen managed to get the ear of Maryland Senator Barbara Mikulski and made the case for a second flight for HUT and the Astro UV telescopes. Mikulski was a senator from Maryland and chaired the committee that oversaw the NASA budget, so she had clout. She was also largely considered the champion of Hubble and its recovery from the spherical aberration problem after its launch.

And so it was that in late May 1991, an event was held at JHU where Senator Mikulski came and announced that *there would be an Astro-2 mission!* Never one to miss a PR opportunity, Arthur and the rest of us worked with the JHU communications office and put together a full press kit (see D25), including information about HUT, the personnel, and science teasers from Astro-1. This was available at the time of the Astro-2 announcement. There is a famous photograph from this event showing a broadly smiling Sam Durrance, Arthur Davidsen, Barbara Mikulski and JHU president William Richardson, with NASA administrator Lennard Fisk standing to one side with a somewhat resigned look on his face. (See P62 – P64). Astro-2 was initially announced for some time in late 1993, but by this time it should not surprise anyone to know that STS-67 did not occur until March 1995.

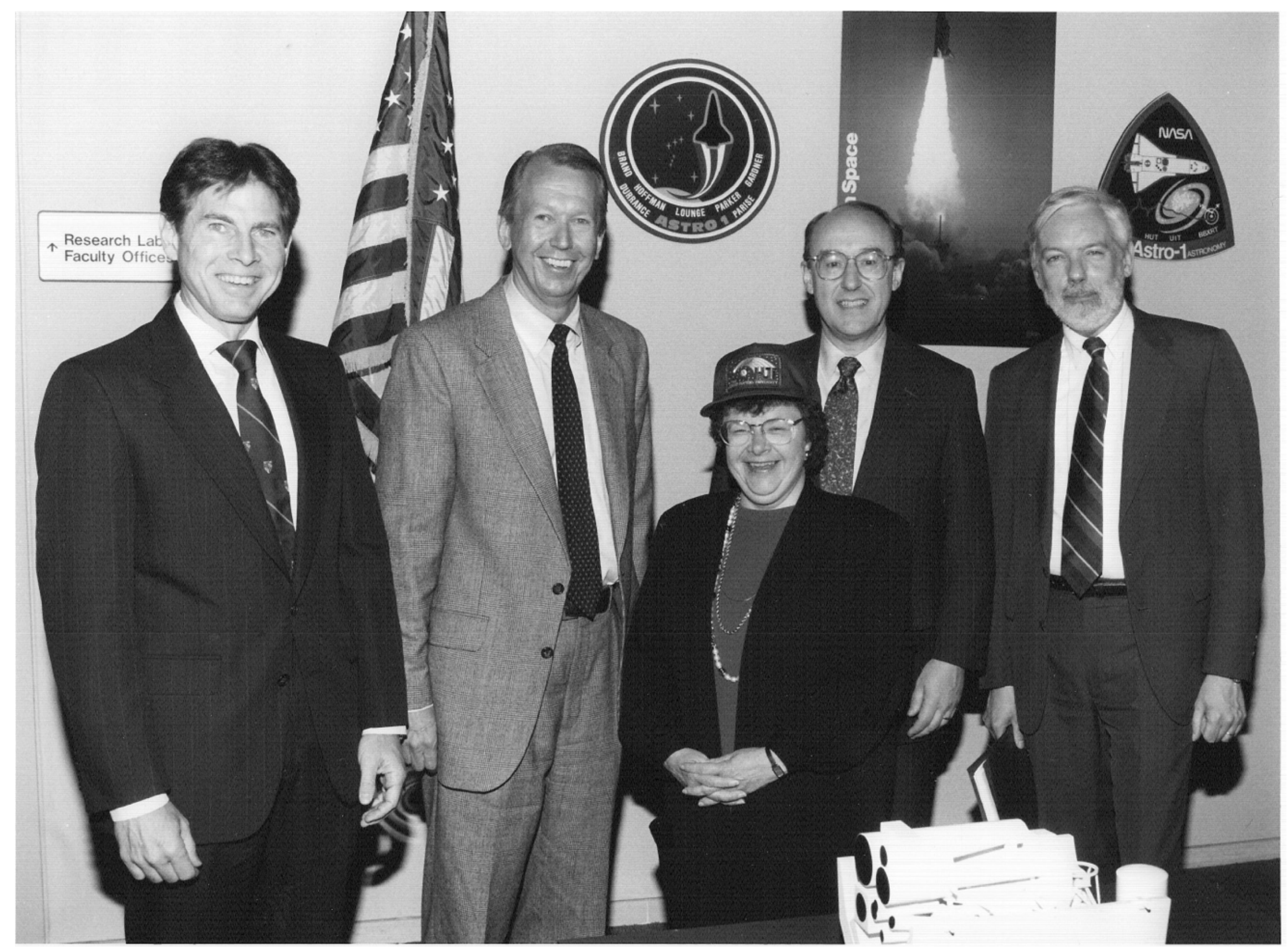

Astro-2 is announced in a press conference at JHU on May 26, 1991, just over five months after the end of the Astro-1 mission. Shown are Sam Durrance, Arthur Davidsen, Sen. Barbara Mikulski, JHU president William Richardson, and NASA Administrator Lennard Fisk. (Photo: JHU)

The Astro-1 mission was the experience of a lifetime. I have often reflected on the fact that I recorded some of this while it was relatively fresh in my mind, because I certainly would have forgotten many of these details otherwise.

# Chapter 5: Between the Missions

## The Post Astro-1 Period

As the dust settled after Astro-1, the mission managers from NASA Marshall requested our feedback. I solicited the HUT team for their input and edited it into a fairly lengthy response. (See D21, which also contains a memo from Jerry Kriss specifically about the training we received at Marshall to support the real time mission activities.) I think this was an honest assessment, including some of the feelings of mistrust and poor communications we had experienced, as well as the training and the facility at Marshall. I honestly think some of this must have been taken into account, as one of the major outcomes was the formation of a "Continuous Improvement Mission Planning Team" that went a long way toward improving communications and processes for Astro-2. (More on that below.)

As of mid-1991, it was clear there would be a second flight of the payload (see the end of Chapter 4), which was made official by a letter from Lennard Fisk (NASA Administrator) dated June 24, 1995 (see D26). However, the first order of business was the analysis and publication of the data that had been garnered by the heroic efforts of both the astronauts and ground controllers (including the instrument science teams) during the Astro-1 mission. Looking back, it is amazing to me that only a few weeks after the end of the Astro-1 mission, the HUT team presented a dozen papers containing early results at the American Astronomical Society meeting in Philadelphia in early January 1991. By April, we had submitted a preliminary science summary from the mission to NASA (D24 and the appendix D23). However, that is different from actually publishing peer-reviewed science papers, which is the real currency of scientific success. That takes time and effort.

As far as HUT was concerned, since so little was known about the 900-1200 Angstrom spectral range,*[19] every spectrum (even fairly ratty data) was unique and worthy of publication at some level. Each person on the science team had one or more main areas of scientific interest, and so we prioritized the projects and supported each other as we went through the process. In January 1993, Arthur Davidsen published what amounts to a "review paper" of the highest priority science results from HUT on Astro-1 which was published in Science (see P67). It was about this time that all 13 refereed HUT papers from 1991 and 1992 and 10 papers that were in advanced readiness were reformatted and published together in the famous HUT "Yellow" book, which was released in March 1993. Numerous other articles were published after that date that as well, with a compilation from early 1996 showing 35 refereed articles from Astro-1 HUT data, and many non-refereed articles in various conference proceedings. (Of course, by 1996, results from the Astro-2 flight of HUT began to take precedence.) Arthur attended a COSPAR meeting in July 1994, and the team provided him with an excellent set of summary science viewgraphs, which I have added to the archive (D39).

*[19] A previous mission, OAO-3 (also known as *Copernicus*) had observed this spectral region, but only for a handful of the brightest stars in our Galaxy, not for many different kinds of astronomical sources. Also, since HUT coverage extended up to 1800 Angstroms, it provided a way to connect the unique 900-1200 Anstrom coverage with other instruments like IUE and Hubble at longer UV wavelengths.

It was also during the second half of 1992 that Arthur and the HUT science team crafted a NASA proposal in response to a Small Explorer mission Call for Proposals. The Quick Ultraviolet Exploratory Survey Telescope (QUEST; see D30) was a smaller telescope with capabilities very similar to HUT but with improved silicon carbide (SiC) coatings, that could be launched on a Pegasus rocket rather than a conventional launch vehicle. The two year free-flying mission would include a year of PI team observations followed by a year where Guest Investigators could propose observations. The entire concept was built on the success of HUT on Astro-1, and predicated on the idea that, even with an Astro-2 flight on the books, the existing opportunities would only scratch the surface of the science that could be done in the 900-1200 Angstrom region, even at the relatively low spectral resolution of about 3 Angstroms.*[20] The proposal was submitted to NASA in January 1993, and while it was unsuccessful, two of the QUEST emphases, the SiC coating on the optics and the idea of offering a Guest Investigator program, both ultimately made their way into the Astro-2 mission.

*[20] We have yet to mention that a major mission to observe the 900-1200 Angstrom region at much higher spectral resolution of roughly 20,000, the Far Ultraviolet Spectroscopic Explorer (FUSE) MidEx mission, had its PI at JHU as well, Prof. Warren Moos, who was also a Co-I on HUT. FUSE was still in its design phase at this point.

It was perhaps inevitable that there would be some attrition in the HUT team ranks after Astro-1. The majority of the HUT team had been in place since the early to mid-1980's, and had joined the team with the understanding that the payload would fly multiple times before 1987 (with the possibility of additional flights under discussion). It is remarkable that nearly everyone hung on through the long hiatus and uncertainty caused by the *Challenger* disaster, but after the devastating delays through 1990, the travails of the Astro-1 mission, and the prospect of several more years (at least) before Astro-2 would fly, several key team members were ready to take the next step in their careers. Randy Kimble actually took a position at NASA/GSFC in Sept. 1990, before Astro-1 flew, although he worked the mission and came back to JHU regularly to work on the data through much of 1991. Knox Long left JHU in mid-1991 for a position at STScI, and Chuck Bowers took a civil service position at NASA Goddard in Greenbelt before the end of 1991. (Knox would return to support Astro-2 operations, however.) On the plus side, two new team members, Wei Zheng and Richard Buss, joined JHU/HUT in December 1991 to support Astro-2 preparations, and Tom Brown and Brad Greeley joined HUT as a graduate students at JHU in August 1992.

Moving quickly to shore up the team management structure, in January 1992, Arthur announced new team assignments and titles for HUT team members (see D27). Jerry Kriss took over as HUT Project Scientist replacing Knox, Jeff Kruk was appointed as Deputy Project Scientist for Instrumentation, and Bill Blair was named Deputy Project Scientist for Mission Planning and Operations. Additional changes happened on the JHU/APL side as well, with Ben Ballard taking over as HUT systems engineer, replacing Steve Conard, and Tom Zaremba becoming HUT Program Manager in place of Glen Fountain.

There was also a significant change in personnel on the NASA side. Ted Gull from NASA/GSFC had been the NASA Mission Scientist since the inception of the project. He had the unenviable job of overseeing all three instrument teams while trying to also coordinate Astro-related activities between NASA HQ and the NASA/MSFC contingent who were given management responsibility over the Astro missions. There had been friction between GSFC and MSFC with this arrangement over the years, and Ted took himself out of the running for Astro-2 Mission Scientist. It was not until well into the fall of 1991 that Charles "Chip" Meegan from NASA/MSFC was named Astro-2 Mission Scientist. The main down side, of course, was the loss of corporate memory of the project, and the fact that Chip had no particular previous involvement with ultraviolet astronomy, having been previously involved with high energy cosmic ray experiments.

The original plan for the payload specialists was to select one from each instrument team, but with multiple flights, two of the three would fly on any given mission. Obviously, on Astro-1, it was Ron Parise (UIT) and Sam Durrance (HUT), with Ken Nordsieck from WUPPE taking a lead communicator position in the POCC (along with John-David Bartoe on the other shift). With the long delays from the Challenger disaster and several more years until a possible Astro-2 flight, Ken Nordsieck decided to pursue other challenges and so took himself out of the running. Hence, by late 1993 Sam and Ron were once again designated for Astro-2 (see P68). Scott Vangen, an engineer from KSC with Astro-1 experience with WUPPE, was appointed as alternate and trained with the crew.

As to the hardware, the integrated telescope package was maintained at KSC in the O&C building. In Oct-Nov. 1991, we were able to de-integrate HUT and transfer it to the adjacent "ATM clean room" where we could access and remove the spectrograph to permit a post-mission calibration in the lab. Jeff Kruk was primarily responsible, and spent significant time in December 1991 at the SRF (Synchrotron Research Facility) at NIST performing this calibration activity. At some point during the first half of 1992, it was decided to fully build out the back up spectrograph, depositing silicon carbide (SiC) on the grating for improved FUV performance and making a new slit plate with a different set of apertures. HUT meeting notes from late-July 1992 indicate that this work was well underway by that time.

This is significant in that HUT was the only of the three UV telescopes that actually underwent significant improvement between Astro-1 and Astro-2. Improving and re-flying the instruments was one of the long lost expectations from the "multiple flight" strategy adopted in the early years of the project. Not only was the coating of the spectrograph grating with SiC an improvement, but ultimately the HUT back-up primary mirror was also coated with SiC (see below), resulting in HUT being roughly 2.5 times more sensitive at FUV wavelengths than on Astro-1. In retrospect, this increased sensitivity led to the successful detection and characterization of the intergalactic medium, Arthur Davidsen's pet science project and one that garnered tremendous attention during and after the Astro-2 flight. (Of course, this increased sensitivity benefited ALL observations on Astro-2, not just the IGM program.)

---

## The Continuous Improvement Mission Planning Team

With the seemingly endless number of launch delays for Astro-1, we had learned a lot about how to plan observation timelines and adjust them for various launch slips to cover a period of time. However, it wasn't until the actual mission that the importance of the near-term replanning activities became fully into focus. The replanning process, involving the horse trading between the three instrument teams as well as the interactions with the NASA/MSFC staff and ultimately NASA/JSC and the shuttle crew themselves, was cumbersome at best on Astro-1. In the pre-Astro-1 mission simulations, we were always restricted as to how many observations could be rescheduled and how quickly it could be accomplished. Astro-1 stressed this system almost to the breaking point, but what we learned was, when push came to shove, much more could be done. The MSFC planning staff provided this on the fly over and over, so it could be done. The question was, how to accomplish that goal without stressing everyone out.

At the April 23, 1992 Investigator's Working Group meeting at NASA/MSFC, a team was set up to address the replanning process for Astro-2. This group became known as the CI/MPT (Continuous Improvement Mission Planning Team). "Continuous Improvement" was short for Continuous Improvement/Total Quality Management (or CI/TQM), which refers to a management protocol that was in vogue at the time, and this team was to be trained in the protocol and then apply it to improving the mission planning process for real-time operations during Astro-2. The team was comprised of a facilitator, planner representatives from each instrument team, and a handful of key MSFC staff who had worked relevant POCC planning positions during Astro-1.

The CI/MPT met for the first time in mid-June 1992 at NASA/MSFC. We met with a CI training person who went through the protocol with us and helped us through some exercises to see how it was supposed to work. Additional in person meetings were held in July 1992 (Wisconsin), December 1992 (Huntsville), and March 1993 (Huntsville), with much of the work being accomplished between meetings via e-mail (which by 1992 had become available) and telecons. Presentation and updates to the larger project were discussed at several IWG meetings during this time.

While the CI process seemed a bit hokey and cumbersome, it came with sanctioning by MSFC management, and as time went on and the team came up with a number of changes and improvements, the managers basically had to accept most of them despite some anxiety on their part. By the time of the March 1993 meeting, word on the grapevine had it that the mission managers at MSFC wanted to shut the group down due to the many changes invoked by the team's efforts to streamline the MP process. The culmination of our efforts were finalized and published about a year later, in the MPHIRD (Mission Planning Handbook and Interface Requirements Document, see D38), and yes, the group was disbanded shortly thereafter.  Somewhere in my files is a modest certificate of acknowledging the work of the CI/MPT, representing over two years of hard-fought improvements to both the planning and replanning process.

While I have relatively few notes from the CI/MPT meetings, I did manage to archive a long memo I wrote in response to a request from a graduate student who in late 1994 was studying CI/TQM for a class project (see D43).  That memo provides a good summary of my perspective of the process and the outcome, and the reader is directed there for details. However, here are a couple of excerpts from that memo that capture the main points:

---

> By far the most beneficial part of the process was getting knowledgeable people from all aspects of planning in the same place to discuss the problems from their perspective, INDEPENDENT OF THE PROCESS by which those discussions took place. The personal relationships and improved direct communications that were fostered from our discussions have had an intangible, but very real, impact on the planning process for Astro-2.
>
> The primary improvements coming out of the CI/MPT were:
> a) much improved communications between the teams and MSFC,
> b) somewhat better communications between MSFC and JSC.
> c) a much better understanding by PI planners of the constraints and procedures necessary on the MSFC side of the house, and vice versa,
> d) a much better flow to the planning process, resulting in reduced time to generate mission timelines,
> e) a complete document called the "Mission Planning Handbook and Interface Requirements" document, that specifies much of our collective memory about the process and the details of the interactions between the planners,
> f) much more flexibility in the mission planning process than was in place for Astro-1, and
> g) much improved software for mission planning, an improvement that affects both pre-mission and real-time replanning capabilities.

---

I will mention two specific changes made by the CI/MPT that were extremely beneficial to the planning process for Astro-2. The first was the establishment of a POCC position called the File Manager (FM). This position provided a single point of contact for all file transfers between the teams and the MSFC planning system. In combination with newly defined file contents and formats that were imposed via the MPHIRD (and enforced by the FM), this single interface was a huge step forward.

Secondly, we adopted the concept of "block scheduling," which arose during Astro-1 and was formalized and used for Astro-2. In pre-mission planning, we agreed to assign a rotating set of 2-3 orbit blocks of time to each team. Once the blocks were assigned, there were no ongoing "equity" issues to be negotiated. Once "Guest Investigators" were selected, their times were added to each teams' time proportionately as appropriate, and the block assignments took them into account. The observations planned within those blocks were done at the discretion of the assigned team, who also had the freedom to replan within those blocks in 12-hour chunks of time during the real mission (in consultation with the other teams so they could plan coordinated observations as well). This removed most of the angst from the replanning process that was a continual struggle during the Astro-1 replanning process. Each team could replan within their time block as necessary to meet their priorities (with a few caveats for targets of expressed joint interest).

The eventual success of the planning during Astro-2 was a testament to the work of the CI/MPT and the revised processes put in place from their discussions.

---

## Coating the HUT Back-up Primary Mirror with SiC

The coating of the HUT grating with silicon carbide (SiC) by optical engineers at NASA/GSFC in the fall of 1992 was a resounding success. Tests of the grating indicate a 70% improvement in FUV reflectivity over the Astro-1 version. With that success, the GSFC folks were interested in trying to coat a larger optic, and the idea of attempting a SiC coating on the HUT back-up primary mirror was floated.

The only reason this idea was not acted on right away had to deal primarily with schedule. With any assumed launch date, a schedule would be built that showed when all hardware had to be delivered to KSC and when the integrated payload needed to be handed over the NASA to support shuttle launch preparations. While these major delivery dates maintained some contingency times within the schedule, the real problem was that the assumed launch date (or "manifest date") kept moving around every time a new manifest was released.

To give some idea of the volatility, in July 1992 we were still working to a nominal launch date of Sept. 14, 1993, with a Jan. 1993 ship to KSC and a July 1993 turnover to NASA. But by the time of the IWG meeting in Sept. 1992, we were assigned to a Sept. 1994 launch. In early November 1992, a manifest update now showed Astro-2 on Jan. 10, 1995, making it look like there was plenty of time to possibly swap out the mirror (assuming the SiC coating idea panned out). But by just a month later, in the Dec. 7, 1992, we were hearing about uncertainty in the launch manifest, with dates of June 1994, Oct. 1994, and Jan. 1995 all being discussed, depending on which shuttle and any accompanying deployable payload. By late Jan. 1993, HUT meeting notes indicate a manifest of Jan. 18, 1995, on *Columbia* with the Wake Shield experiment, again indicating there would likely be time for the mirror swap. But less that two months later, the launch manifest was showing a Nov. 3, 1994, launch and an extended 13-day mission on *Columbia,* paired with the Capillary Loop Experiment. Pulling the launch back earlier again removed planned contingency time, causing concern about the possibility of a mirror swap.

It was not until the June 1993 timeframe that the decision to try and coat the HUT primary with SiC came to a head, and when it did, things happened very quickly. It helped that back in April, APL had completed an assessment of what it would take to change out the primary mirror (time, cost, and risk). When it became clear in June 1993 that launch would not occur until at least early 1995, things began to happen. HUT was disassembled and the existing primary mirror was measured as a baseline. On or about June 24th, MSFC made an official request for a detailed plan, and by early July (exact date unknown) we submitted a short proposal with all the details (including schedule and cost--(see D35.)

I've not found any follow-up "approval" message, but by mid-July, HUT meeting minutes note that Sam had taken the back-up primary from JHU down to GSFC, and the July 26th minutes state that the back-up mirror was being coated "even as we speak." Jerry sent a memo to KSC on July 30th describing the re-integration plan and schedule, and noted, "The backup mirror has been coated, and we are now making calibration measurements to qualify it for flight." This was followed up by a memo to the MSFC management on Aug. 11, describing the recoating and verifying the back-up mirror had been flight qualified (see D36). The HUT meeting notes from Aug. 16 indicated the mirror was now at KSC and the witness mirrors were "clean." By the Aug. 30 HUT meeting, notes say that the mirror swap was successful and the reassembly of the telescope was in progress. Notes from Sept. 13 state the back-up (now primary) spectrograph and the HUT TV camera were being shipped to KSC. By the end of January 1994, notes indicate that the reassembled HUT had been turned over the KSC and that installation and alignment on the cruciform structure were underway.

---

## The Astro-2 Guest Investigator Program

With the approval of Astro-2, the idea of having outside astronomers, that is "Guest Investigators" (GIs), participate was part and parcel of the decision. The idea of other astronomers using HUT was mentioned in the original proposal in 1978, although the concept was for many flights at that time. The idea only made sense for all three instruments, as the observatory's capabilities were broad enough to support a wide range of science experiments well beyond those of interest to the scientists on each team. Engaging the broader community was also a way to demonstrate to NASA that the asset was one that the community at large wanted to use.

This is not to say that there was no tension in giving some of the precious observing time away. One of the ways this tension was de-fused was to encourage the selected investigators to put skin in the game, to invest time into the instrument teams' work and to effectively become team members if you will. Ultimately, a number of the selected GIs even participated in the mission SIMs and the actual real-time mission operations support at MSFC. The other "protection" was that each team could specify a selection of targets to be reserved for the PI teams' use in scheduling, thus protecting their primary science interests.

Over several months in mid-summer of 1992, the science teams worked with Mission Scientist Chip Meegan and Robert Stachnik from NASA HQ to design the GI program and draft the NASA Research Announcement text. The concept for the GI program was approved by NASA UV-MOWG (UV Mission Operations Working Group) in early September 1992. After removing calibration activities, roughly half of the available observations would go to GIs and the other half (split three ways for each instrument team) would go to the PI teams. No more than 15 GIs were expected to be selected.

The actual NASA Research Announcement (NRA 93-OSSA-14, see D33 ) was released on Apr. 30, 1993. In the interim time period prior to the release, each instrument team had to develop a detailed instrument handbook for their instrument for use by potential GIs in the proposal process. The teams also had to develop and finalize their Reserved target lists, which involved setting science priorities as well as both intra- and inter-team negotiations. Proposals were due July 30, 1993, and HUT meeting notes from Aug. 9, 1993, indicate 37 proposals were received, many requesting use of more than one instrument.

Bill Blair and Wei Zheng were responsible for HUT technical reviews of the proposals, and they also supported the NASA peer review near the end of September 1993. HUT meeting notes from Oct. 11 indicate 10 GIs were selected by NASA and assigned to the various teams. HUT GIs include Claus Leitherer (STScI), Nolan Walborn (STScI), David Findley (UC Berkeley), and John Raymond (Harvard-Smithsonian CfA); Brian Espey was split between WUPPE and HUT (but technically assigned to WUPPE). GIs became involved almost immediately after selection, with attendance at an IWG meeting at MSFC on Nov. 2, 1993, and HUT GIs attending an orientation meeting at JHU on Nov. 18, 1993. At this time, the manifested launch date was Dec. 1, 1994, so we were (nominally) only about one year from launch at that time. (Of course, the launch subsequently slipped into March 1995.)

For context, it should be noted that STS-61 *Endeavour* was launched on Dec. 3, 1993, and spent the next 10 days capturing and repairing the Hubble Space Telescope. This mission included the installation of COSTAR*[21], the device that was responsible for correcting the Hubble mirror problem for several instruments (see P69), as well as installing the WFPC-2 camera with corrected optics built in. The success of this mission (see P70) went a long way toward repairing not only Hubble, but NASA's reputation in the public eye.

*[21] JHU professor Holland Ford was the PI of COSTAR; Jim Crocker from (Ball Aerospace) was the technical manager and subsequently worked at JHU on the SDSS project in the 1990's (see P91). The concept for COSTAR was largely Jim's doing.

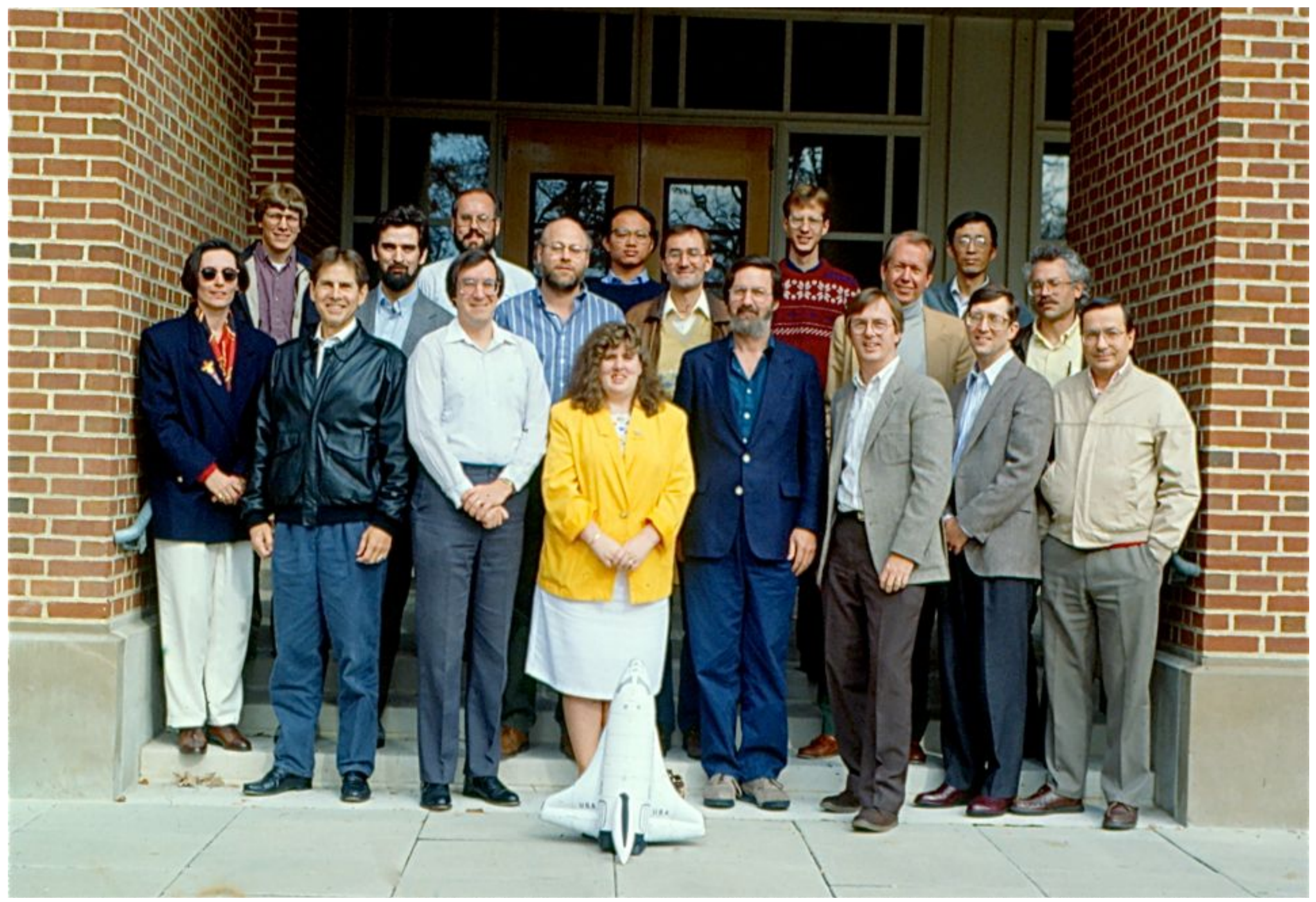

A group picture of the Astro-2 Guest Investigators with the HUT team in November 1993.

---

## Preparations for the Astro-2 Mission

Early 1994 was a busy time on both the hardware and planning sides of the project. HUT meeting notes document that HUT was officially turned over to NASA/KSC by the end of January, although team members continued to participate in alignment and other integration activities. The Level IV Interface Verification Test occurred on 4/20-21/1994, and the Level IV Mission Sequence test followed on May 19-20, 1994.

On the planning side, the time and targets for each GI had to be added to each team's fraction and then the blocks for use in block scheduling needed to be assigned for the upcoming timeline planning activities. (For instance, since HUT had more GIs and more GI time, more HUT blocks were needed for planning.) Since some GIs had time on multiple instruments, this negotiation was not as straightforward as it might have been. By mid-February 1994, POCC position assignments had been made and relevant training began shortly thereafter, including several GIs.

The document archive contains a good overview description of the somewhat daunting list of constraints that mission planning had to accommodate (see D37). There was one late-breaking twist that the planning activity had to take into consideration: there was concern that residual oxygen at shuttle altitudes could impact the HUT primary mirror at high velocity and oxidize the new SiC coating. (The previous iridium coating did not have this sensitivity.) Hence, a new planning constraint to avoid pointing within 20 degrees of the Ram vector (i.e., the instantaneous direction of motion of the shuttle) had to be added into the planning software.

With the delivery of the MPHIRD in March 1994 (defining many details of the revamped planning process), we were ready to schedule a timeline of observations.
The question remained, what WAS the launch date? In mid-March 1994, a flight manifest came out showing our flight had moved from Dec. 1, 1994 to Jan. 12, 1995. It was not the last launch day change we would encounter.

---

*(Author's Note: Much of the rest of this chapter comes from personal memoirs that were written in the eight month period during the run-up to Astro-2. They have been edited for clarity and to remove redundancy, so they flow more freely into the rest of the HUT History document.)*

---

## Planning Preparations for Astro-2

The months prior to Astro-2 were a busy time for mission planning. In June 1994, we finalized HUT blocks and assembled team blocks into the Final timeline. WUPPE used the new GROSS (GRaphical Observation Scheduling System) package, in development by MSFC, while UIT and HUT used the JHU planning s/w. WUPPE planners were somewhat cavalier in following the guidelines about limb angles and such, which caused a lot of extra work on our (my) part to get the blocks put together properly. The timeline was delivered to MSFC by the deadline of June 13, 1994.

We then had to turn around and generate accurate coordinates for all of the UIT targets. This included the problem of "finding something to look at" in many of their fields on extended targets and galaxies or star clusters, then getting an accurate coordinate, and finally making sure UIT was OK with any offset in position. This pushed on through the rest of June, with the last updates several days past the desired deadline.

Moving into July 1994, the next big job was preparing all of the Sequence Database Files (SDFs) in support of the planned timeline. There were 415 pointings planned in the "Final" Astro-2 pre-mission T/L, and each one had to have an SDF filled in with the proper entries to set up the observation. I coordinated the assignments to the HUT team and organized the review meeting as well as filling out about 30 of the files myself. It was a pile of work. We actually reviewed all 415 files in a two-day marathon meeting in late July.

The planning work eased up a little in early August 1994, which was lucky for me because HST proposals were due on Fri. August 12th and I had barely had time to work on the ones I was involved with. The only planning work needed was making minor corrections to the SDFs, handling a few situations that required special guide star work, and turning in the SEQNUM file, which lists the special Target Procedures (TPs) required for each scheduled observation. This was also due Aug. 12. Not coincidentally, I had to leave for MSFC on Aug. 12th in the evening for training exercises (over the weekend!) followed by the first mission simulation (SIM) the following week.

Everything had been going very smoothly on the hardware side. The integrated payload had been tested in a vertical configuration at KSC during the week of August 1st, followed by the Level III/II Mission Sequence Test on Wednesday - Thursday Aug. 10-11. The Wednesday test had been described by many as the smoothest day of Spacelab testing that they had ever seen at the Cape, which should have told us that something was about to happen.

## The DEP Problem and Uncertain Launch Date

When I checked my e-mail Friday morning Aug. 12, I got the first inkling that we had a problem at the Cape. Evidently mid-afternoon on Thursday Aug. 11, about an hour before the end of testing, HUT had encountered a serious problem. Initial assessment seemed to indicate a problem with the Spectrometer Processor (SP, one of two computers in HUT), but possibly with the Dedicated Experiment Processor (DEP) or with the interface board between the two. Since the payload was integrated and mounted on the Instrument Pointing System (IPS) at this point, this was a serious problem, if for no other reason than to fix it would probably require the payload to be partially disassembled to get at the offending parts. This, of course, would have an important impact on schedule. Luckily, we had insisted on having money budgeted for the construction of flight spares of these components, and work was nearly complete on these units at JHU/APL. However, while the units were built they were not completely tested and flight-qualified, still needing to have a "shake test" and undergo thermal vacuum testing.

The immediate impact, while less important than the hardware problems, was that all of the hardware and engineering folks were going to stay put in Florida to do troubleshooting instead of coming to MSFC for the training and SIM. Hence, we were about to go into a SIM, the purpose of which was largely to test the new ground commanding procedures that were being put in place for Astro-2, without any of our people who actually would be involved in this activity during the mission!

I flew to Huntsville Friday evening, Aug. 12th, and we met a portion of the Florida contingent in Nashville on the way down. The folks who stayed behind in Florida would be writing a test procedure over the weekend, going over it with KSC personnel on Monday, and running the test Tuesday. This required getting an RS232 cable to the Electronics Module (EM) while the payload was still integrated. If the test confirmed the problem, they would begin the process of taking things apart.

For the rest of us, the training at MSFC began Saturday at 8:30 a.m. and ran all day Saturday, all day Sunday, and Monday morning. Some people had to work into the evenings, while some (like myself) wrapped things up about 4 p.m. With this being about the 10th mission run out of the MSFC POCC, one would hope that they would have their act together by now, and it seems that they did. While parts of the training were boring or redundant, none of it was as bad as anything we experienced during Astro-1 training, and parts of it were actually pretty good. The worst part was that they got us in there at 8 a.m. on Monday morning for mostly redundant training, and then expected us back at 5 p.m. for a pre-SIM briefing, followed by a call to consoles at 7 p.m. for an 8 p.m. SIM start. That shift would then run until about 5:30 a.m. the next morning!

The SIM start was rather hectic, but then it settled down. We were not sitting in the part of the POCC where we would be eventually, and we didn't have any of our own Ground Support Equipment (GSE), but even so the SIM was pretty good for the purposes of refreshing our memories. I had John Raymond and Richard Buss working replanning on my shift, and we had a lot of cross training to do. We were able to use GROSS on the local MSFC computers to check some visibilities for replanning, so we got along OK without our computers. Arthur et al. staffed the air-to-ground and engineering station and actually managed to learn how to send commands to the instrument, having done some unanticipated cross training over the weekend. The SIM ended 8 a.m. on Wednesday. Overall it was a pretty good exercise, certainly better than expected ahead of time.

Wednesday was more or less a recovery day, with a SIM debrief at 2 p.m. followed by a couple of ancillary meetings. I had a nice dinner with Arthur and Brian Espey at the Outback Steak House, the first decent meal in several days. That evening, I watched NASA Select in my room, and saw crew interviews with the astronauts flying the mission to be launched the next morning at KSC. The Space Radar Lab was scheduled for a 5:54 a.m. launch aboard *Endeavour*. I gave a passing thought to the fact that the next time *Endeavour* was to fly, Astro-2 was supposed to be in the payload bay.

The next morning I was busy getting packed and, frankly, forgot that a shuttle launch was scheduled. Hence, it came of a surprise to me when one of the other team members asked me if I had heard what happened at the Cape. *Endeavour* aborted less than two seconds before launch due to an overheat reading in a temperature sensor in one of the main engines! Even if the problem turned out to be a minor one, the main engines have to be changed out if they've been fired, which would delay things by at least several weeks, maybe more. This cast a rather familiar pall over the Investigator Working Group (IWG) meeting later in the morning, where even the most basic schedule-related issues were thrown into uncertainty. Between the HUT instrument problems and this, I began to get the feeling that some replanning was going to be in order. We dragged back into Baltimore Thursday night (Aug. 18) with some pretty big question marks hanging over our heads.

It took quite awhile for a new launch date and schedule to be announced. By mid-September the new launch "assumption" seemed to settle in, and the date Feb. 23, 1995, at 6:37 GMT became the new planning target. This is only 21 days after the first launch of 1995 (scheduled for Feb. 2), and hence we would still be quite susceptible to further small slips.[*22] This amounted to a six week slip of our current timeline, which had been planned for Jan. 12, 1995. While we did some preliminary analysis of the impacts of this slip to the timeline, no one (myself included) wanted to devote much energy to analyzing the situation until we actually saw *Endeavour* get off the pad; we wanted to have some confidence that the late-February timeframe would hold before putting in significant effort. This was consistent with the CI/MPT's stated philosophy to "not chase launch slips" this time around.

[*22] Note: 21 days was nominally the NASA-mandated minimum between launches, although this rule was violated for Astro-1.

Much of September was spent in making deliveries of timeline products for the January 12, 1995 timeline, which was still the baseline. The reasoning was that this timeline was to be used for the rest of the SIMs and that many of the products (i.e. revised target procedures, the target book, etc.) would be usable for the new launch date as well, thus minimizing any wasted effort. In addition, we generated new spectral histogram files for use in the HUT simulator in support of the SIMs.

On the hardware front, the problem detected in the level III/II MST was tracked down to a short in an I/O board in the DEP. There was also concern that the corresponding board in the SP had been "stressed" by the short. Both boards were returned to APL and thoroughly tested. It was decided to fix the existing boards and re-insert them rather than replace them with the recently completed back-ups that had yet to be fully flight-qualified. Analysis of the problem indicated that a work around could have been implemented on-orbit that would have prevented this particular problem from being a catastrophic failure, but it easily could have been a problem without a work around that would have shut us down. Since the test that showed the problem was static, we were quite lucky that the problem just happened to occur in the last planned hour of ground testing before flight. Still, the fact that no obvious cause was identified for the problem and the fact that the HUT electronics had just been sitting around for a long time left most of us with an uneasy feeling. By late September, HUT was repaired, back on schedule, and again flying through the various integration tests at KSC. In addition, the rescheduled flight of *Endeavour*, carrying the Shuttle Radar Lab, got off the ground on September 30. It was time to start getting serious about replanning yet again.

## Final Preparations for the Flight of Astro-2

In October 1994, we started holding regular meetings of our HUT planning people to get everyone more familiar with our planning software and to get organized for the upcoming SIMs and mission. Also, this would help us verify the software and clean up our documentation. Even as we began replanning for the Feb. 23, 1995, launch date, we started hearing that a launch date in the first week of May was being considered, but that no official decision would be made until mid-November. This is par for the course--we need to replan, and they couldn't even give us a believable launch date to plan to!

We had another big hardware scare about mid-October. Jerry Kriss got a call from KSC that a technician had made a mistake in the procedure switching from the internal vacuum pumps to an external pump. The HUT spectrograph had to be kept under vacuum at all times to maintain its ultraviolet sensitivity. If air (and especially water vapor) got into the spectrograph, it could oxidize the photocathode and destroy the ultraviolet sensitivity of our detector. While some air leaked into the spectrograph, it was probably not bad enough to hurt anything, but it was possible that other contamination, perhaps from the vacuum pumps themselves, had gotten into the spectrograph during the incident. We spent a very uncomfortable week and a half between the incident and when we were able to get to KSC to test things. Luckily everything seemed to be OK. Another bullet dodged.

Another thing we did in October was to finalize the layout of the new HUT brochure with our graphic artist as well as finalizing the text. We also produced the basic framework for a HUT page on the newly available World-Wide Web (!). Much of the PR material I had written was incorporated into this new resource, thanks largely to the technical skills of Mary Romelfanger.*[23]

*[23] As strange as it may seem, back in those days one had to write the HTML code directly since there were no canned programs to accomplish this in the early days of the WWW.

We spent a lot of time prior to PI/Cadre SIM#2 in late October trying to reorganize into three 8-hour shifts instead of two 12-hour shifts. While this appeared feasible for much of the team, the 12-hour replanning cycle made this awkward for the replanners. We finally settled on a hybrid system whereby we basically stuck to 12 hour shifts with three replanners on each shift. However, only one planner stays the entire 12 hours on a given shift, with one planner coming in for the first eight hours and the other coming in for the last eight hours. The logistics (car arrangements, rooms, etc.) were complicated, but it seemed like it could work. We all agreed that straight 12-hour shifts (which are actually longer than this including handovers, SOPG meetings, etc.) would be unworkable for the nearly three weeks we expected to be on console at MSFC during the mission.

PI/Cadre Sim#2 was held from Oct. 24-27, 1994. We flew down to Huntsville on Monday Oct. 24 in the morning. As I was driving to the airport, I got rear-ended on the beltway. I was the front car of a four car pile-up, but I just got bumped in the back (no damage to car). The guy behind me got mashed pretty good. (One car length away from disaster--that's cutting it pretty close.) By the time I got to the airport, my back had a "catch" in it, and it bothered me the rest of the day. But luckily, it didn't develop into anything more serious.

The pre-SIM briefing was at 2 p.m. and we spent the rest of the afternoon in the Science Operations Area (SOA) of the POCC checking out our newly installed GSE (ground support equipment, which is NASAese for our computers) and getting organized for the SIM start the next morning.

I started the SIM with a 12-hour shift, beginning at 7 a.m. Tuesday. We had some problems, mostly related to the logistics of transferring files between computers, remembering (without total success) how to do things on the MSFC "VMS" computers,[*24] and getting acclimated again. Replanning activities started out slowly, but ramped up in a deafening crescendo from time to time during the SIM. One interesting twist they threw at us was a "door" problem that caused us to have to replan some targets very quickly via OCR (Operations Change Request--the normal replanning procedure is called a Replanning Request, or RR, which takes 12 hours to implement.) The SIM was good from a replanning-practice standpoint, and really pointed out a lot of places where we needed to be prepared better than we were. The split 12-hour, 8-hour shift schedule seemed to work OK. We returned to Baltimore Thursday evening after attending de-briefing meetings at MSFC.

[*24] By the early 1990's, distributed computing had become the rage and most of us on the HUT team were using Sun Unix workstations instead of Vax VMS mainframes.

Emil Venere, the JHU science writer assigned to cover HUT and Astro, published a nice article in the Nov. 7, 1994, JHU Gazette that provides a good summary of how the mission was being presented to the public. This article also announced the availability of the HUT web page for the first time in public (see P71).

We spent some time right after PI/Cadre Sim#2 trying to get going on some of the problem areas identified during the SIM. For the planners, this meant getting the RP Console Notebook organized with so-called "cheatsheets" for many of the tasks that we had to do during one aspect or another of replanning, as well as putting together some new software to help us better extract information from MSFC planning files. Even with the holes in our procedures, it was clear we were going to be so much better prepared this time around compared with Astro-1.

Most of my effort had to go into organizing a new round of timeline planning for the new launch date, which was still officially February 23, 1995. No one really believed this date because it was exactly 21 days after the first scheduled launch of 1995, and of course we might also get delayed by weather or some other problem. I finally convinced people that we should plan the timeline to a March 2 date (i.e., one week later than the official date) and if the official date stayed February 23, we could slip the timeline back to that date with our launchslip program for delivery. As luck would have it, after we had started planning the new T/L, NASA reviewed the manifest yet again, and assigned a new launch date for Astro-2: Mar. 2, 1995!

This mission planning cycle was very hectic because we were under a lot of time pressure. We had to get the new Mission Timeline (MTL) file put together very quickly, and consequently, there were additional changes needed even after the file was supposed to have been "finalized." In addition, the timeline itself was a difficult one to plan. It's funny how each of the timelines we generated seemed to have it's own personality; we seemed to have to "wrestle" this one into position because of the changing visibilities of the targets. Many of the "orbital day" targets in the previous timeline now had night-into-day visibility because of the slip, which make poor use of the orbital night time (since the setup for those observations occurs in orbital night). In a related issue, the day/night breakdown of the overall program had shifted so that the night observing time was even more oversubscribed than usual, so a lot of bargaining and target swapping had to occur in real-time as we were planning the timeline. The bottom line turned out to be a solid, high efficiency timeline (almost 80% on target efficiency!), but it didn't just fall into place; it took a lot of work.

In the portion of the timeline that the WUPPE team had generated with GROSS and sent to me, they had bent some of the planning rules that we normally apply, mainly lower earth-limb avoidance angles at the start of observations. I had assumed these were planning errors generated by the software they were using since we had seen this sort of thing before, both from WUPPE and UIT inputs. Running their inputs through our s/w caught and corrected these problems. However, it turned out they actually wanted them that way; since they were often observing very bright stars, they assumed acquisitions could start at lower limb angles than we normally allowed.

When I put the corrected file out for review in late November, I got accused (electronically of course, and forwarded around to the rest of the planners and MSFC folks) of screwing up their pointings, (i.e., the implication was that I had done this on purpose). The funny part is, they bent the rules, didn't ask permission or consensus from the other teams in doing it, didn't tell me that they'd done it, and then they accused me of screwing things up when I “corrected” the problems! In actuality, it wasn’t very funny to me. Such was life on the planning front, where no good deed went unpunished.

PI/Cadre SIM#3 took place from December 4-7, 1994. It was a full 48-hour SIM that would start as if we were 5 hours after launch and ready for activation activities. I left the house at 5:45 a.m. on Monday to catch a ride to the airport with a neighbor who was also flying out that morning. Speaking of which, we flew Southwest Airlines this time, which had a new direct flight from BWI to Birmingham, AL. Birmingham was about 90 miles south of Huntsville, but the tickets were almost $500 cheaper per person ($199 instead of $688) and fully changeable. Of course, we had to rent cars anyway, so there was just a small inconvenience from having to drive an hour and a half to get to Huntsville. With University overhead taken into account, we figured this saved the project $17,000 on this one trip alone!

Monday was spent getting the Payload Operations Control Center (POCC) ready for the SIM. We had a new workstation for the replanners this time, so we had to fool around to get everything working right. We tried to get the MSFC planning software to run, but it kept bombing out. Chris Anderson, who was the main proponent of GROSS on the science teams, even came over and tried to help us get it running, without success. We had a terrific dinner that night at The Outback Steakhouse restaurant, which was particularly appreciated since the schedule I was working would prevent me from having a decent dinner for the rest of the week.

The SIM started at 8 a.m. on Tuesday, but I didn't have to come in until about 10 a.m. I pulled a LONG shift from then until midnight, but then got a shorter shift the next day in return. This SIM was covering the activation period, and no observations had been "missed," so the replanning activities were fairly mundane. This allowed us to concentrate more on cross training at the replanning desk as we continued to familiarize ourselves with processes. Again, overall, a decent training SIM compared with anything on Astro-1.

After returning from the SIM, we had to get right back to preparations for the March 2 timeline. We basically only had two weeks to work up any coordinate or roll angle updates, submit them, get finalized roll angles back from MSFC, and finalize our sequence files and Target Procedures. This usually takes us a month, and that's without the complication of the holidays thrown in. Except for a three day Christmas weekend, I worked straight through. Tuesday, Dec. 27th was memorable, as we came in early, ran BUILDSEQ (the program that processes the database, checks for certain errors, and generates the sequence files), and prepared the materials for a review meeting, which took all of Tuesday afternoon. Then we had to turn right around and generate the information for the new target book pages by the end of the week and finalize the text of new target procedures in support of the JOTP (Joint Observation Target Procedures) book. Oh yes, and while we were at it, we also had to get ready for the next SIM, which was right after the first of the year!

A highlight in this period was the delivery of our new HUT brochure. It arrived on Wed. Dec. 28th, and reaction was quite favorable. I rushed to get some mailed out to family and friends to beat the price increase on first class mail, which (historical note) was going up from $0.29 to $0.32 on January 1st.

## JIS#1

Since New Years day was on a Sunday, Monday Jan. 2, 1995 was technically a holiday. I say technically because I went in to work in the afternoon for a few hours only to find several of my cohorts had been in most of the day trying to finish up their support activities for the new timeline. I came in mostly to get ready for the SIM, for which we had to leave bright and early the next morning (7 a.m. flight!). What a way to start the new year!

This SIM was the first of two Joint Integrated Sims, or (you guessed it) JISs, which involve not only the NASA/MSFC and PI teams, but also the flight operations people from NASA/JSC in Houston. Actually, they were so slow in getting us the specifics about when things were happening that we had to make the reservations before we knew. Hence, we were locked into a 7 a.m. flight to Birmingham, and then had to cool our heels until the pre-sim briefing, which occurred at 4 p.m. While we can always find things to do, most of us would have taken an afternoon flight down if we'd had a choice. Then it turned out that the pre-SIM briefing contained essentially no useful information anyway (about 90% the same as previous SIM briefings), so we were not real happy. A nice dinner at the Outback Steakhouse that evening helped ease our angst.

The JIS was scheduled to start at 8 a.m. Wednesday. They seemed to have a hard time getting everyone sync'ed up, and things started late. This SIM was scheduled to start at L-9 minutes, so we expected a launch delay (at least) and possibly a non-optimal orbit. Although the SIM started late, when they started it, the purported launch actually occurred "on time." Only one catch: due to a JSC simulator problem, the "shuttle" aborted to a landing in Africa instead of to a "low orbit," which is what they were trying to simulate! Hence, they backed it all up, reloaded their computers, and tried it again. This time they made it to orbit, but just barely, to a 106 nautical miles. (Our nominal orbit was 195 nautical miles.)

This turned out to be a pretty good exercise. There was a lot of confusion about just how much they could/should raise the orbit and how much fuel it would use to do so. This in turn could affect the orbital period how many shuttle maneuvers we could do for science operations. On the "science" side, figuring out how to adjust our science plan for the new (shorter) orbit was something we hadn't done since Astro-1 days. Instead of a "timeline slip", it was more of a "timeline contraction." We had to scratch our heads a little bit, but we finally got the adjustments made.

We tried a different schedule for the replanners this SIM, which was basically a cascade arrangement with one person handing over every 4-5 hours and two people on duty at any one time. It worked all right, although the SIM wasn't really long enough (36 hours) to get into a pattern and see how it worked. I worked a long shift, starting at 7 a.m. on console and not leaving until almost 7 p.m. (Paul Feldman arrived at 5 p.m., but he had so many questions it took me two hours to hand over.)

The rest of the SIM went OK, ending at 8 p.m. on Thursday. They had a lot of down time (simulator problems) and, although lots of fake problems occurred, most of them were nothing to do with HUT. Also, since we were mainly doing activation, the science replanning was almost nil. Still, it was one step closer to the real thing.

We had a post-SIM debrief Friday morning before driving to Birmingham for the flight home. We had changed reservations to a mid-day flight to get home at a reasonable hour, but a weather system over the eastern half of the country took care of that. Our flight connected through Louisville. We got to Louisville OK (although the takeoff from Birmingham experienced some very interesting cross winds), but our incoming flight was late. We took off from Louisville about a half an hour after we were supposed to have arrived in Baltimore. When we got to Baltimore it was raining right along, but it could have been worse: it was a couple of degrees above freezing. I gave Arthur and Jerry a ride home, and got home myself about 8:45 p.m. Like I said earlier, what a way to start the new year!

The couple of weeks following the SIM were very busy. We de-briefed ourselves after the SIM and generated a number of action items for various of us to work. We had to finish and deliver the Target Book materials the week after the SIM. And with that done, we finally had completed the deliveries necessary to support the March 2 timeline.

Next we put effort into rewriting the JHU press kit materials to make them as good as possible. It had become quite obvious that the NASA/MSFC PAO efforts were going to be sub-par, at least as far as describing the science was concerned, and so it became more important to have the JHU press kit just right. We finalized these materials and gave them to Emil Venere (our JHU press person) on Jan. 20, just before we left for our final SIM.

## JIS#2

We left Baltimore early Monday morning on the SW flight to Birmingham. When I left the house at 5:30 a.m. it was still dark and the sky was clear. The moon was third quarter and Venus and Jupiter were shining brightly in the eastern sky. Being the planner that I am, it struck me that they were getting close to where they were supposed to be for the way we had planned them in the March 2 timeline. It gave me a funny feeling, connecting squiggly lines on a timeline plot to the actual naked eye sky, and for the first time in a long time I got that feeling of anticipation, that it really was going to happen and that it was getting close. Launch was scheduled just over 5 weeks away.

Another treat occurred on the ride up from Birmingham to Huntsville. I was driving Arthur's rental car because he had had a very short night and wanted to be able to doze off. It was cloudy and gray in Birmingham (it always seemed to be...), but as we drove north it began to break up. There had been a light snow the night before and, although the roads were clear and dry, the snow clung to every little twig and branch of the trees. For about 25 miles it was like driving through a fairy land. Very pretty.

I dropped Arthur off at the PAO training in building 4207 at MSFC just in time for his 10 a.m. "class." In the mean time, I went to the POCC and got things ready for JIS#2, our last SIM prior to the mission.

After lunch, I went to the PAO training exercise myself. It was useful, but would have been more so if they had told us how to prepare for it ahead of time. After talking about techniques for focusing our messages and presentations, we went through the exercise of actually making a presentation. I did this with Peter Vedder (NASA HQ), Ted Stecher (UIT PI), Harry Ferguson, and Regina Schulte-Ladbeck (WUPPE). Then Bob Stachnik asked us some questions in "press conference" format. All in all, an interesting exercise.

Preparations for the JIS were getting all too familiar. At least there was no pre-SIM briefing this time (which was no great loss judging from the past ones). Still, there was a lot of organizational work to do. Since most of us had on earphones at our stations in the POCC, we wanted to have as much as possible within easy reach of our positions. With this task done, I checked in at the Residence Inn, which was where we were considering staying during the mission. It was pretty nice and also good to have different surroundings from where we had stayed on so many previous trips.

The JIS started at 8 a.m., and we were on console by 7. There was supposed to be a pre-SIM briefing on the Loops at 7:45, but it never happened: the SIM just sort of "started". I was on shift until noon, then handed off to Paul Feldman. I hung around until about 2 p.m. working a task related to the potential use of a partial door opening on HUT if the 50 $cm^2$ aperture was broken. I had lunch with Arthur and Tom Brown at Shoneys. I rested a little and worked some in my room. Then went to dinner at Green Hills Grille with Jerry and Warren Moos. I watched some of Bill Clinton's State of the Union Address before turning in about 9 p.m.

I had to be in for my shift by 4 a.m., which was the shift I would be working during the mission. They were throwing lots of problems at the MSFC Cadre, and we were making very few "observations", but they really weren't throwing many difficulties at "us" (i.e. the instrument teams), which made the SIM kind of boring for us. The replanning was restricted to using the first 66 targets in the timeline, as it had been in past SIMs, which made the replanning fairly unrealistic. Still, we always think of new things that need to be addressed. The bottom line was that we still felt like we had a lot of work to do to feel completely prepared before the mission.

That night, as the SIM wound down at 8 p.m., we all met at a place called the Green Bottle. It was quite a fancy restaurant, and we didn't get out until after midnight! We had about 18 people there, and it somewhat unexpectedly (to me anyway) turned into a bit of a blowout. The wine started flowing and didn't really stop. I guess since this was the last sim, people were in the mood to party. Arthur unexpectedly picked up the tab for everyone, so I don't know how much it cost, but I'm guessing $40-50 per person![*25] Some folks were actually flying back on the 9:30 a.m. flight from Birmingham. Must have been a short night for them.

[*25] In 1995 dollars; that was a lot of money back then!

I met Jerry the next morning about 10 a.m. We took my car back to the Huntsville airport, and then went to the POCC. The post-SIM debrief was at 1 p.m., and was quite straightforward. A real good rapport had been developed between the MSFC Cadre and the PI teams this time around, building on the positive CI/MPT interactions. The feelings between MSFC and JSC, however, seemed much less satisfactory. They were at each other's throats much of the time, which was an indicator of a “control” issue between the two NASA centers. (Maybe they needed a round of “continuous improvement” discussions to work this out!) We left about 3:30 for Birmingham, with Jerry, Jeff and I on the same flight home.

In early February 1995, after the JIS, we started weekly RP meetings to whip the Replanner Handbook into shape as well as preparing the various files and materials needed to support the Mar. 2 timeline. We also had to get the various computer accounts straightened out and consolidated. Between preparing and running these meetings and doing my share of the work (as well as the fact that there were 10 other things going on at the same time), this was a busy time. Writing and editing PR materials, including web page support and a "science overview" article I put together, getting phase II submissions done for a Cycle 5 Hubble proposal (with a new proposal submission system), and getting my Vela Voyager paper submitted to the Astrophysical Journal all joined to make things very hectic. Plus, I had several presentations to make a local schools during this period, and I had to prepare a “job talk” to present at STScI on Thursday Feb. 23rd! (Yes, like so many on the project, we had to be looking ahead to what was possibly next.)

An interesting flap developed during the PI telecon on Tues. Feb. 14, 1995. The word was coming down from "on high" at NASA that items such as pins, patches, color brochures, etc., were being frowned upon as frivolous, and could be construed by congress as wasteful expenditures and might hurt future NASA budgets. This was just crazy! NASA had been pushing education and outreach activities very hard up to this point, and they were spending who knows how much in a rush to get Astro-2 "on-line" via the World Wide Web (for which a relatively few privileged early adopters would be able to use) and then they complain that educational brochures are frivolous! In reality, this sort of outreach keeps the public engaged and helps NASA justify its existence to the tax payers. The next day, HUT team member and professor Dick Henry provided a "new" brochure as a joke, printed on a piece of tissue paper!

As with Astro-1, the launch itself was expected to be a big event, and JHU planned a “launch party” in the Schafler auditorium of the Bloomberg Physics and Astronomy building, even though the launch was scheduled for the wee hours. The Maryland Space

Grant had hosted the Astro-1 launch party back in 1990. Since all HUT personnel would be at KSC or MSFC for launch, a host was needed locally at JHU. Luckily, Nolan Walborn from STScI had been selected as a HUT Guest Investigator and was not heavily involved in the POCC operations plan, so he hosted the launch event at JHU (see P72).

The L-14 day press conferences were held on February 16, 1995. I couldn't really afford to, but I spent a good part of the day watching NASA-TV and screwing around with VCRs to tape the proceedings. Some of the themes we had developed came in handy. There was some "memory" in the press about the problems from Astro-1, and they were definitely after "how many pointings" we were planning so they could start their bean counting. The science briefing went well, and the crew briefing was good, too.

A nice memory was from Saturday, Feb. 18, 1995: Arthur and his wife Frauke hosted a very nice "send off" party for the team at their house in the Pimlico area of Baltimore. Sam Durrance was back in town for the last time before launch. My wife Jean had left the day before to visit a friend for the weekend, so when I showed up alone for the party the joke became that she "left town" when I told her I was writing a song parody again. In actuality, I wrote three songs: *Sittin' in the POCC and SOA*, *Still Planning After All These Years*, and *Out on the Loops*.*[26] I even managed to coerce Brian Espey and Van Dixon into being "pips" in the background on the last number (*Billy B. and the RPs*). A small excerpt from the first one:

> *I saw an old timeline on the desk last night.*
> *I flipped the pages and I smiled.*
> *But the date was November and it had the wrong year.*
> *Still planning after all these years,*
> *Oh, still planning after all these years.*
>
> *I ran a new launchslip on the plan last night.*
> *It did some damage, I must say.*
> *But I'll patch it back together even though it pains my rear.*
> *Still planning after all these years,*
> *Oh, still planning after all these years.*
>
> *(Etc.)*

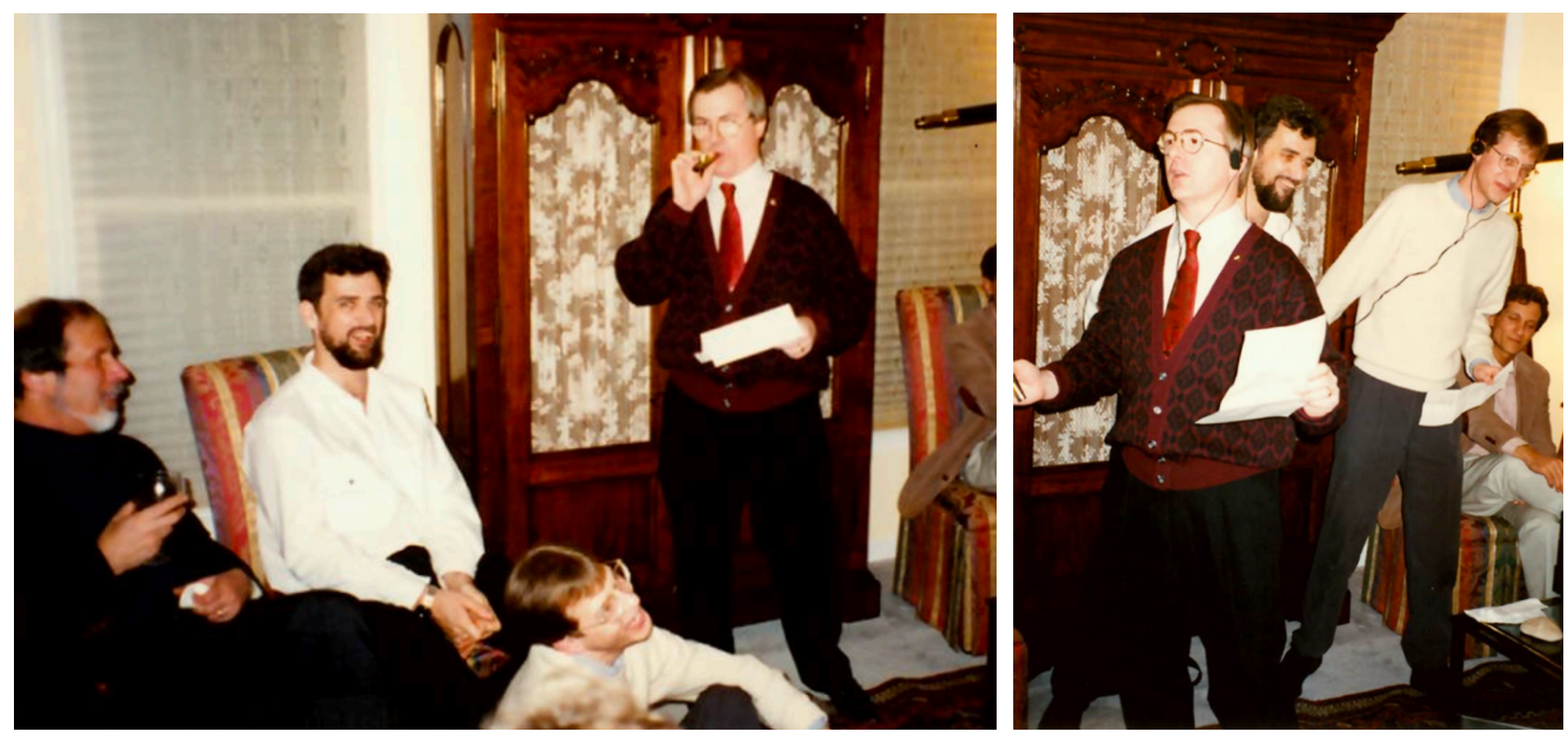

Revelry at the Feb. 18, 1995 gathering, including a rendition of “Out on the Loops” (right).

The songs were well-received, given that a number of bottles of wine had been emptied prior to the singing. *Baltimore Sun* reporter Frank Roylance was at the party, but thankfully had left BEFORE the singing started. The food was very nice indeed, with shrimp cocktail and salmon, and crackers and cheeses for appetizers, plus a very nice London broil done on the grill as the main course. A good time was had by all.

*[26] After the mission, as with Astro-1, there was an Astro-2 songbook along with an appendix of limericks that resulted from another HUT party; see 1995-08_A2_songbook_and_App.pdf in the **Diaries-Memories** portion of the online archive.

The pace really picked up the week of February 20th, the last full week home before we left for Huntsville and the flight. Things were complicated by the fact that I had a job interview at STScI on Thursday and had to prepare a science talk as well as continuing mission planning preparations. Tuesday was wild. I had a replanners meeting from 10-noon, a HUT team meeting from 1:15 - 3:00, the Astro PI telecon from 3:15 - 4, and a special telecon with our cataclysmic variable collaborators from 4:30 - 5:30! And my collaborator John Raymond was visiting from the Center for Astrophysics for the day. John and I met Jean and the kids for dinner at Chilis, then I came back into work to put my talk together, staying until 2 a.m. When I came back into the office after dinner, I checked my e-mail, which had gone unattended to all day, and I had 32 new messages, mostly related to mission planning!

After another late night on Wednesday to finish getting my talk ready, the interview at STScI almost seemed like a relief. I really wasn't nervous (I think I was too tired to be nervous). I didn't feel like my talk was up to my usual standards, but under the circumstances, I think it was OK, and IT WAS OVER! When I came back to the office in the afternoon, we had begun hearing various rumors about concerns with the Shuttle processing at the Cape. An article in the Feb. 27 *Aviation Week and Space Technology* noted,

> "The mission's targeted Mar. 2 launch date was under review late last week as NASA officials debated whether to remove and inspect one of *Endeavour*'s three main engines for cracks in its high-pressure fuel turbopump. The concern arose after inspectors found cracks on a pump run on a developmental engine at Stennis Space Center, Miss., on Feb. 22."

Since the mission did not end up being delayed, NASA must have resolved their concerns about this situation. We were never "in the loop" on this discussion.

For details, see the Aviation Week article P73; this article also gives a good pre-mission summary. It notes that the mission is expected to set a "duration record of nearly 16 days." After outlining some of the technical improvements for this mission, the article notes:

> *"Perhaps most importantly, mission planners developed detailed target observation plans for only the first few days of the flight. [NASA's mission manager Robert] Jayroe said the remaining observations will be conducted from a menu of about 600 target UV sources, which should simplify replanning if Endeavour's launch is delayed."*

While not entirely accurate (we had a full timeline planned, but yes, we had full flexibility to replan the entire mission on the fly if needed), I particularly appreciated the call out of the improvements to the mission planning process in this article.

# Chapter 6:  The Astro-2 Mission

*(Author's note:  As with earlier chapters, much of this chapter was first compiled and written in the first few months after Astro-2.  I have left it in the "day-by-day" format in which it was first written.  I refer the reader to the article P73 which provides a good published pre-mission perspective for Astro-2 from Aviation Week.)*

**Monday Feb. 27, 1995:**
It was a gray, icy morning in Baltimore; school was two hours late for the kids.  I went into my JHU office in the morning to finish getting things together for the trip to Huntsville for the mission and to tape a TV interview for Channel 24, a small local station.  Emil Venere was the JHU News and Information person assigned to cover HUT, and his wife, Jennifer Streisand, did the interview.  Although I didn't see it until much later, she did a very credible job putting together a 5-6 minute report on HUT and Astro-2.   It was aired numerous times during the next few weeks, and a follow-up report was even done in late April.

In addition to getting ready for the trip, I was trying desperately to get a Cycle 5 phase II HST proposal submitted before I left town.  (This is the second submission after a Hubble proposal has been accepted, that specifies all of the details needed to perform the actual observations.)  I had submitted a version by the initial deadline (Feb. 15), but there were outstanding questions that had yet to be dealt with.  They had sent it back to me and said to re-submit it by Mar. 15.  I knew if I didn't get it in before I left town it would probably be late (assuming of course that the mission went off as planned on March 2).  But alas, after meeting with Howard Lanning, Tony Keyes, and Antonella Nota of STScI, there were STILL loose ends about how to execute the observations.  I thus felt compelled to take a big stack of HST support materials along to Huntsville to finish up the proposal from there.  Silly me. I never looked at it, and in reality, I never got the proposal re-submitted until sometime in August!

During lunch I watched NASA-TV (previously called "NASA Select"), which was showing the arrival of "our" astronauts at KSC (a tape really--it had occurred the previous night).   Sam looked good--all smiles-- and he had on his blue jumpsuit and "astronaut swagger" (as did all the others).  Parts of this were very reminiscent of scenes from Astro-1, although the comments and Q/A with reporters was VERY brief and very superficial.

I left the office about 2 p.m.  On my way home I stopped at Radebaugh's florist and ordered two bouquets of flowers to be delivered to my wife Jean at various times while I was gone (which included her birthday).  I got home just as the kids were getting home from school and was able to see them for about 15 minutes before leaving for the airport.  I rode to the airport with Richard and Mary Buss, who picked me up about 3:35 p.m.  Mary dropped us off about 4:15 p.m.  We checked our bags and went down to the gate, where a host of other HUTsters were already clustered.  The flight left about 30 minutes late due to the weather, and it was a rather crowded, bumpy flight. It was gray and drizzly, about 60 degrees in Birmingham when we arrived.

I drove up from Birmingham with Richard Buss and Warren Moos. We stopped about half way to Huntsville and had dinner at a "Cracker Barrel" restaurant. The chatter was a combination of Astro-1 recollections, comparisons to Astro-2 (so far), and mission planning issues and concerns. Warren dropped me at the Huntsville airport to pick up a rental car—a nice, red Buick Century that served me well over the next two and a half weeks. I drove to NASA/MSFC, stopped at security for my badge and POCC (Payload Operations and Control Center) access card, and then stopped out to the POCC to drop off some of the support materials I'd hand-carried down. Staffing of the POCC was to begin the next morning, and I wanted to be sure the first shift had what they needed on hand. Of course, I signed on to the computer and immediately got embroiled in "doing stuff." I finally tore myself away at 11:30 p.m. (CST from now on) and checked into the Residence Inn, just off University Drive and Jordan Lane. Room 712 was home away from home for the next few weeks, but I didn't see any too much of it. I did some quick unpacking and got to bed about 12:15 a.m.

**Tues. February 28, 1995:**
I was up at 5 a.m., showered, and was out to the POCC by 6 a.m. John Raymond was on duty, having come in at about 2 a.m. to start getting his sleep schedule adjusted. It was the first time I had seen my good friend on this trip. John had been selected as a Guest Investigator for Astro-2, and was really one of the GIs that participated in most activities (including the SIMs and now the mission) as a full team member. John was also working on the replanning desk, and I was glad that the schedule allowed us to overlap for at least half of a nominal 8-hour shift. The weather prediction for KSC was not sounding good for the night of launch, but it was still two days off, and besides we still had plenty to do to get ready for the mission.

The work today was mostly on checking "functionality." We exercised a lot of the software, plotting, printing, etc., and checked connections back to JHU. I contacted all of our collaborators working on cataclysmic variable stars (CVs) to make sure we really were in direct contact. CVs are binary star systems that can go into outbursts at irregular intervals. We had a set of CVs that were being monitored from ground-based telescopes, and if an outburst occurred during the mission, we had the flexibility to quickly get that object onto the observation schedule.

Since Paul Feldman and Brian Espey were at KSC for the launch, we were down two people on the replanning desk. To cover the bases, I stayed on through the afternoon and we adjusted our schedules temporarily to close the gaps. (My "normal" shift time was 4 a.m. to noon, but there were few days during Astro-2 that I left before 5 p.m.) I left the POCC about 5:15 p.m. and had dinner at the Outback Steak House, sitting with Jerry Kriss, Ji-Cheng Liu (graduate student) and Nemo Nguyen (ugrad). After a nice dinner, I got back to my room about 7 p.m. and did a little "housekeeping." I got into bed about 8 p.m. and luckily fell asleep fairly quickly.

## The Mission Itself...

**Wed. 3/1/95 and Thurs. 3/2/95: (These two days just blurred together, so I might as well write them up that way...)**
I woke up about 3 a.m. and headed out to the POCC by 4 a.m. for my normal shift time. We continued preparations for the mission. We were less than 24 hours from launch, IF the weather cooperated. It was not sounding good, however. Because they can only try for launch on two consecutive days, there was a lot of discussion as to whether they should even try for tonight, or just delay a day and go for Friday and Saturday.*[27] In any event, I was still at the POCC at 4 p.m. when they made the final decision to tank it up, even though they were stating the odds at only 20% for acceptable weather. (Someone apparently had a crystal ball...)

*[27] Saturday was sounding best, so if they tried for Thursday and Friday and it was bad, we would have to stand down Saturday and try again on Sunday.

I left the POCC at 5:30 p.m. and literally swung through the Wendy's near the motel to get a bite to eat. If the launch was on for tonight, the window opened at the nominal launch time of 12:37 a.m. local (1:37 EST), and I was due on duty beginning at 4 a.m. I found NASA-TV on channel 55 of the cable TV at the motel, and tried (mostly unsuccessfully) to doze from about 7 to 10 p.m. As they showed the astronauts suiting up, then leaving for the pad, then getting in the shuttle, I kept jumping up to try and take pictures off of the TV screen! I finally gave up trying to sleep, and just got up, showered, and headed out to the POCC.

It was a cool, damp, foggy night in Huntsville. The light drizzle and fog and the nearly empty roads made for a rather eerie drive out to the POCC. I had brought some Windham Hill tapes from home for playing in my room and in my car, if indeed the car had a tape player. This one did, and the instrumental music made the drive even that much more surreal. Could it be that we were about to launch? Even now, it didn't seem real—the Astro-1 experience of getting so close so many times combined with the negative weather report kept me from getting too excited. Still, the firecracker was ready to be lit, and it could happen. Maybe luck would be with us this time.

I arrived at the POCC about midnight. They had moved a 12 foot model shuttle out in front of the POCC and had it lit with big spotlights. As I crouched down to steady my camera for a time exposure picture, I could almost imagine that this was the real thing and I was standing there waiting for the launch! I finally started getting excited.

I had decided to get dressed up for the launch. Shirt and "shuttle" tie, navy sport coat with Astro-2 lapel pins, slicked back hair—the works. What the heck, if not now, when? I hung up my raincoat and headed for the HUT section of the Science Operations Area (SOA). As I entered, there was an audible cheer, applause, and ooo's and ahh's, mostly from the WUPPE group whose area was adjacent to the entrance. After a brief bow, I joined my colleagues in the HUT area.

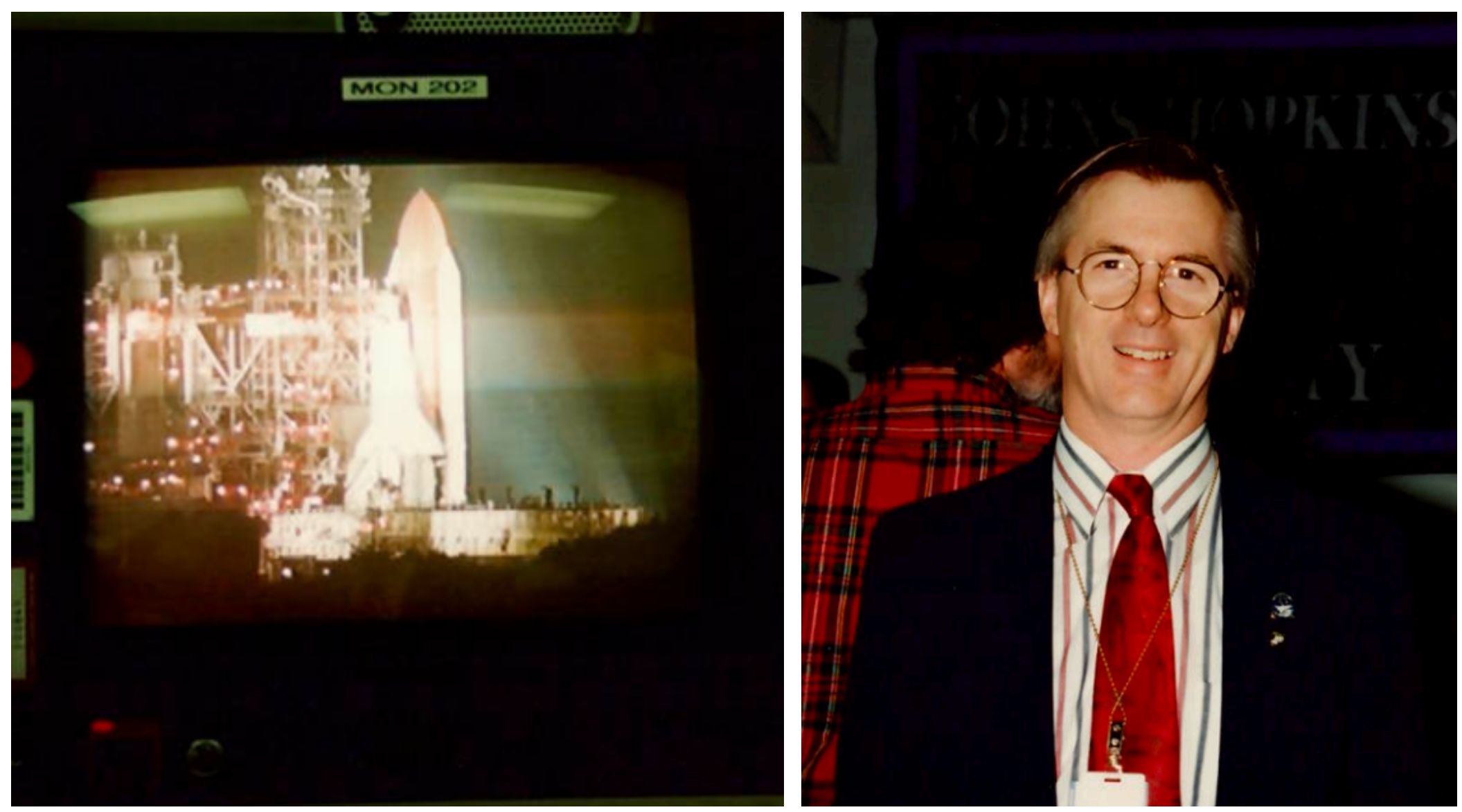

Left: *Endeavour* on the pad as seen from the POCC. Right: Spiffy Bill Blair.

The POCC was packed. It seemed I was not the only one who couldn't sleep. The countdown was proceeding smoothly and everything was sealed up and ready to go. Looking at the view of the shuttle on the pad on our monitors, I couldn't help but have flashbacks to the Astro-1 launch, including remembering thoughts that had escaped me in the meantime. Could it be that there were really people in that thing, let alone an actual friend and colleague, Sam Durrance? What if something happened? Could any amount or quality of science be worth the risk? Were we really ready if it went tonight? What if we launched late or went into a non-optimal orbit? Passing thoughts, but at a time like this, one's mind tends to race through a lot of venues.

As we passed within a half hour of launch, the countdown continued to go smoothly and the weather reports started sounding more hopeful. I picked up a spare headset and listened in to myriad conversations taking place simultaneously in my ear. It's kind of funny with all the high tech apparatus, communication links, etc., that were in place to support the POCC once the mission was underway, it was very hard to get the "straight poop" on conditions at the Cape! I remember this from Astro-1 as well…it was even hard to monitor the countdown accurately. At T-9 minutes they went into a planned hold in the countdown while final checks were made with all of the launch operations personnel. Low and behold, after a few minutes the word came back that we were ready to resume the countdown!! Even the balky weather was cooperating! We went into a very brief hold at about 5 minutes, and then the count resumed. It was really going to happen!

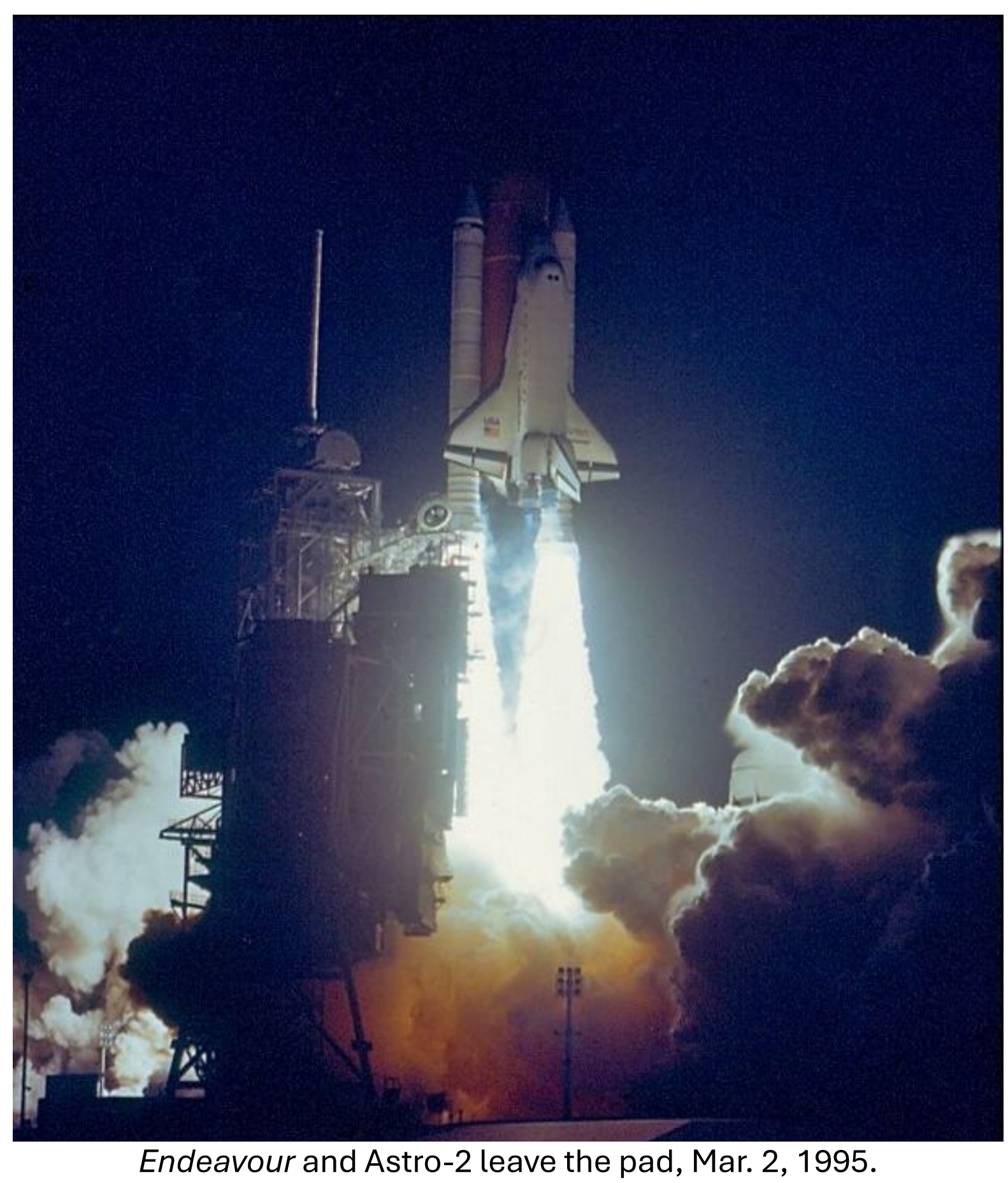

*Endeavour* and Astro-2 leave the pad, Mar. 2, 1995.

The launch and assent were picture perfect. The brief hold in the final countdown pushed us to launch 1 minute 13 seconds late—completely inconsequential from a replanning standpoint. The TV images of the launch were somewhat less spectacular than Astro-1 due to the clouds in the general vicinity of the Cape, but we were still able to follow it all the way up past SRB separation until it became as one of the stars we were about to observe. After holding our breath for 8 minutes (try it sometime...) we could breathe again. Astro-2 was in orbit! (See P74 for the AP story about launch, and P75 for the *New York Times* story.)

It's not easy to describe how I felt over the next hour; perhaps a pleasant version of "shock" is the best I can do to describe it. Of course, there had been a big cheer at liftoff and lots of cameras popping off, but now we were on orbit, and there was a lull in the activity until payload activation activities kicked in. An on-time launch into a nominal orbit was not something we had spent a lot of time considering, the possibility seemed so remote. The replanners on duty, Van Dixon and John Raymond, sort of looked at me as if to say, "Shouldn't we be doing something?" but there really wasn't much to be done...nothing bad had happened to us! After the initial hand shakes, hugs, and back-slaps, most of us with Astro-1 experience sort of sat around and shook our heads with silly grins on our faces. We really almost couldn't believe what had just happened. Could it be we were really charmed this time?

After I recovered slightly, I got on the e-mail and sent out a message to all of the collaborators who were making (or attempting to make) joint observations to let them know of our good fortune.  We then started getting our first round of "replanning requests" (RRs) ready for submission by 6 a.m.  Even in "nominal" operations, there is a steady stream of ongoing activity—a flow to the replanning—that takes place in 12 hour cycles.  The replanners are always looking out 12 - 24 hours ahead in the timeline, making adjustments as needed or simply submitting the nominal pre-mission plans, as appropriate.  As I picked up my normal shift of duty at 4 a.m., it struck me that this was the real thing, and we were in it for the long haul.  I had better be careful not to "burn out", at least before the end of the mission.  As it turned out, I just about made it.

Several people, including PI Arthur Davidsen, Paul Feldman, and Brian Espey, had been at the Cape for the launch and had arranged to have a private plane flight to Huntsville after the launch.  Arthur and Paul came directly in from the airport and arrived at the POCC about 7 a.m. all smiles.  They related a few particulars about the launch, not the least of which was the remarkable clearing that permitted the launch to occur.  There wasn't much time to waste, however, as we had to get Arthur up to speed on local status in preparation for the first Science Operations and Planning Group (SOPG) meeting at 8 a.m.  This group would meet every 12 hours to discuss RRs and any other matters needing attention and coordination between the science teams, NASA mission planners and operations folks, and mission management.

Except for some very minor glitches, activation proceeded very smoothly during this first shift.  The IPS (Instrument Pointing System) activation went well, and HUT came up like a champ.  WUPPE had to do some minor trouble shooting that pushed us somewhat late on the timeline, but otherwise things were going well.  One brief glitch I will relate had to do with activation of one of the DDS's (Data Display Systems--basically "smart terminals") onboard the shuttle.  There are two such units, one primarily for use by the mission specialist on duty to set-up the IPS, and the other for use by the payload specialist to set-up the telescopes.  On Astro-1, one of the DDS's (called DDU's back then) had over-heated and shut down at 10 hours into the mission, never to return.  While the MS and PS can share one DDS, it causes many normally parallel operations to be done serially, slowing things down considerably.  (If you recall, on Astro-1 the second DDS also over-heated, leaving us temporarily "stranded" after 4-1/2 days.)  On Astro-2, one of the first things we heard from the astronauts was that they couldn't get one of the DDS's to come to life!  After about 20 minutes with our hearts in our throats, the balky unit turned on and to my knowledge caused no further problems during the mission.  Charmed.

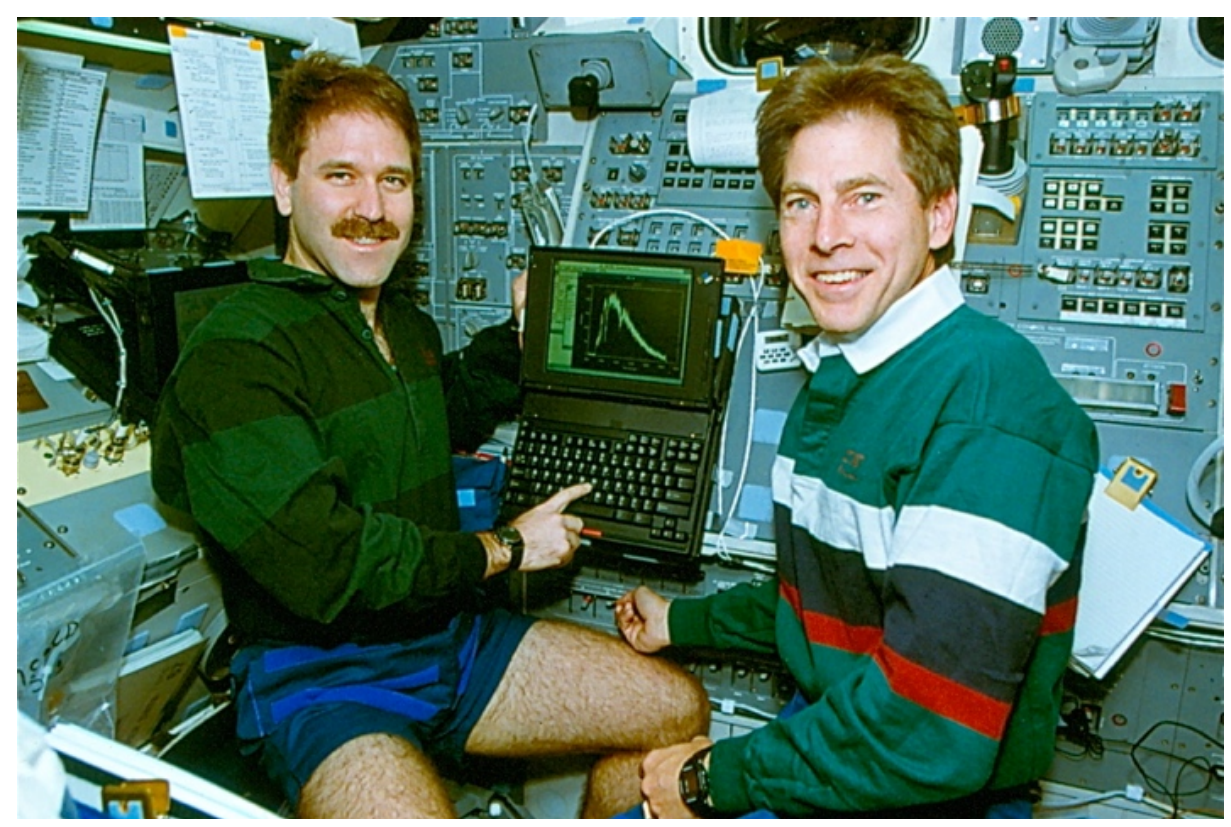
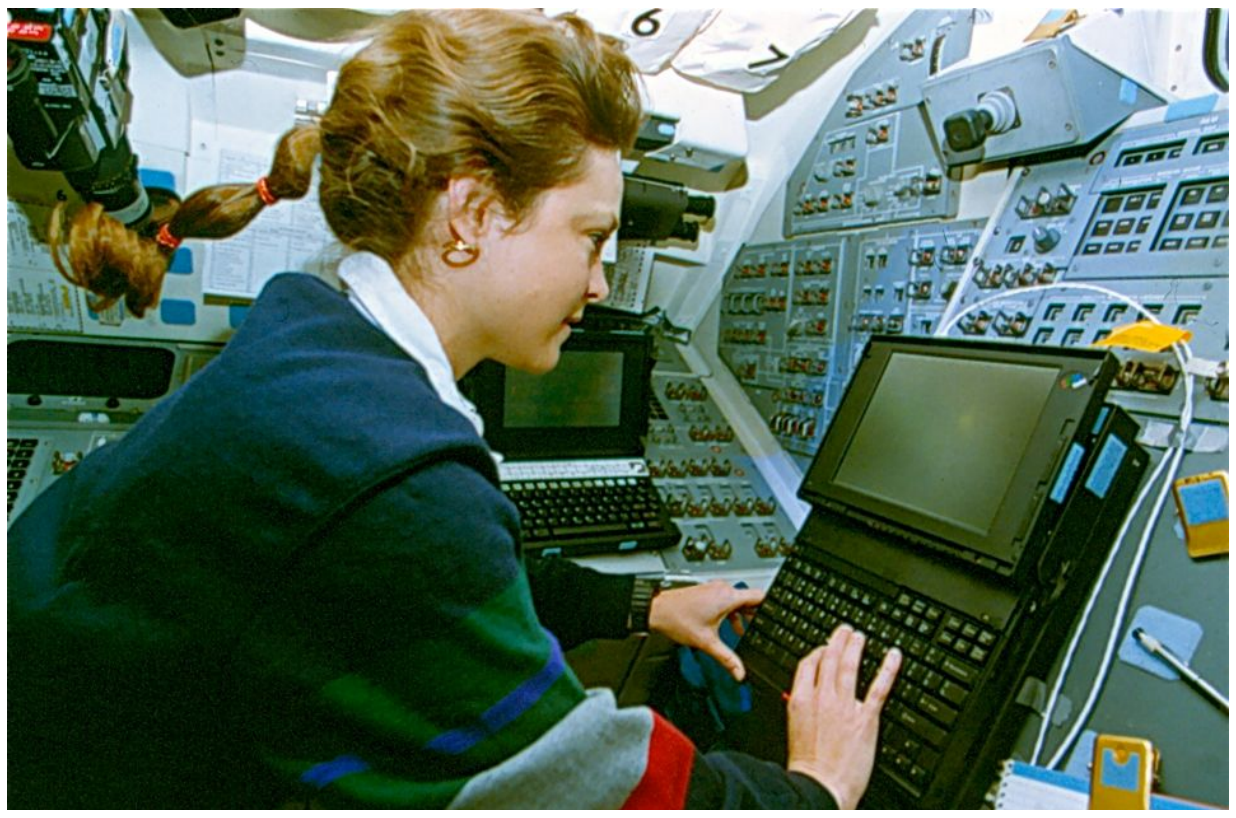
John Grunsfeld, Sam Durrance, and Tammy Jernigan with DDSs on the aft flight deck of *Endeavour*.

Paul Feldman returned about noon to relieve me. However, since he had been at the launch, he hadn't settled into things in Huntsville yet, and it took a long handover to get him up to speed. I finally left the POCC about 2:30 p.m. after (ahem!) a rather long shift.

I decided to stop and get some groceries on the way back to my room. I hadn't dared stock up the refrigerator before we launched, lest I doom us to another series of launch delays for sure! (Not that I was superstitious...) The main thing I wanted to get was some cereal and milk so I could have something easy and quick before going into the POCC by 4 a.m. every day. I still don't know how I did this, but I bought lots of stuff none of which was cereal (I discovered as I unpacked things in my room)! This little faux pas came back to haunt me a few days later (see below). Anyway, I had a snack and laid down about 4 p.m. to get some sleep. I was dead tired already, and we had weeks of effort ahead of us.

**Friday March 3, 1995:**
I initially awoke about 11 p.m., but I made myself sleep some more, finally getting up about 2:30 a.m. I felt rather headachy, but just couldn't lie in bed any longer. At 25 hours into the mission, NASA-TV seemed to indicate that we were into the science portion of the timeline in the first block of HUT time! After a shower and an OJ, I got out to the POCC at 4 a.m. and handed over with Van Dixon. It turned out that the IPS was having more problems than NASA-TV let on, with both discrete "jumps" in the pointing position and large drifts when supposedly locked onto a target. Several acquisitions in a row had been lost and we were considerably behind the nominal activation timeline. They set up a "tiger team" to work the problem and report back at the SOPG at 33 hours MET. On the other hand, all three instruments were basically functional, and all of the hardware was in good shape, so things could have been a lot worse. Especially with the longer mission, there was certainly no panic in the air. But people were clearly getting antsy about getting on with the science!

We had done a lot of work in the pre-mission planning to build "redundancy" for activation activities into the early part of the timeline. Nearly all of the HUT objects in the timeline before 40 hours could be used for science or as back-ups for some aspect of the activation activities (for instance, spectrograph or TV focus, calibration, etc.) Hence, while there were considerable discussions about how best to use the upcoming pointings, the targets themselves were in place, making things relatively easy still for the replanners. We did, however, have to start refining our methods for tracking the success or failure of each pointing and keeping tabs on the status of each of our science programs (of which there were about 20), a time consuming task that would be important for making replanning decisions later in the mission.

I had another rather long handover with Paul Feldman starting at noon. Then I hung around to get caught up on some e-mail and reading of stuff in our "in" box, and doing a little PR work on the phone. We took our first crack at a Cygnus Loop position that afternoon, but it was a faint one and tough to dig out of the airglow emission. I left the POCC at 4:40 p.m. and had dinner at Tim's Cajun Kitchen with Knox Long and Jerry Kriss. During dinner we even used my cellular phone (as they were called in the mid-1990's) to call out to the POCC and talk through an operational issue that had come up in discussion. (I only mention this because it was one of the few times it really came in handy. As it turned out, the main use I had for it in future days was to call the "dinner group" to see where they were going since I was usually out at the POCC much later than the end of my shift!) I got back to my room about 6:30 p.m., had a nice chat with my wife Jean, and got to bed about 8 p.m.

**Saturday March 4, 1995:**
I got up about 3 a.m. and made it to the POCC about 4 a.m. Things seemed to be going much better, at least for HUT. We were about 50 hours into the mission and counting. Although the pointing was still rocky at times, if sources were bright enough HUT could still get good data. However, WUPPE and especially UIT were more dependent on a pointing error feedback signal from the Image Motion Compensation System (IMCS) to correct bad pointing, and the IMCS still had some problems. But, hey, we were in a HUT block and getting some good stuff!

One of the highlights this shift was a pointing at Jupiter's moon, Io. Shortly after launch, we had received a notice of some kind of infrared brightening on Io, possibly indicating increased volcanic activity on the little moon. We had a pointing at Jupiter scheduled at 55 hours in the pre-mission timeline. As the first scientific demonstration of our "charmed" nature on this mission, not only was Io visible at this time, but it was at greatest elongation from the planet, and calculations showed that the "hot" side of Io was pointed directly toward Earth! Combine this set of circumstances with the fact that we actually made the observation as planned and got excellent data, and you know we were charmed! Paul Feldman was floating on air, and it wasn't the last time, for him or for me!

We also made our first pointing at a cataclysmic binary star (VW Hyi) and another shot at a faint Cygnus Loop position, this time with good pointing. A quick analysis showed very faint emission, so we quickly replanned an upcoming pointing that was similar to a brighter position in the Cygnus Loop, to give us more time to analyze these data. This kind of "quick" change to enhance the science return would have caused all kinds of contortions back in Astro-1 days, but we were able to pull it off smoothly numerous times on Astro-2 without ruffling a feather, thanks to the work of the CI/MPT in the interim time between the missions.

We also made our first pointing at the high-redshift quasar HS1700+64 during this shift. This was our highest priority science target on Astro-2, to be used in a search for so-called diffuse intergalactic medium (IGM), formed in the Big Bang and predicted to exist, but never clearly detected. This was the highest priority program for our PI Arthur Davidsen and one that had been highlighted in PR going back to pre-Astro-1 days. Sadly, the poor pointing during Astro-1 had negated our attempts to get more than a cursory observation of HS1700 at that time, but in retrospect, it may have been too faint to permit a good result anyway. With 2-1/2 times more sensitivity on Astro-2 (due to the silicon carbide coatings on both the grating and the primary mirror) and stable pointing, we planned to nail this faint quasar with as many as 10 full night pointings, more time than devoted to any other single target in our plan (almost 20,000 seconds total).

The idea with this first pointing was simply to see if we could actually see the quasar on the HUT TV so that we could verify its position relative to the selected guide stars; if we actually got any observing time on the object in this first pointing, we would consider it gravy. If we were to spend up to10 precious pointings on one object, we wanted to make sure it was centered in the spectrograph aperture! Special procedures were needed to crank up the HUT TV gain all the way and then integrate the video data in order to see such a faint target. We did see it, right at the limit of what we could see, and we even got a few hundred seconds of data -- a drop in the bucket, but still a good first shot! We were elated, but of course really had no hint of the ultimate result at that time. Interestingly, the phone rang almost immediately with calls from various reporters wanting to talk with Arthur about his "result"! (See P76, for example.)

We also experienced some of the "old" NASA during this shift. A "PAO" event (Public Affairs, basically a crew interview with a TV station) was scheduled right during a HUT observation. We had inquired the previous day about this (it was scheduled by JSC pre-mission without any consultation with the MSFC planners), especially about whether it would impact the acquisition scheduled at the same time. (It was a "night" pointing at a time-critical cataclysmic variable star, and we really didn't want that one messed up.) The word had come back that it would be "no impact" on operations. Well that turned out to be a crock. Not only was the interview with the mission specialist who was on duty (and should have been helping with the target acquisition), but the bright TV lights prevented Sam Durrance (the PS on duty) from proceeding with the acquisition by himself. I was not amused when they were able to "suddenly" identify the target about two minutes after the interview ended. We lost 20 minutes of night pointing (out of about 30 available minutes) because of an interview that could have (should have!) been scheduled in orbital day! I was really angry, especially since we had asked about this ahead of time. There were numerous such events scheduled in upcoming days, and I definitely did not want to see this sort of stunt repeated.

After registering my feelings with various appropriate parties (such as Mission Scientist, Charles "Chip" Meegan), I also voiced my opinion of the situation in no uncertain terms (shall we say) during the following SOPG meeting.  It no doubt helped that I was known from all of the mission planning improvement work as a fairly level headed sort of guy--not the kind to fly off the handle easily.  Also, I had been as supportive as anyone on the science teams of PAO activities, and had done a lot of work in this area in the months approaching launch.  I was not against PAO activities, but there was just no reason to allow them to impact the science activities when they could be scheduled at other times.   I think I got my point across, as there was a "heightened sensitivity" to this issue after this point. We still had a couple of tussles with JSC on certain PAO events, but in general the problem did not recur at a significant level.

I got out of there about 4:30 p.m. and was able to get back to the hotel before dinner for a change.  After I washed my face, I picked up my glasses to put them on and the earpiece fell off.  Great!  With my schedule, how was I going to get this fixed?  With some effort I managed to find the little screw that had fallen out.  As I walked over to meet the dinner group, I ran into Van Dixon, and I described my plight.  He said, "That happens to me all the time.  Just a minute, I've got a little screw driver in my room."  In a few moments the glasses were fixed!  Charmed again.

Knox Long and I broke off from the rest of the group and had dinner at Fratelli's (Italian), which was relatively nearby.  It was a nice time to compare Astro-1 and Astro-2, talk strategy on our joint science programs, and just kibitz with an old friend and close collaborator.  I got back to my room about 7 p.m. and called Jean to get an update on the home front, and made the common error of turning on NASA-TV as I crawled into bed. There were two pointings scheduled on HS1700 that evening, and both went well.   We also got a wild spectrum of an extremely hot white dwarf star, KPD0005.  It had unexpected absorption from molecular hydrogen (apparently) along the line of sight somewhere that caused a very peculiar and unexpected spectrum!  The acquisitions seemed to be going very well now.  I also knew there was a Cygnus Loop position scheduled about 10 p.m., and I kept one eye open as I tried to snooze.  Unfortunately, it turned out that there was no K-band TDRSS coverage then, so I couldn't see the data in real-time anyway.  I had stayed awake for nothing!  Ack!  I only slept from about 10:30 p.m. to 3:15 a.m.

**Sunday March 5, 1995:**
I managed to stagger out to the POCC just about on time, with a hint of a scratchy throat. Things continued to go quite well today, and they also seemed to have made some progress on the IMCS problem (which again had affected mostly UIT and WUPPE).

We started making some progress on some of our 'extragalactic' programs today, getting a pointing at a bright Seyfert galaxy, NGC 4151, and a couple of other active galaxies.  We observed NGC 4151 during Astro-1, but the idea this time was to observe it every couple of days through the mission to look for variability on timescales of days to weeks.  This first observation just about knocked our socks off -- the thing was *five times brighter* than when we observed it during Astro-1, near it's all time brightest level!  Combined with the increased sensitivity of HUT, this spectrum showed incredible detail compared with the Astro-1 data.  Nature cooperated again!

We observed a second high redshift quasar today, another possible candidate for the IGM program.  We thought this one was fainter than HS1700, but you really don't know until you observe these objects and see.  Indeed, it was fainter and we decided to continue putting most of the time for this program into observing HS1700. Again, a near-real time science decision that enhanced the science result.

A protocol issue had cropped up overnight that caused some concern.  There were seven crew members aboard, but it only took six to really run things (three each on two-12 hour shifts).  That left our commander, Steve Oswald, without much to do.  He latched onto a middeck experiment, called MACE (Middeck Active Control Experiment) and started hogging the air-to-ground communication loop.  This caused problems with some of the acquisitions, where the MS or PS wanted to talk to the ground (or vice versa) but the link was "busy" (so to speak).  The ironic part of this was that there are two air-to-ground channels, but flight Commander Oswald had earlier decreed that only one would be used. (One of his previous flights had had a problem with too much unnecessary "chit chat".) Now his "chit chat" (which was not time critical by any means) was affecting our ability to observe!

Evidently it is not proper NASA protocol to tell the commander of a space shuttle to "shut up."  Since air-to-ground goes out over NASA-TV, a more indirect approach had to be taken. Hence, meetings were held and faxes were sent back and forth from MSFC to JSC, and finally after many hours a discreet message was uplinked to Commander Oswald that effectively told him to (ahem) "shut up" during acquisitions!

Another unexpected issue arose today that was along the same lines: the crew exercise period.  Now how could something this mundane become an issue, you ask?  Well, it turns out that whoever made up the list of what to take along on the shuttle forgot to add a device that isolates the vibrations from the exercise bike from affecting the rest of the shuttle.*[28] Hence, whenever one of the crew was exercising, the shuttle, still bound by the laws of physics after all, would merrily bounce around causing the pointing to wobble.  Since the crew had to exercise a certain amount each day (and there were seven of them), someone was almost always on the bike.  After much discussion, we got them to try and stay away from exercising on night passes, and each of the instrument teams was allowed to select one target per 12 hour shift to "protect" from crew exercise.  What a way to do science!

*[28] This was realized shortly before launch evidently, but the weight limit had been reached, and that was that. The science teams were of course never informed.

In reality, it was really a sign of our good fortune that such things mattered at all. Here we were putting all of this effort into optimizing things to glean the last little bit of efficiency out of the operation, when the bottom line was we were doing science and actually getting incredible data! I had a period during the afternoon, perhaps brought on by a combination of fatigue and a hormonal imbalance caused by my eating habits (!), where I had this funny feeling—I'll call it a "satisfaction feeling" for lack of a better term. I had spent much of the previous several days subconsciously waiting for a bomb to drop on us, no doubt a sign of PTSD from Astro-1 where incendiaries were falling with great regularity.

I had put huge effort into improving the planning procedures since Astro-1, and I was extremely pleased to see things running smoothly on that front. And not only were we racking up some wonderful observations, but *Nature* was cooperating as well. I had remarked in a telephone interview with a reporter the previous day that I was glad I had "hung around for Astro-2," and it actually made it into the papers. I was there when Arthur Davidsen happened to read it, and he just looked over at me, smiled, and shook his head. We had a long way to go, and the "bomb" feeling would still haunt me from time to time, but I'll remember that feeling of satisfaction for a long time.

I left late again this day, nearly 5 p.m., and met Harry Ferguson, Knox Long, and John Raymond at the Camino Real Mexican restaurant adjacent to the hotel. Perhaps it was the giant Margarita I had, but I was actually able to get to sleep about 7:30 p.m. and had a badly needed decent night's sleep for a change.

**Monday March 6, 1995:**
Once again, I arose about 3 a.m., in time for a quick shower and some breakfast before heading out to the POCC. Almost 100 hours into the mission now and things were continuing to go quite well, although in some ways we came back to earth today. If there was a fly in the ointment, it was that the "acquisitions" (that is, getting set up on targets and ready to observe) was still taking longer than expected and in many cases longer than they did on Astro-1 (when things were actually working that is). For most observations it was not a serious problem, but for certain objects only planned for relatively short intervals we were basically missing them. While some improvement had certainly occurred over the previous couple of days, it was looking like we simply needed to stop planning short observations. This was not a problem in principle (i.e., just a change in operational strategy), but it couldn't help but cause ramifications downstream. There were only so many observing opportunities, and if some took more time than we had planned, something would have to give. (Orbital observing time was a "zero-sum game".)

Another operational growing pain involved the focus. We had made enough observations now to realize that we had to position the mirror differently to focus the TV camera and to focus the spectrograph (for the best science data result). This indicated that "something" had shifted a little bit, presumably during launch and orbital insertion. While this was not surprising (there are tremendous vibrations associated with a shuttle's launch and climb to orbit), it was not supposed to happen. This forced us to choose a compromise focus whereby the TV camera (which was crucial for identifying the targets and guiding on them) performed well enough in most cases while having an acceptably small effect on the spectroscopic data quality. The TV performance issue caused us to look ahead and try to flag those observations where we might have a problem due to faint guide stars. Just one more little detail to worry about.

Some team members had also had time to start looking back at the data we had obtained so far to make judgments as to where we stood. It turned out that some observations that had appeared successful at the time were only partially so. In particular, white dwarf stars, which were observed both for calibration and science purposes, needed to be "photometric" (that is, we didn't just need a spectrum, but we needed to measure the precise amount of light as well). For some observations poor pointing affected the photometry, while in other cases it turned out that poor target centering in the HUT spectrograph aperture (perhaps in conjunction with the focus problem mentioned above) had compromised the data. Hence, the list of things that needed to be replanned began to grow.

As a replanner, one has to be somewhat of a jack-of-all-trades. One has to keep a finger on the pulse of real time operations, being aware of any operational problems that may affect planning, and getting a first order impression of which observations were successful and which were not. One also has to deal with tracking the past observations by science program and keeping a status list up to date. However, we work mostly in the future, planning things that will occur some 12-24 hours downstream. The really tough calls are on things that will happen in the near term, including checking the results of the last replanning cycle and if errors are found, deciding what to do.

One of these planning errors occurred on one of my favorite targets called the "SM Star" ("SM" for Schweizer and Middleditch, the two astronomers who had discovered it). This star sits directly behind a young galactic supernova remnant which allowed us to use it to probe the insides of this exploded star.*[29] This star provided the only such alignment known at the time, making it a unique measurement for HUT and a high priority for me. Because the star was quite faint for HUT, we needed to maximize the amount of night observing time. Unfortunately, the visibility for the object was from about half-way through orbital night until about half-way through the daylit part of each orbit. Hence, we needed to "get on it" as soon after it rose above the earth limb as possible to maximize the night observing time.

*[29] Absorption lines in the spectrum of the star would be caused by the intervening material from the supernova.

Objects with such visibility are actually quite painful to schedule because they force one to leave a previous night target and spend orbital night time slewing to and setting up on the target on interest. It is something we really only did for the highest priority targets. While glancing over the replan products for the current shift, I found that the first observation of the SM star was coming up in a few hours. However, MUCH to my chagrin, the acquisition began some 10 minutes after the object rose above the earth limb. By the time we acquired the target and began the observation, it would hardly get any night observing time at all!

I got on the loops and started checking into what had happened and whether anything could be done about it. Part of the problem was that the slew time (time to maneuver the shuttle from the previous target to the next target) was much longer than we had estimated with our software. This was due to a number of constraints that only the MSFC planners could check that lengthened the slew (including such things as maximizing TDRSS contact time and keeping us from pointing into the "Ram" direction, or direction of motion). In such cases, the MSFC planners were just splitting the difference between the two affected targets. In previous cases it had only amounted to a few minutes at most and it hadn't made a difference. I was unlucky in that it was more than a few minutes, and more than 'half" had been taken out of the SM Star observation! After conferring with team members including Arthur, I was given the go ahead to put in an "OCR" (Operations Change Request) to get the time put back on the SM Star.

Again, at some level such changes are "risky" because they need to be incorporated into the observing plan on short notice, without all of the usual checks of a full replanning cycle. Our newfound confidence in the replanning procedures was not misplaced, however, and several hours later the changes were in place for the observation. Still, we didn't want to have to do this every shift, and I worked behind the scenes to encourage more communication between the MSFC planners and the science teams when slews needed to be changed by more than a few minutes from what we anticipated.

The SM star observation was one for which I had waited a long time. The object was too close to the sun during Astro-1 and so this unique observation could not be made. The primary purpose of the observation was to look in HUT's unique spectral range for a predicted broad absorption or "dip" in the spectrum caused by twice-ionized iron atoms in the expanding remains of the supernova explosion. The presence of this dip would go a long way toward confirming our theoretical ideas about so-called Type 1a supernovas, which are important objects for estimating distances in the Universe. It wasn't so much whether we expected to see the feature or not, but rather how deep and strong (or not) the absorption would be.

The observation occurred when I was "on duty" and I was supposed to go to one of the regular SOPG meetings, but there was no way I was going to be sitting in some meeting when this observation was going on! Richard Buss (one of the other replanners) kindly agreed to cover for me. The star field was easily identified on the HUT TV, but the acquisition seemed to take forever. (In actuality, it took no longer than any of the others, but such is one's state of mind at a time like this.) Finally, the guide stars were locked and we were observing! The count rate was just as expected, and the spectrum slowly built up on the HUT TV screen. This updating display only shows every fourth point in the spectrum, and is just to provide a means of monitoring the progress of each observation in a crude way. At first, as the data from this faint star crawled in, it looked as if the expected feature was present, but as the spectrum continued to build up with time (and improve in quality), it was obvious that the feature was very weak, if indeed it was there at all! It would take a second pointing and the star, and a pointing at a similar comparison star that was not behind a supernova remnant before I could be sure, but it appeared I was being left with a quandary instead of an answer.

As was becoming all too usual, I stayed late again today, largely to plan strategy for the cataclysmic variable star program with Knox. We had been receiving regular (sometimes twice daily) reports from the amateur astronomers' variable star group, the AAVSO, about the optical status of our highest priority variable stars. This needed to be integrated into our priorities and, in some cases, coordinated with collaborators planning simultaneous observations with other satellites, such as ASCA (X-ray) and EUVE (the Extreme Ultraviolet Explorer, which observes in the EUV). All of this took a lot of thought and communication (mostly electronic), and there was no time to work this in during my replanning duties.

Knox and I left from the POCC and met John Raymond and Harry Ferguson for dinner at the Green Hills Grille. After dinner, Knox and I returned to the POCC because a couple of Cygnus Loop pointings were on the docket. When Knox went on shift at 8 p.m., I headed back to my room and got to bed about 9 p.m. Another very long day.

**Tuesday March 7, 1995:**
I arrived at the POCC at the usual 4 a.m. shift time, and things became hectic very quickly. After handing over with Van Dixon, I reviewed the replan products for the upcoming shift and found several glaring errors! We had generated a full mission timeline several months prior to the mission which formed a "baseline" plan from which we intended to work. However, in subsequent preparations, we had discovered several problems with some of the planned targets that caused us to change our minds about the intended observations. Of course, we then had to track these changes somehow, just one of what seemed like an endless stream of details that had to be kept in mind. But now, there in the timeline was an observation we had no real intention of making; in addition, of course, the "replacement" target we really intended to observe at that time was nowhere to be found in the timeline. Perhaps fatigue was building as the mission wore on, but this was an error that should not have occurred. In addition, we had another big chunk of time taken out of a target for a longer-than-expected slew, so my work behind the scenes the previous day had not entirely percolated through the ranks. Correcting each "little" problem like this took a large amount of concentration and attention to detail on someone's part (mine, in this instance) to push it through the system successfully. We managed to get the changes made, but we were clearly beginning to fray some nerves with these last minute changes. My fatigue level was increasing as well, and I had a harder time remaining level-headed with my teammates who had made the blunder.

The other activity (in addition to ongoing operations) that made things hectic today was self-imposed, but necessary.  From a replanning standpoint, we were pushing toward halfway through the mission.  While we could make some judgments about each program's status, we really needed feedback from the individual scientists as to where they felt things stood for their programs.  Hence, we wrote up and distributed our (the replanner's) estimate of each program's status (which meant we had to get our notes up to date first) and requested inputs back ASAP.   Consequently, over the next 24 hours we had a steady stream of people coming to talk with us about their science programs (and resolving differences between their accounting and ours).

On top of all of this activity, things were heating up on the PR front.  Arthur did a live CNN interview this day, and NASA-TV was becoming more aggressive about doing some live interviews with scientists in the POCC.  In addition, JSC had PR people in Huntsville doing tapings for their daily "Mission Update" program.  And twice a day we received draft "Shift Reports" from MSFC PAO that usually needed a careful reading and editing.

About the middle of the afternoon, Richard Buss started complaining about not feeling well, and said he thought he was coming down with something.  We encouraged him to get out of there lest he infect someone else!  Working in close contact with people day after day, this is a real concern, and we could ill-afford (!) to lose very many people and maintain our operations at the necessary level.   Richard called a doctor and arranged to have someone give him a ride.  Little did we know at the time that as Richard staggered out the door it was the last we would see of him until we were back in Baltimore!  He came down with something akin to pneumonia and never returned to the POCC, leaving us short-handed on the planning desk.

As 5 o'clock approached, it was clear I had too much to do to leave yet and they had ordered out for Chinese, so I decided to hang around into the evening.  Although we had no outside windows in our area, people coming on shift had mentioned that it looked very stormy and that there was a tornado watch up.  Sure enough, about 5:30 p.m. an alert was called and the MSFC operations people (known collectively as the POCC Cadre) came streaming into our area, which was "tornado safe."  We had actually done this exercise during a simulation.  Key people from NASA's upstairs operations room would come down and commandeer prearranged terminals in the science team's areas to reconfigure them and carry on operations (not as simple as it sounds, since different communications loops are available to different Cadre members).  Interestingly, the guy delivering the Chinese food arrived in the middle of the alert, and we sat there eating Chinese food off of paper plates surrounded by hungry looking Cadre members trying to carry on in the middle of a tornado drill!

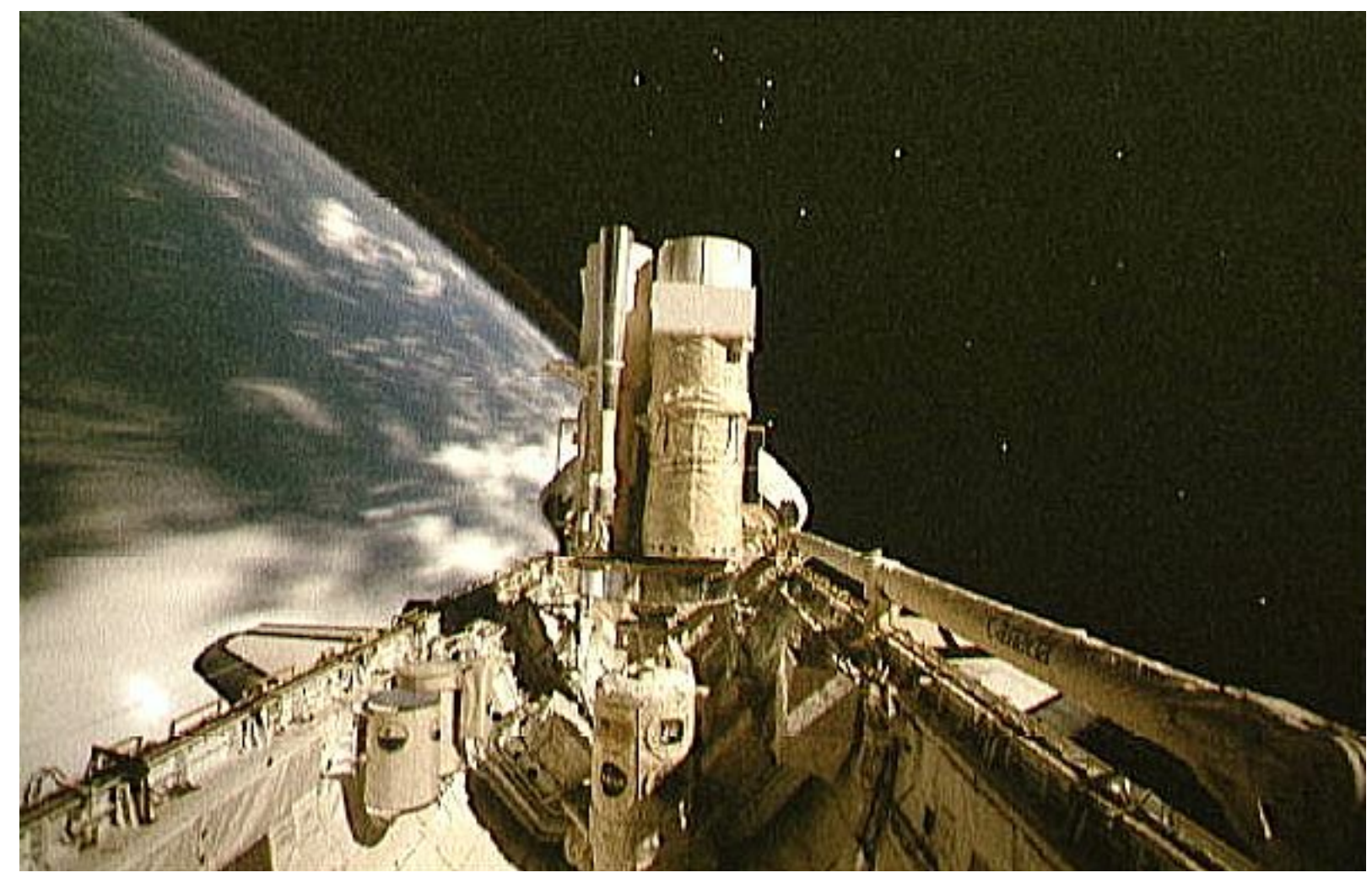
**Moonlit view of the payload in the cargo bay (taken from a TV monitor in the POCC).**

An interesting coincidence was that the shuttle was passing over the Gulf of Mexico at the time and sent down very impressive video from above of the storm that was pounding us. (Lightning storms viewed from orbit are spectacular!) The shuttle had different video cameras on-board than they did on Astro-1. The color video was really very impressive, and some of the earth views were just spectacular. However, another set of cameras that just provided black and white were also on board, for use especially on the night-side of the orbit. These cameras could be cranked up to show the earth in moonlight, city lights from space, and lightning storms. After showing us the storm we were under, the shuttle flew over Florida, and it was like looking at one of those "Earth at night" posters in live downlink video! At other times in the mission, these cameras showed views of the Earth limb, the atmospheric layers, and stars in the sky that were truly breathtaking (see photo above).

Shortly after things got back to normal, I cashed in my chips and headed home (about 8:30 p.m.). I called Jean from my cellular phone and talked while I drove so I could get caught up with her but still get to bed soon after I got home. While getting ready for bed, I turned on NASA-TV (of course!) and heard something about a "HUT detector problem." Now you have to understand that the announcers on NASA-TV never let on when there is a problem, but just keep droning on in their normal monotone. However, it was clear from the pictures and a few comments on air-to-ground that something was not right with HUT. But it was 9 p.m. and I was dead tired. I turned the TV off and said to myself, "They'll call if they need anything."

I must now make a brief aside for the next section to make sense. You may recall I mentioned having forgotten to buy any breakfast cereal when I went to the grocery store right after launch. On Monday afternoon, I had given Scott Shelton (one of our engineers) a ride home from the POCC, and had related the story to him. After chuckling, he said, "I go shopping each morning with Sharon to stock our refrigerator at the POCC. What do you want? We'll get you something." Hence it was that a box of Honey Nut Cheerios arrived at the POCC Tuesday morning with my name on it. In all of the day's activities, I completely forgot about the cereal and ended up leaving it at the POCC.

Returning now to the HUT detector problem, I turned over in bed about 11:30 p.m. to see that the message light on my telephone was blinking. I thought to myself, "This is it! It's all over! They fried the detector and they left me a message saying 'Don't bother to come in tomorrow'!" I picked up the phone and with trembling hands dialed the front desk…

> *"This is Mr. Blair, room 712. I have a message?"*
> *"Yes, just a minute," a man's voice came back. After some shuffling of papers, he replied, "Your Cheerios are here."*
> *"What?" I said incredulously.*
> *"Your Cheerios are here," he repeated, not knowing my jaw had dropped onto my chest at the other end of the line.*
> *"Thank you very much. I'll pick them up in the morning," I said dumbfoundedly.*

It was probably a half an hour before I could get back to sleep; not bad considering the amount of adrenalin coursing through my veins at the time I made that call. It is indeed unfortunate that when I turned over about 12:30 a.m., I saw my message light blinking, and it started all over in my mind. "The detector is broken, and our beautiful mission is down the tubes! I might as well call and find out…I can't sleep not knowing what's going on."

> *"This is Mr. Blair, room 712. I have a message?"*
> *"Yes," said a perky female voice at the front desk. "I have a box of Cheerios here for you!"*
> *"What?"*
> *"A box of Honey Nut Cheerios," she said.*
> *"I know! I already got this message!!" I stammered angrily.*
> *"Oh," she said. "Well, I just came on duty, and the note was lying here, so I thought I'd better let you know."*
> *"Well, I'll be sure to pick them up tomorrow. Please make sure you throw that message away this time."*

Sheesh. Nothing like a good night's sleep to perk a fellow up. The worst part was it was late the following afternoon, on my way back from the POCC, before I could stop by the front desk, and no one could find the box of Cheerios!! (It turned out the cleaning person had already put it in my room!) At long last, the following morning, I was able to have bowl of Honey Nut Cheerios before I went on shift.

**Wednesday March 8, 1995:**
Despite the previous night's escapades, duty called and I was back in the POCC at 4 a.m., after another "eerie" ride through drizzle and fog. (It'd been like that practically every morning so far.) I found out that we had indeed dodged a bullet the previous evening when a procedural error had caused the HUT doors to open before we left a bright target, nearly frying our detector. We were saved by a "safing" mechanism that had been built into the HUT computer long ago, a software check that basically said, "If you see something too bright, turn off the high voltage and shut down." This was one of those things, however, that had never been tested because it would have meant intentionally putting the detector in jeopardy. Luckily, it had worked as planned. It took them a little while to recover things and we missed a couple of pointings (including one at HS1700, which would be replanned), but otherwise things were back to normal when I arrived.

In general, things were more "in control" today. Luckily, we had a respite from replanning mistakes requiring OCRs, and things went relatively smoothly. We had requested people's inputs back by today, but I hadn't even had an opportunity to give more than a cursory glance at the data on supernova remnants and CVs. Van Dixon kindly stayed a couple of hours late, which gave me a chance to look at some of the data from my programs and basically get my own act together for the ongoing planning effort. It was FUN and exciting to actually see some of the data in more detail.

Speaking of CVs, we obtained some wonderful data on "dwarf novae" today. These binary stars, which consist of a white dwarf star and a normal star in a tight orbit about each other, undergo occasional and semi-periodic outbursts, getting brighter by typically a factor of 100 or more. One such object, U Geminorum, was of particular interest to us. It had been observed during Astro-1 about 10 days past an outburst. This one had a typical outburst period of 120 days, so this was relatively soon after an outburst. At the time we launched, our amateur astronomer friends were telling us it had been 180 days since the last outburst, the longest inter-outburst period ever observed for this system. Hence, if we could get a spectrum before its next outburst, we would have a very valuable comparison to our Astro-1 data and would be observing the binary under conditions never before observed. Well…we got it! A very nice, stable pointing with high signal-to-noise ratio. The spectrum was much different from that obtained on Astro-1, and indicated the white dwarf (which dominated the UV light in this system) had cooled dramatically from our earlier data. Thus, the white dwarf gets heated during an outburst, and cools down with time after the outburst. This was neat stuff! NASA-TV could tell that something was cooking in the HUT area. They sent a camera down and did a brief live interview with me, which was fun.

Also on the PR front, since things were quieter today, I finished gathering some materials (pretty pictures of HUT targets, "popular" level descriptions, etc.) and gave them to the NASA-TV people. I had meant to do this much earlier, but it had gotten lost in the shuffle until now. The NASA-TV commentary had gotten pretty repetitive already, and we weren't even halfway through the mission! They seemed appreciative when I gave it to them, and I think some of it was used from time to time during the rest of the mission.

Sam Durrance did a live interview from the shuttle to a TV station in Baltimore today. They had Becky and the kids in the studio, and they talked briefly. It was touching and cute, and came off well I thought. Sam looked terrific and seemed to be having the time of his life…for the second time!

On the other side, we heard from Richard today, and he sounded terrible. The doctor had told him to stay in bed until at least Friday. Since I had been staying through the afternoons anyway, it really wasn't that much of an adjustment for me to fill in, with others coming in an hour early or staying an hour late when needed to fill the gaps. In some ways, it was a relief because some personality conflicts had been developing between Richard and some of the other replanners, and were almost at the boiling point the day before Richard cashed in his chips. These things are unfortunate, but almost bound to happen when people are working this closely and under such pressure. In some ways, it's surprising there wasn't more of this sort of this thing. All in all, we worked extremely well as a team.

I left the POCC at 5 p.m. and ate with a large group at Tortellini's, an Italian restaurant due east of MSFC. When I got back to the room, I had a CARE package from Jean (cookies and fudge!) and a phone message from my mom, who was visiting my aunt and uncle in Florida. I called her back and had a nice chat. Then I caught up on some note taking, keeping an eye on NASA-TV, and didn't get to bed until about 10 p.m.

**Thursday March 9, 1995:**
Despite going to bed late, I woke up at 2:45 a.m. Arriving at the POCC about 4 a.m., I got Van to stay until 5:30 a.m. so I could catch up on other things (PR and e-mail mostly) and have a little time to regroup. We had been up a week now (about 170 hours MET), and despite the horrendous schedule and lack of sleep I was feeling pretty upbeat.

We took another crack at the SM star today and got sufficient data (in combination with the earlier pointing) to put a very strong limit on the expected iron absorption feature. We also made a joint HUT/Hubble Space Telescope pointing today at Jupiter, which caused quite a stir. Hubble was used to image Jupiter in a passband that showed the aurorae near the planet's north pole. At the same time, HUT obtained a spectrum of the auroral emissions. It came off without a hitch, which really excited the PR and NASA headquarters people as much as the scientists. Combining these data sets permitted a more quantitative analysis of the magnetic processes involved in the Jovian system.

After putting in a long replan shift, I stayed on to try and compile a list of missed targets from the beginning of the mission, and reasons why they were missed (i.e. bad pointing, shuttle waste water dump, whatever). Each team was to do this and come together for a meeting at 9 p.m. to even out the "relative pain" experienced by the three teams. We really hated this kind of 'bean counting' especially because it took a lot of effort to compile the information, and took someone (like myself) away from other duties that were more important (or in this case just gave me more duties on top of what I already had to do!). But it had to be done or we could unnecessarily lose pointings that we wanted/needed for HUT. I blew off going out to dinner and just ate at the POCC while I worked. When Arthur arrived about 7 p.m., he was clearly on the ragged edge; perhaps the lack of sleep was catching up with him, too. As I "compiled", I could hear some rather heated replanning discussion going on behind me, but I mostly stayed out of it. I was having a hard enough time concentrating on the task at hand, and after all, I was supposed to be "off duty" hours ago. I got done just shortly before the 9 p.m. meeting was to begin, and just barely had time to go over it with Arthur before the meeting.

Earlier in the evening there had been two pointings at the binary star AM Herculis, which was another "cast off" from Astro-1 that Knox Long and I were particularly interested in. A glance over my shoulder during the observations told me that the first pointing had gone well and the second had gone OK, but had some pointing problems. As tired as I was, I had to at least have a look before I left. The spectra were beautiful, with lots of variability evident even within a single observation. This was one of the observations that was supposed to have simultaneous coverage with other satellites, and I found out later that it did.

I was finally feeling pretty burned out. I was supposed to be back in 7-1/2 hours and I still had to drive home. Van Dixon, who had just come on shift an hour or so earlier, took at look at me and told me that he didn't want to see me at 4 a.m. and to not come in a minute before 6 a.m. I was in no position to argue. I drove home and crawled into bed about 10 p.m.

**Friday March10, 1995:**
I initially woke up about 4:30, but didn't get up until 5 or so. I was very congested when I got up, but no place was open to get anything. I actually stopped at the Kettle (a 24 hour restaurant) for actual "breakfast food" before heading out to the POCC. I watched the sun rise as I was driving in, and it struck me that it was the first time in days that I had seen the sun. I got there about 6:30 a.m., feeling pretty good, all things considered. Luckily, there was some decongestant in our supply cabinet at the POCC, and when our support people found out I was looking for medicine, I soon had an assortment to choose from! I think they were really afraid that I was coming down with what Richard had. Luckily, it just lasted a couple of days and was nothing serious. (The word on Richard was that it would be several more days at least before he could return to duty, but that never happened.) At roughly 200 hours into a nominal 360 hour timeline, we were definitely more than halfway.

Today we made our second observation of a nova, which is a binary star system where one of the stars has brightened by many orders of magnitude due to some sort of outburst (possibly a nuclear runaway on the star's surface). It was yet another demonstration of Nature's cooperation, for nova outbursts typically last only a month or so, and often times they are quite faint as observed from earth. Not only was this nova fairly bright, but it was one of three bright novae which had gone off in the month and a half before launch! Hence, once again we had a unique opportunity to study not one but three novae in the far-ultraviolet at different times beyond the actual outburst.

In reality, these objects were of more interest to the WUPPE team; these stars are in the plane of out Milky Way galaxy behind a lot of dust and gas. This makes them difficult targets for HUT because dust really absorbs out the UV light. Still, no one had ever observed a nova in the 900 - 1200 Angstrom range unique to HUT, so if we saw anything it would still be new and interesting. Also, these novae were relatively close to the sun, not too close for HUT, but too close for other satellites like the Hubble or IUE telescopes. Hence, our spectra longward of 1200 Angstroms would also be of interest and would not duplicate other observations. It turned out that at least one of the novae did show some emission in HUT's unique range as well.

In addition to replanning today, I did some PR work. Peter Vedder from NASA HQ asked me to do an interview on the Cygnus Loop for the Mission Update program (which I did), and I talked with several reporters on the phone, including Faye Flam of the Philadelphia Enquirer and a guy from a radio station in northern California. I talked with him for a long time, but if anything ever came of it I never heard. There was a brief piece in the Philadelphia Enquirer the next day, so at least that one amounted to something.

I arranged to have dinner with Knox and John Raymond tonight at TGIFriday's. We looked over data obtained so far on the CV program and planned some strategy as to how to proceed. One of our main ideas for the CV program was to follow one of the "short outburst period" objects with multiple pointings to monitor its behavior before, during, and after an outburst. However, we definitely liked certain objects better than others, and so far the ones we liked best were not the ones that had gone into outburst. Since we only had a certain number of expected pointings to play with, we couldn't afford to just keep observing these objects hoping that they would cooperate, because the outburst intervals were not strictly regular in these objects. We went over information sent by the AAVSO and made a revised plan.

After all of the work I had done the previous night to put together the "missed target" information, it turned out that the meeting to reconcile differences amongst the teams never occurred because one of the team reps didn't show up! They were regrouping again tonight, however, so after dinner I returned to the POCC to update the list and finish typing it into the computer. I called it the "HUT H.O.M. Page", for "Hit Or Missed", an obvious take-off on the idea of a "Home page" on the World Wide Web.

Speaking of the internet, ours was the first shuttle mission for which NASA tried making real time information available on the Web. The idea came along very late in the game, and really only got put together in the six weeks or so before the mission (much of it was barely in place at launch). It turned out to be a rousing success, with over two million "hits" by the end of the mission, from over 200,000 sites in 59 countries! People could even ask questions of team members or the crew, and some were even answered by the crew on NASA-TV. Each day a few of us would take some of the questions and try to answer them by e-mail, a time consuming (but I thought worthwhile) task. (See P77 for context.)

I left the POCC about 9 p.m. after watching an observation of the bright CV called Z Cam (another that had joint pointing with the ASCA X-ray satellite), and called Jean before turning in. It was about 10:30 before I got to sleep.

**Saturday March 11, 1995:**
I got up at 4:15 a.m. today, and felt pretty awful when I did. After I was up and around, I was OK, at least for a while, but I could really tell I was dragging today. For the first time I had a hard time concentrating and "multiplexing" tasks the way I was usually able to do.

We got a nice pointing at the supernova remnant position we called "VELA-H" today, one of our first pointings at the Vela supernova remnant. We had waited to do this object until later in the mission because Vela had better orbital night visibility then. Mike Herrington of the Mission Update program came by in the late morning and did an impromptu interview about it with me, for broadcast on Monday.

By afternoon I was clearly losing it, however. I would sit with my hands poised over the terminal and completely lose my train of thought. I couldn't handle having the normal five or six conversations going on in my headset and pick out the important one to listen to. And a couple of times it was almost like my speech was slurred. I had a relatively early dinner tonight with Knox and Harry at an Indian restaurant near the hotel, gave my brother John a quick call to say hello (a rather incoherent conversation, as I recall), and crashed about 8:30 p.m.

**Sunday March 12, 1995:**
I awoke about 2:45 a.m. and was in to the POCC by 4 a.m. for the first time in a few days. I felt pretty bad again when I woke up, but did much better during the day. Since it was Sunday, I got dressed up, and tried to remember to give my friends back at my home church a special thought during the morning. I was sure they were doing the same for me.

Things seemed to be under control again on the planning front. We had come through a period where replanning of early missed targets had had a "snowball" effect, knocking out other targets which subsequently had to be replanned as well. In some ways, the second half of the mission plan was relatively pristine, involving a number of targets planned in this time frame pre-mission for better visibility. Practically the entire elliptical galaxy program, for example, was still to be done, and you're not going to plan over too many pointings of a program that hasn't had it's place in the sun yet (or its place out of the sun yet, as the case may be). Roughly half the pointings at this point were as they were planned pre-mission, which is pretty incredible in comparison with our Astro-1 experience.

By the end of the day, I was interested in a quick dinner and a good night's sleep. I met Harry Ferguson, Knox Long, and Jerry Kriss at the Mexican place right next to the hotel, and got to bed by 7:30 p.m.

**Monday March 13, 1995:**
I woke up at 3:15 a.m. Wow! Almost 8 hours of sleep for the first time in two weeks, and I felt...awful! I felt like I'd been hit with a sledge hammer, and no, I didn't have a Margarita the night before!  Once again, after I was up and around I felt somewhat better, but the effects of stress and poor sleep were catching up to me.

We were now at about the 270 hour mark, and for the first time, I had the fleeting thought that maybe we were really going to do this mission all the way to the end, just like it was supposed to be done.  I was almost able to put the Astro-1 "boogie man" to bed, but in the back of my mind I still knew things could go wrong in a hurry. The Astro-2 mission was so long that even during the mission, some reports started appearing that described some of the success that we were having (see P78).

One of the key science observations today was a reprise on AM Her.  This was an "extra" pointing at this target, added at John Raymond's suggestion.  The motivation was to cover the portion of the binary orbital phase that had been missed in the previous two pointings. And cover it we did.  This observation gave us a big surprise.  An eagle-eyed Arthur Davidsen noticed in real-time as the observation was occurring that it looked like there were fluctuations in the counting rate with time.  At first there was concern that the object was near the edge of the spectrograph aperture so that small pointing errors were making us lose some of the light.  But on closer inspection it became clear that the count rate was going UP for brief intervals of 10 seconds or so, not down!  It was not a small effect, sometimes amounting to a 20% increase in brightness.  We were seeing real short term time variability, possibly due to discrete "blobs" of material crashing onto the magnetic pole of the white dwarf in the system (which was pointed toward us at this phase).  This was absolutely unique and unexpected data from an observation that we hadn't planned to make.  Charmed again!

On the planning front, since we were planning at least a day ahead, we were even closer to the end than it seemed, and I could see we were getting into a bind.  For a couple of days at the beginning and end of the mission, the time was distributed evenly between the three instrument teams, but through the rest of the mission HUT had more total time to specify because a larger number of guest investigators were aligned with our team (and hence more observing time was needed to accommodate them).  We were quickly approaching the time where we would be more constrained.  Also, we had few "holes" in which to replan priority targets; many targets were planned late in the mission NOT for reasons of priority, but because they had better visibility at that time.  It was crucial that we again status our overall program, set careful priorities on the targets that needed replanning for whatever reason, and look carefully at the preplanned targets between now and the end of the nominal mission to figure out what had to give.  This had to be done in as objective a way as possible, which was difficult since everyone had a bias.  I was aware of the overall program as well as anyone (and was also aware of most people's biases!).  Although it took a lot of work again today, I basically laid out the options and sat down with Arthur in the late afternoon and we hammered out the plan for the remainder of the nominal mission.

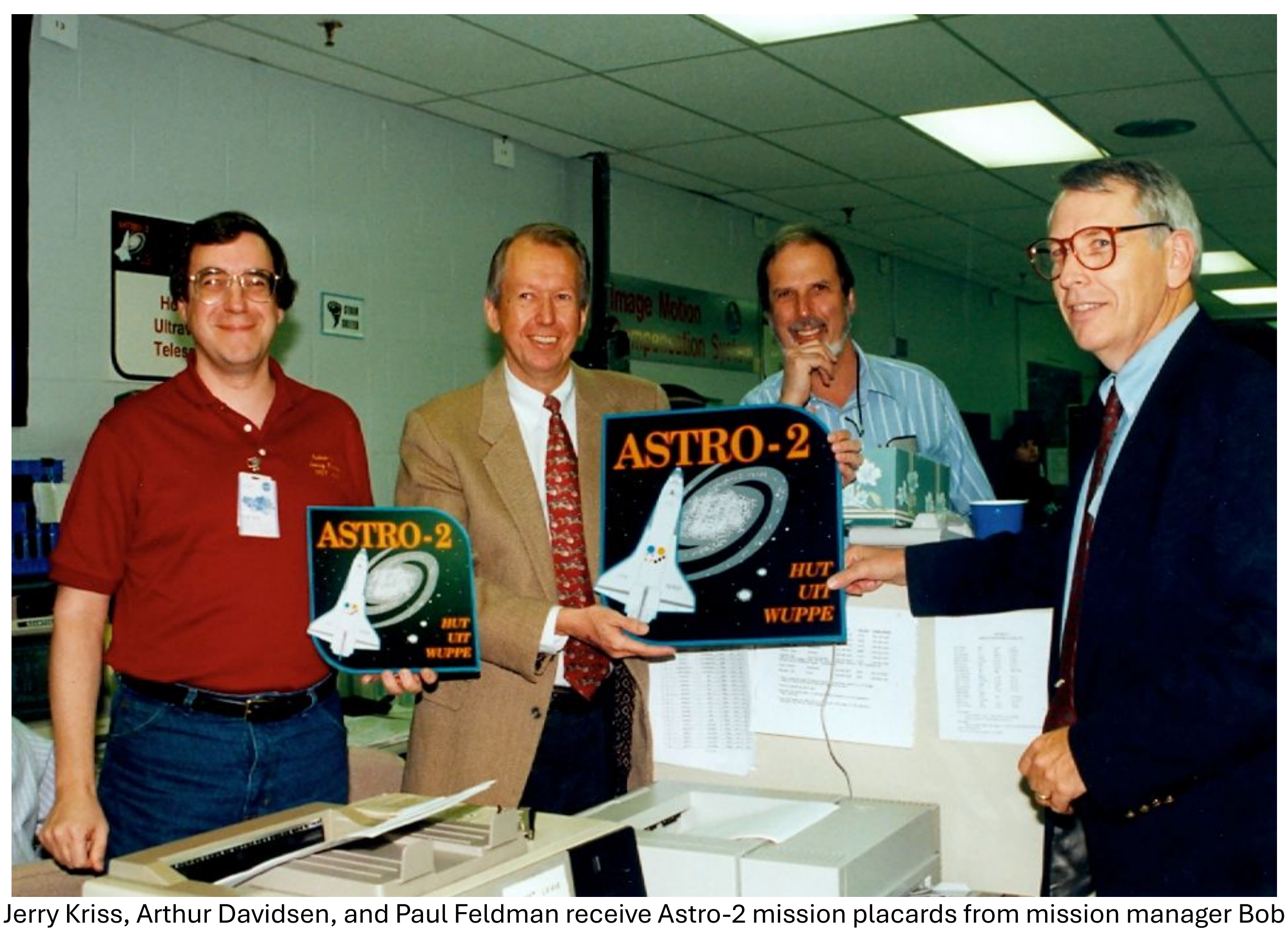

Jerry Kriss, Arthur Davidsen, and Paul Feldman receive Astro-2 mission placards from mission manager Bob Jayroe.

With this task done, the replanning from here to the end of the mission would be pretty mechanical. (Figuring out what you want to do is more than half the battle!) However, we weren't quite done yet. The nominal landing was set for Friday and science operations were to end late Thursday evening. But the weather was not sounding good for a landing at KSC, and the lobbying began today to get an extension day for science, especially if the landing was delayed for weather. This of course involved science justification from the teams and lots of discussion at the mission management levels. To their credit, the MSFC folks, including our mission manager Bob Jayroe, went to the mat for us on this and were all for it. We had a tremendously improved working relationship with the Cadre and mission management at MSFC this time, but the rest of NASA was an entirely different matter. It took until the next day for a decision to be made. Just the same, we had to look ahead and have plans for what we would do IF we got lucky. You'd hate to get caught with your pants down on an opportunity like this.

I didn't wrap things up until about 6:30 p.m. at the POCC, but luckily there was one other person who was hungry and ready to get something to eat. Nolan Walborn, one of our guest investigators, was working a loose shift schedule and stepped out with me for dinner at TGIFriday’s. I got to bed about 9 p.m.

**Tuesday March 14, 1995:**
I arose at 3:15 a.m. and got out to the POCC a little after 4 a.m. Van agreed to stay on again until about 6 a.m. so I could have a chance to look at some data. I looked at the AM Her data to convince myself that the variability we saw yesterday was real. It was.

This was a very interesting day, despite the fact that there were fewer HUT-primary targets in the schedule. Of course, we observed on every possible target, whether it is one of our team's choosing or not. But often times, UIT may observe a galaxy that does not have a specific target of interest (say an active galaxy nucleus or a bright H II region full of hot stars) for us to point at and get a spectrum. WUPPE tends to observe very bright stars, but often they are cool stars or moderately hot stars with lots of interstellar dust in the way, again causing them to be boring targets in the far ultraviolet band where HUT specializes.

We observed a planetary nebula today that was part of the WUPPE program. This was the only planetary nebula we observed, which is too bad because these objects can be fairly bright in the UV. We had toyed with the idea of putting some of these objects in our own program (one was observed on Astro-1), but it wasn't anybody's "primary" interest on our team, so we hadn't. This observation turned out to be very interesting. We spent part of the observation pointing at the central star and the rest offset into the glowing nebula. Both spectra were very peculiar, defying any quick-look understanding.

Another interesting pointing came from UIT. They were observing a number of fields in the Large Magellanic Cloud, the next nearest galaxy to the Milky Way (and hence, large in angular size, requiring multiple pointings). In one field I had suggested an object of particular interest to me, a young supernova remnant called N132D, as the object for HUT to observe. I had an HST program to investigate this object in detail. While HUT could only get a '"global" spectrum of the object, in effect smearing the emission from many filaments of differing chemical composition, HUT data were still unique in providing information on the farthest reaches of the UV spectrum. As with many objects observed by HUT, I was particularly interested in whether the object showed emission from "O VI" (O-six, which through a peculiarity of atomic nomenclature is five-times ionized oxygen). This is one of the highest ionization lines in the UV spectrum, and unique to HUT's spectral range. The model predictions for this object were uncertain, predicting everything from very weak or no O VI to fairly strong emission. The observation not only showed O VI, but it was by far the strongest line in the spectrum, a very surprising result! It just seemed like we could do no wrong, whether we were looking at "our" objects or one from the other teams.

The scuttlebutt on the extension day was not too good. It sounded like they would not officially extend the science mission even though all of the consumables (maneuvering fuel, food and supplies. etc.) were all in great shape. They were still holding out the possibility that we might do science if the landing was delayed for weather, but only if the call on this came early enough (that is, before deactivation was scheduled to begin). We continued plans for an extra shift of observations, including negotiating with the other teams as to who got what times, but it was sounding more like "much ado about nothing."

A small group of us ate at Tim's Cajun Kitchen tonight. We had been almost everywhere at least twice by now. We were longing for some home cooking. I did some catching up on notes and communications, before hitting the sack about 8:30 p.m.

**Wednesday March 15, 1995:**
I dragged out of bed at 3:15 a.m. It seemed like the more sleep I got the worse I felt and the harder it was to get up. I arrived at the POCC about 4:20 a.m. after finishing up my infamous Honey Nut Cheerios!

Things were really winding down as far as mission planning was concerned. At about 320 hours into the mission now, we had things planned very carefully through the end of the mission. While we were still anxious to get those last observations in the bag and really finish out our program on a high note, truth be said, if there had been a reason to shut us down right then nobody would have cried too loudly.

There was some excitement today about the CV U Gem possibly starting an outburst, but it turned out to be a false alarm. I managed to sneak away and give my wife Jean a birthday call and get caught up on happenings at home. With planning winding down, I had more time to realize how long we had really been at this.

One activity that turned out to be very useful was the construction of an "as flown" timeline that included as many of the changes we had made as possible. It was important to do this while it was fresh in our minds as it had been very difficult after the fact to reconstruct the Astro-1 timeline. I also found time to type in a couple of mission planning "songs" that had been kicking around in my head and partially drafted. These would be needed in a couple of days for our splashdown party. After all, I had a reputation to keep up. Thus came to be “Billy Boy” to thte traditional tune by the same name, and “Sammy Boy” (to the tune of the traditional Irish tune “Danny Boy” since it was almost St. Patrick’s Day). Billy Boy in particular had a verse specific to each on the replanners on the team:

Oh what have you planned, Billy Boy, Billy Boy?
Oh what have you planned, charming Billy?
"I have planned a QSO. It's a faint one as you know.
Maybe next shift I'll plan a couple others."

Oh what have you planned, Pauly Boy, Pauly Boy?
Oh what have you planned, charming Pauly?
"Jupiter and Io's plume, and some pointings at the moon!
Maybe next shift I'll plan a couple others."

Oh what have you planned, Johnny Boy, Johnny Boy?
Oh what have you planned, charming Johnny?
"SNRs and CV stars, and a planet they call Mars.
Maybe next shift I'll plan a couple others."

Oh what have you planned, Brian Boy, Brian Boy?
Oh what have you planned, charming Brian?
"Just a symbiotic star. It's the brightest thing by far!
Maybe next shift I'll plan a couple others."

Oh what have you planned, Vanny Boy, Vanny Boy?
Oh what have you planned, charming Vanny?
"Quasars, clusters, and CVs, and some Seyfert galaxies!
Maybe next shift I'll plan a couple others."

Oh what have you planned, Richard Boy, Richard Boy?
Oh what have you planned, charming Richard?
"Not a thing since last Tuesdee...I've pneumonia can't you see?
I'm in bed, now, and underneath the covers."

Oh, Sammy boy, alas the mission's over!
The IPS is stowed and we are packed to go.
But here's to you,
Since you are still on orbit.
Oh Sammy boy, oh Sammy boy,
We love you so!

Oh, Sammy boy, you've flown the longest mission,
That on the shuttle they have ever known.
And now we wait
For you all to de-orbit.
Oh Sammy boy, oh Sammy boy,
Oh please come home!

----------------------------------------------------

Arthur's wife Frauke was around today and for the next few days. She certainly had a way of brightening up the place with her boisterous, almost "cheerleading" manner! Just about everybody was running on empty at this point, so we needed something to keep us going.

Late this afternoon, final word came down on the extension day, and the result was predictable but disheartening: No. No science, that is. If they had to stay up an extra day for weather, they would just look out the window (see P79). To our knowledge, the crew was "for" doing science (a sentiment confirmed later by John Grunsfeld), and the only dissenting opinion was from the "flight surgeon" at KSC who refused to agree to an extension. He was concerned about crew fatigue since this was already going to be the longest shuttle mission on record. And the MSs and PSs had worked 12 hour shifts for the entire mission.

From a NASA bureaucrat's point of view, the decision made perfect sense:
the mission was already a success. Why take a chance that something could go wrong during an extra day that would taint that perception? But from the scientists' point of view, your country has spent half a billion dollars to launch the shuttle and put you up there, the telescopes are working, the Universe is calling, and the astronauts are willing, so *why not observe!?!* Believe it or not, we finished preparing the observing plans anyway, just in case the situation changed.

The MSFC PAO people had arranged for me to do a live interview with WOOD channel 8 in Grand Rapids, Michigan, at about 4:45 p.m. (5:45 EST) so I toddled upstairs about 4:30 to get ready. As I was combing my hair, it felt somewhat greasy, and I realized that I was so tired that morning that I had taken a shower but never used any shampoo! It turned out that after a couple of questions from Susanne Geha (the anchor), they opened the phone lines for questions! I didn't know this was coming and people were asking about everything from the space station to whether I thought NASA should be "privatized"! I was shooting from the hip on most of it, listening to this little voice in my ear and trying to look straight into the camera. I guess it came off OK, but I took a pretty good ribbing from some of my cohorts who had been listening in on NASA-TV.

A fairly large group met at the Bombay Cuisine restaurant for dinner. I left right after dinner and headed back to my room and to bed at 7 p.m. The long hours had really caught up to me.

**Thursday March 16, 1995:**
I awoke at 3:30 a.m. but was in no particular rush to get into the POCC since replanning was winding down. I had a real stick-to-your-ribs breakfast at the Kettle before going in about 5 a.m. We finished planning out the extra shift just in case it was needed, but by mid-morning the word came down that we would pack up on schedule late this evening (even though the weather for a landing at the Cape was not sounding good and they would probably be staying up an extra day. It became "the shift that was never flown", or perhaps the *shifta non grata*. A disappointing but predictable decision. But things had gone so well throughout the mission, it was hard to be too angry about it.

About that time an interesting development occurred. We missed the last and most important "moon" pointing for UIT. Now, the moon was a target for a guest investigator of UIT, and it was actually a difficult observation technically because the moon is so relatively nearby. As the shuttle moves in its orbit, the moon appears to move rapidly against the background, much like watching a nearby tree pass by as you drive around a curve in the road. This made it quite a challenge to latch onto and track with the Instrument Pointing System. We had spent a lot of time during the mission SIMs practicing moon observations, which was only partially useful because the real hardware (that is, the IPS) was not being used.

The idea was to get two pointings at the moon at two different phases, one a few days before full moon, and then one as near to full moon as possible. Comparing the brightnesses (the reflectivities really) of different regions was supposed to say something about such things as surface roughness and composition, or some such. Actually, no one had ever verbalized to us the real purpose behind the observation. It was not an investigation that most of us felt was of crucial scientific interest. It struck us more like, "Gee, no one's taken a picture of the moon in the UV before, so let's do it."

UIT had planned three observations of "Luna" in the timeline, with the idea being that the first one was sort of a practice shot with the real hardware. Lo and behold, the first observation worked (at about 242 hours MET), roughly according to plan, and so did the second (at 291 hours MET). Hence, it was somewhat surprising when the third and most important observation came up at 343 hours MET (the one near full moon) that they flubbed it! It wasn't even close--a coordinate error (I believe) left them one and a half degrees from the moon, which is a huge error in this business. (Even the moon is only a half degree across!) And with essentially no time to replan it, at least in a normal replanning mode.

I remember attending the SOPG meeting following this and listening to the rambling discussion about what had happened and what the options were for this "crucial" observation. It was viewed as crucial, I guess because it was for a guest investigator, and because they didn't want to leave anyone with a bad taste in their mouths. But we still had about 12 hours of nominal science operations coming up. I finally got tired of hearing all of the bellyaching and suggested that we just do something about it! If it was judged to be more important than something that was in the timeline, make a change! After all, our revamped replanning system could handle it (as we had demonstrated over and over during the mission). As was done so often during the pre-mission planning and even during the flight, the PI team planners were tasked with going away and working out the problem.

We ended up suggesting a "compromise" whereby the moon would be attempted again as the last science observation of the timeline. This required HUT to cut off our last target early (a Cygnus Loop pointing for me!), and required NASA to extend the science period about 20 minutes beyond the nominal cutoff (because of the moon's visibility at the time. Although it took awhile to percolate through the system, they eventually accepted it, which set up a little suspense for the final observation.

I left the POCC at 1 p.m. today to have lunch out--a new experience! So much sunlight! It seemed like I had barely seen the sun for over two weeks. Warren Moos and I had lunch at Shoney's. I went back to my room and dozed a little and relaxed in the afternoon, starting the long decompression. Dinner tonight was at the Green Hills Grille with Paul Feldman, Warren Moos and David Finley. I wasn't really that hungry because of the late lunch, but I went along mostly for the social aspect. I went back to the room with the intent of resting a little while and heading out to the POCC for the last Cygnus Loop observation and the moon, but the rest turned to a sleepy stupor and it was all I could do to crack open an eye and watch it on NASA-TV.

The previous moon observations had been done without HUT in the loop for the observation because we were somewhat afraid it would be too bright for our detector. Of course, the moon's spectrum would look just like the sun's spectrum since the moon shines by reflected light. While the sun is a relatively cool star, it has a bright emission line of hydrogen in the HUT range that would be very bright. But for the last observation, well...why not? As the observation began, there was a "piece" of the moon on the HUT TV! (There had been some doubt that it would show anything more than a bright screen because the dynamic range of the HUT TV is not large.) Sure enough, HUT got a very nice reflected solar spectrum that actually turned out to be interesting for comparison with some of our other solar system spectra like Jupiter, Venus, and Mars.

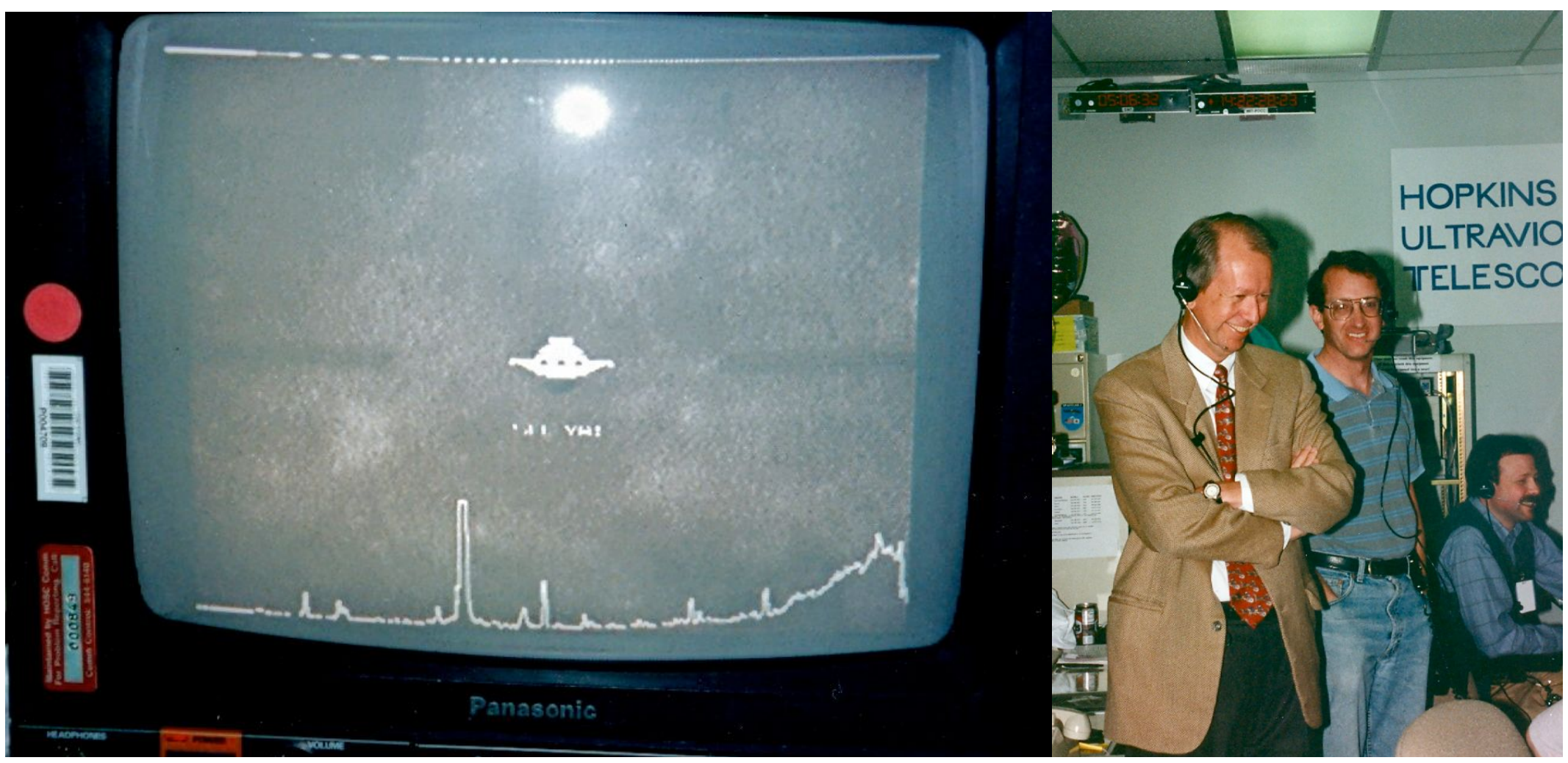

Left: The HUT TV showing a small flying saucer and the words “See ya!” (right) Arthur Davidsen with Ben Ballard watching. Harry Ferguson in the background.

About half way through the observation, all of a sudden the HUT TV display changed from showing the slit fiducial marks to showing a small flying saucer! I knew right away that one of the HUT engineers, Ben Ballard, must have worked some software magic to give the astronauts a special farewell. Sure enough, a few seconds later the call came down:

> *"Houston, this is Endeavour."*
> *"Yeah, Endeavour, we see it, too. Looks like some little green men have stopped by to say good bye!"*

A few seconds later, a message started blinking on the screen below the flying saucer. It said "See ya!" It was a great way to wind up an immensely successful mission.

NASA-TV showed very little, but from a couple of "wide" shots of the HUT area were enough to let me know that everyone was having a good time. It made me wish I had made it back out there for the last observation, but I felt like a 300 lb. Sumo wrestler was sitting on me even as I watched. Very shortly after the end of the observation, the deactivation activities began and I remember watching the IPS stow the telescopes back into a horizontal position before I really went to sleep about 11:30 p.m. The observations were complete.

**Friday March 17, 1995:**
I didn't wake up until 7:30 a.m. and didn't feel any too good despite the sleep. I had breakfast in the lobby at the hotel with a smattering of HUT and WUPPE folks coming and going. I decided to stop out to the Space and Rocket Museum on my way to the POCC. It was a hazy morning, with the sun peeking in and out. It looked like they were getting ready for a Space Camp graduation at the museum, and the gate to the "rocket garden" was open.  I stepped in to take a couple of pictures of the full scale model space shuttle, and when I came back the gate was closed and locked!  Hence, the only way I could get back out was to walk back through the museum.  In all my trips to Huntsville I had never seen more than the gift shop near the entrance!  I didn't spend a lot of time there, but I did walk through (and I did hit the gift shop!).

I got out to the POCC about 10:45 a.m., and packing was in full swing.  It was amazing how much had been torn down and packed already.  NASA was in the process of waving off the landing for today, which meant we could indeed have had an extra day of observations in principle.  But it was clear that folks were pretty burned out, and we did go out with a bang!  In the packing frenzy, a folder of mine containing miscellaneous papers (including the typed versions of some of my "songs" for the splashdown party) was packed away.  Also, the computers had been turned off and dismantled, so I had to try and reconstruct them from memory!

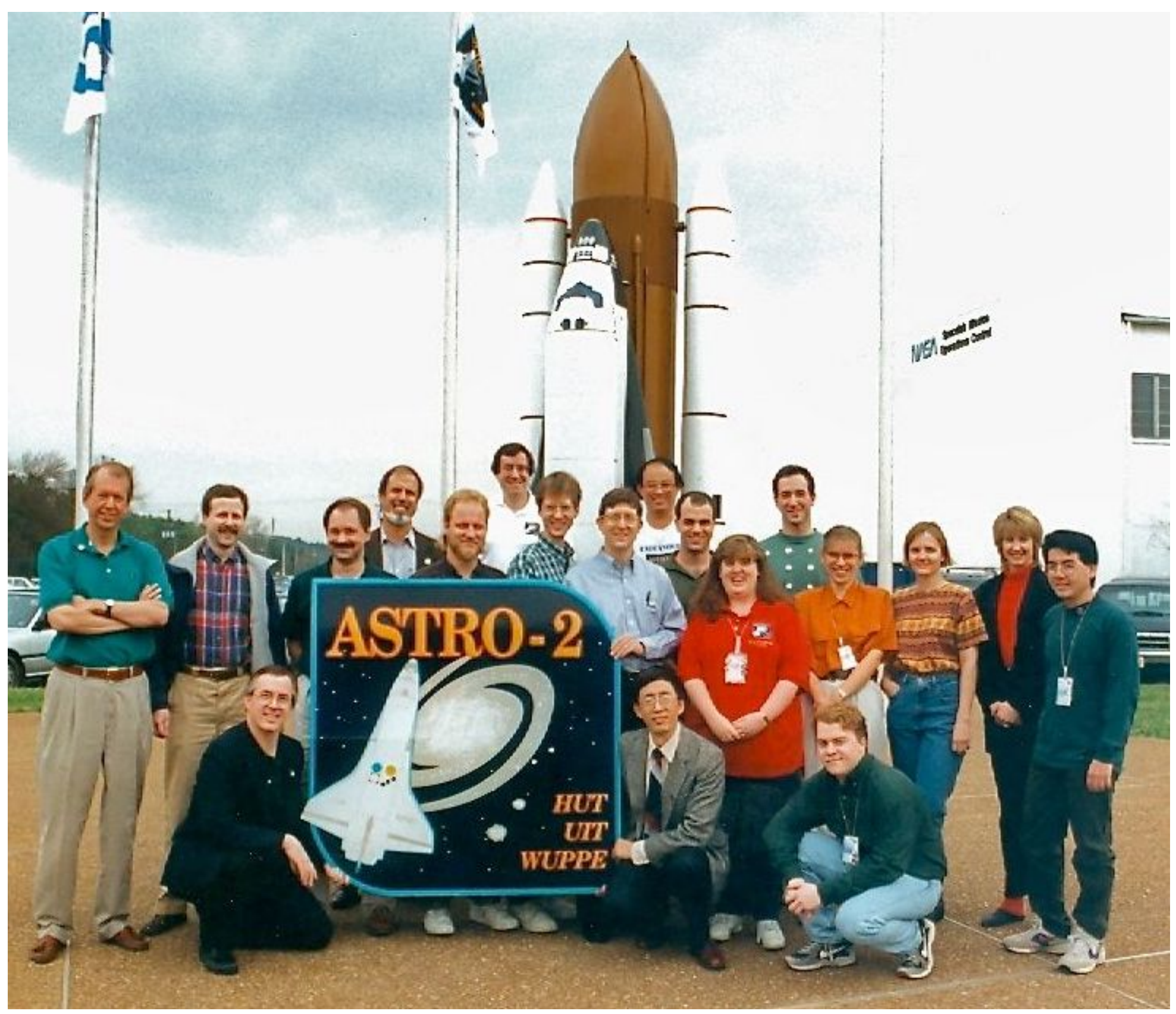

A partial team picture from Astro-2, March 17, 1995.

A big group gathered at O'Charleys for lunch, and then we gathered back at the POCC for a group picture.  Unfortunately, a lot of people had already headed off, so only about 2/3rds of the team was there, but it was better than nothing.  It was quite a production with everybody wanting a picture taken with their own camera.  Frauke very patiently snapped photos with every still and video camera available.  I spent some time in the afternoon checking e-mail back at JHU and finished collecting my thoughts for tonight's splashdown party.

Then at 4 p.m. we drove down to the MSFC splashdown party at the Rustic Lodge down by the Tennessee River. We had a nice time mingling with the MSFC Cadre and the other teams. What a difference from Astro-1! A party instead of a wake! There were nice words and presentations all around, from Bob Jayroe (Mission Manager) to Jim MacGuire (NASA HQ) to the PIs. I felt moved to get up and say a few words about the great job done by the planning team, both pre-mission and during the last two weeks. They even got me up in front to do a song ("*Sittin' in the POCC and SOA, Wastin' Time…*") with a back-up group of Brian Espey, Van Dixon, and Scott Vangen (Alternate Payload Specialist).

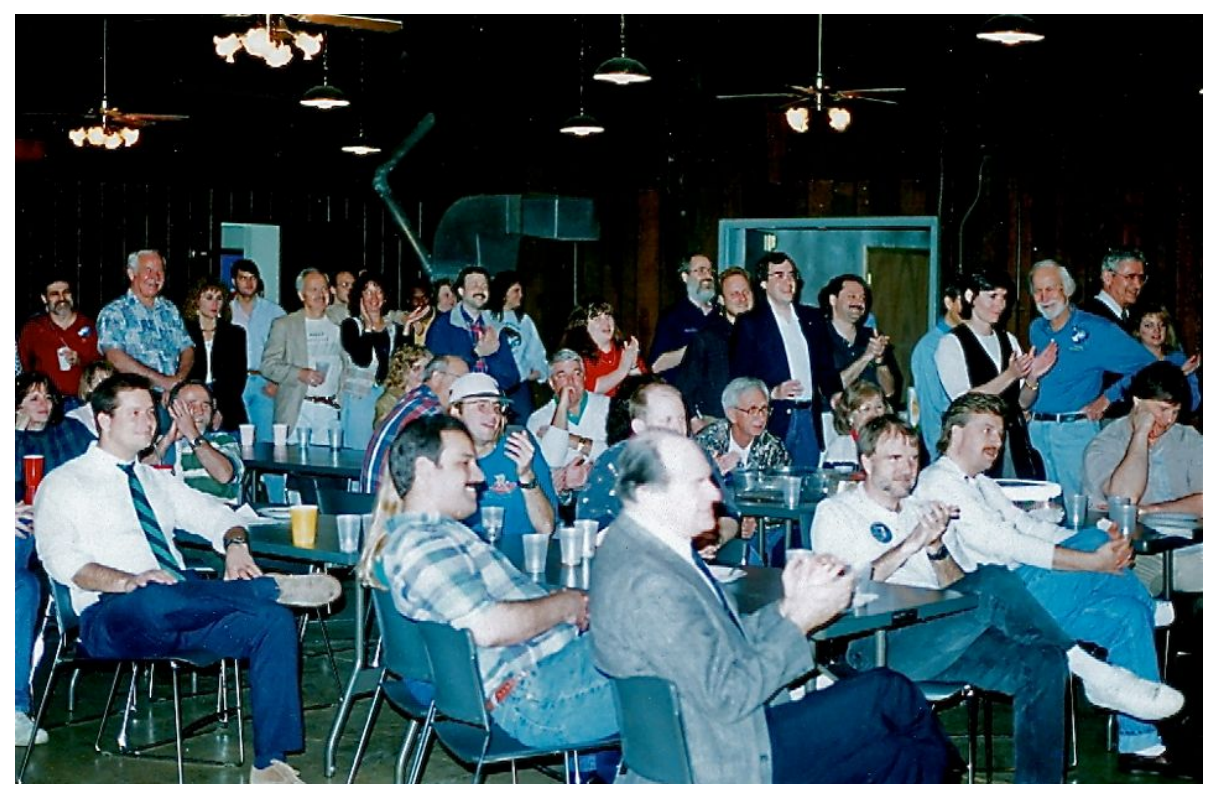

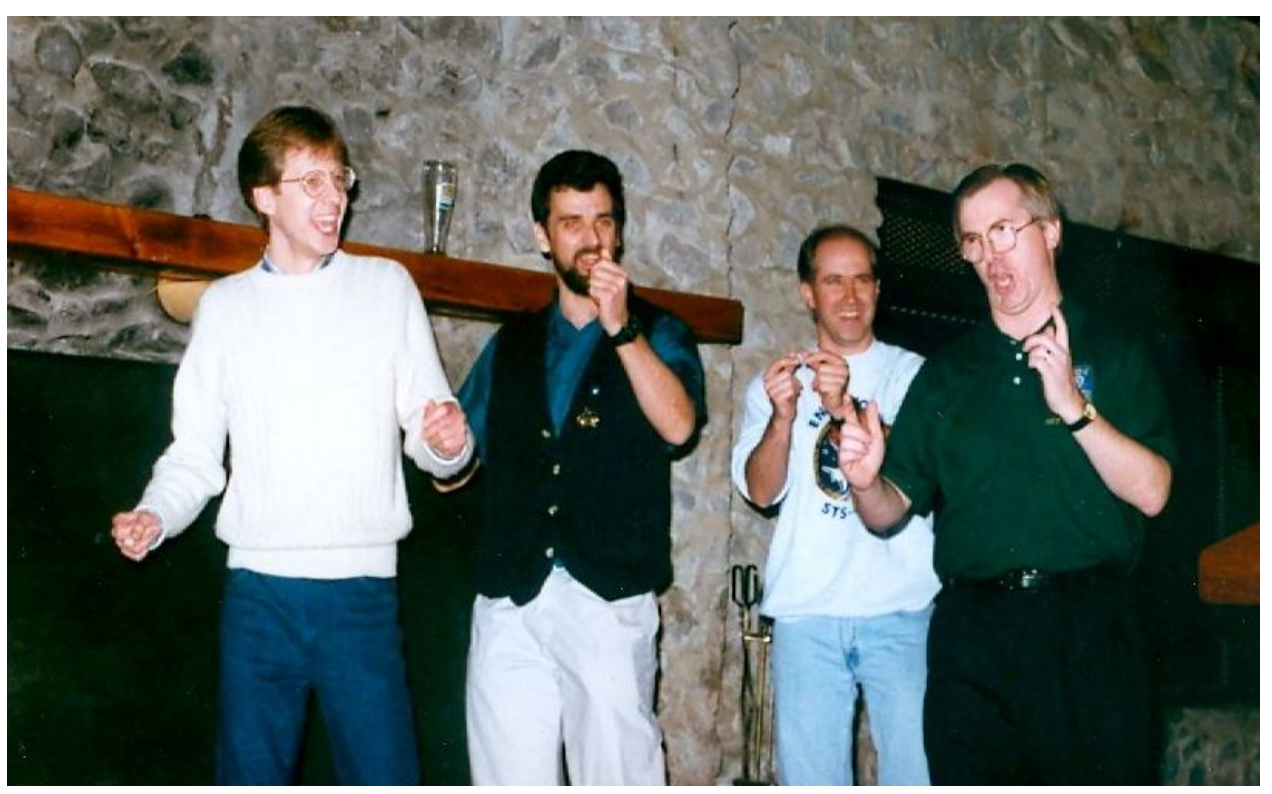

Splashdown party at the MSFC Rustic Lodge.

Later that evening we had the HUT splashdown party at the Green Bottle Restaurant. We had a very nice dinner, but some of us were so tired that we didn't have much pep left to whoop it up. After dinner Frauke Davidsen read a poem she had written to start off the entertainment, and then of course I did a few songs. Since it was St. Patrick's Day, I had to do "*Oh Sammy Boy*". Then there was "*Oh what have you planned, Billy Boy, Billy Boy*", the song with a verse for each replanner. The two biggest hits were the "*Theme from Astro-2*" (Where do I begin…to tell the story of the Astro-2 mission?), and then one that the MSFC File Manager Stephanie Gannaway-Osborn wrote called "*Mission Memories*" (to the tune of "*Memories*" from the musical Cats.):

Where do I begin...to tell the story
Of the Astro-2 mission?
It launched on time when all the clouds got very thin.
For 15 days we really packed the pointings in!
Don't end today...

Sam is quite a guy...he flew the shuttle way up
High into the sky!
To point our telescopes at objects far and near.
And with each spectrum all the scientists would cheer,
"Hip, Hip, Hooray!"

For 15 days...the thrusters fired
Until our MTL expired.
And so have we...of energy
We haven't any more!

And so it's ending, and all our stuff
We must be sending...back to Balto.
And so we go...

Now the fun begins...we'll plot our data
And we'll try and hide our grins!
We'll write our papers and to meetings we will go,
to tell the story of this mission called Astro!
Hip, Hip, Hooray!

---

*To the tune of "The Theme from Love Story," with apologies to Henry Mancini.

Midnight, not a sound from the Cadre,
Has the crew lost its memory?
They are flying alone.
In the darkness the JOT-P sheets collect at my feet,
And the FLMs begin to moan.

Memories, all alone in the moonlight,
I must smile at the old days,
I was timelining then.
When the shift ends tonight will be a memory, too.
Let the mission live again.

Burned-out ends of SRBs,
The stale cold smell of coffee.
A PI mutters, and a PCAP flutters,
and soon it will be morning!

Daylight, I must sleep with the sunrise,
I must think of the new shift,
and I mustn't give in.
When the shift ends, tonight will be a memory, too.
Let the mission... Live again!

Splashdown songs from the HUT party.

All in all, we managed to kill about four hours and left the restaurant about 11:30 p.m. A fair number of folks were leaving the next morning at about 6 a.m. to fly home, so they must have been really hurting!

**Saturday March 18, 1995:**
I got up about 8:15 a.m. after a good sleep. It was a sunny beautiful day. I had breakfast at the hotel, made some calls, and started packing. Then I called Harry Ferguson and we took a ride up to Monte Sano State Park and took a hike through the woods. It was a refreshing change of pace from two weeks of sitting at a terminal for 14 hours a day, and it gave us a touch of spring fever before we got back to Baltimore.

The landing had indeed been delayed from yesterday to this afternoon. We came back to the Residence Inn with the idea of getting cleaned up and going out to lunch somewhere and heading out to the POCC to watch the landing. However, we ran into some WUPPE folks, and they were having a big get together in a room above the hotel lobby to watch the landing, including a buffet, so how could we resist? It was waved off twice for Florida, and they finally decided to bring it home to Edwards Air Force Base in California. By the time it landed quite a crowd of WUPPE and HUT folks had gathered. The landing was smooth as silk, just like the mission, and they broke out the champagne. It was finally over.

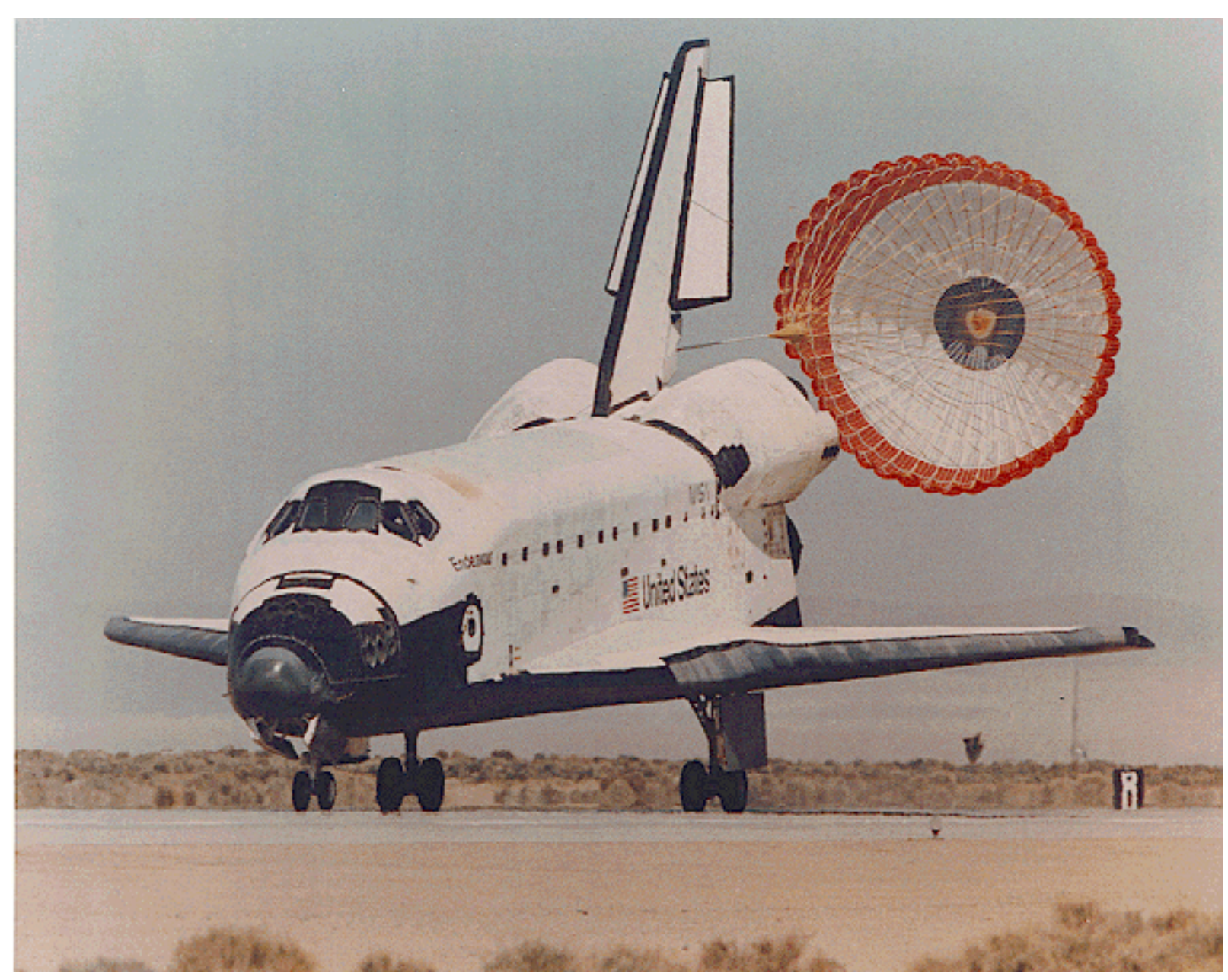

*Endeavour* safely on the ground at Edwards AFB, California.

When things calmed down after the landing, Harry and I ran out to the POCC to see it one last time and turn in our badges. The place was completely cleaned out of our stuff. It was really eerie to see it that way. We gave a passing thought to trying to rip off the Astro-2 flag flying in front of the POCC (decided not to), turned in our badges, and headed out to the airport, where I turned in my rental car.

About 7 p.m., the remnant HUT group gathered for one last dinner out. We decided to go to "Sakura" which was a Japanese restaurant that no one had been to more than once! The WUPPE group ate here quite often and called it eating "bait" (because of the Sushi bar), but I had the shrimp and vegetable tempura. It was a nice time.

I called Jean when I got back to the room, packed a few more things, and went to bed about 10:30 p.m. Tomorrow we were to go home!

**Sunday March 19, 1995:**
I got up at 7:30 a.m. and felt almost human. Even though it had been a long trip, I really was glad to have had a couple of days to decompress and get my schedule turned back around before heading home. Since observations ramped down long before the actual landing, several glowing summary articles about the mission began to appear starting today (see P80 and P81, and even the JHU Gazette, P82 and P83). It was certainly nice to get positive press for a change. Much later, a couple of good summary articles about the mission were published that are also worth a look (see P87 from *Final Frontier* and P88 from *Spaceflight Now*).

I had breakfast at McDonalds (an easy walk now that I was without a car), finished packing, and checked out about 10:15 a.m. Jerry Kriss drove Wei Zheng and me to Birmingham for the flight home. We checked in the rental car, lugged all our stuff into the terminal, and left for Louisville about 1:30 p.m. It was stormy in the Louisville area, and we had a bumpy ride and a very "interesting" landing in Louisville. It wasn't a great connection, and we had an hour and a half to kill in Louisville before finally leaving for Baltimore. This leg of the flight went much more smoothly, and we arrived at BWI a little after 6:00 p.m.

Jean and the kids met me at the gate. (Those were the days!) It was good to be together again as a family. I remember standing around waiting for my luggage, looking around at all of the people. Some were "HUT" families, but some were simply friends and family of people who were on the same flight. Did any of them know that we had spent the last two and a half weeks unlocking some of the secrets of the Universe? Did any of them care? Silly thoughts really. For this had been as much a human adventure as it was a scientific adventure. It was communication and cooperation between people that had made the adventure happen. And now this part of the adventure was over.

## Astro-2 Epilog

While the operational adventure of Astro-2 ended with our return to Baltimore, the scientific adventure continued well into the future. As the data were analyzed and the results from HUT started to come out, the extent of our scientific success even surprised those of us who were involved directly in the flight. The "bottom line" from the mission was that we made 385 science pointings with HUT at about 265 separate astronomical objects. This was something like five times better than we were able to do on Astro-1.

But even this is misleading. The improved sensitivity of HUT, the improved pointing stability, and confidence in the operations really allowed us to do a different "kind" of science—*programs* of investigation instead of "snapshots" of individual objects. The total number of refereed publications was only modestly larger than for Astro-1, but the quality of the data and the systematic studies made from multiple objects in a class or from multiple observations of time-variable objects like AM Her produced a totally different kind of science return compared with Astro-1.

Arthur Davidsen and colleagues did indeed detect the long sought-after intergalactic helium that was created in the Big Bang. It was first presented at a colloquium at the Space Telescope Science Institute in mid-April 1995, then again at our Astro-2 Science Symposium at JHU in early May, with the Astro-2 flight crew in attendance. However, the first public presentation was at the American Astronomical Society meeting in Pittsburgh in June 1995, where the result was one of the headline grabbers of the entire meeting. (See, for instance, the major article in the New York Times “Science Times” section: P84. Also, P85 from the *Baltimore Sun* and P86 from *Science News*.) Other HUT team members made a total of nearly 20 presentations at this meeting, many of which were included in a special dedicated issue of the *Astrophysical Journal (Letters)* from November 20, 1995, that contained 17 refereed papers on HUT results from Astro-2. The UIT and WUPPE teams of course also made numerous presentations at the AAS meeting, with subsequent publications as well.

# Chapter 7: The Legacy of the HUT Project

I wanted to conclude this history of the HUT project with some context about what I perceive as its importance in the bigger picture of both science and future NASA space science missions. Broadly speaking, I see this legacy in three parts: The impact of the project on JHU astrophysics, the science accomplished directly with HUT observations, and lastly the legacy of the people on the project, who went on to have major impacts on numerous NASA missions and instruments over the course of their careers.

## The Legacy at JHU

As mentioned in earlier chapters, the HUT project had an important role in the rapid development of astrophysics at JHU in the 1980's and beyond. Recall that a HUT concept was proposed in 1978 along with the other instruments that ultimately became the Astro Observatory payload. While the road to actual flight of these telescopes was a long and tortured one, the acceptance of the HUT project at JHU became a catalyst for the growth of astrophysics. JHU professors and HUT Co-investigators Paul Feldman, Warren Moos, and Dick Henry, in addition to Arthur Davidsen, were already part of this legacy and participated in both Astro missions.

In 1981, Arthur Davidsen spearheaded the team that worked with AURA (Associated Universities for Research in Astronomy) to submit what was viewed as a long-shot proposal to NASA to host the Space Telescope Science Institute on the Hopkins campus. Princeton was the presumed front-runner in that competition, but when the dust settled, the AURA/JHU proposal was the one that was accepted. While there are many aspects to any such decision, the standing of JHU as a burgeoning powerhouse in UV space astronomy, including the expectation of numerous HUT flights on the space shuttle, was an important factor in the decision.

It might almost be sufficient at this point to say "the rest is history." The presence of STScI on the JHU campus led directly to the expansion of the astronomy side of the department, which officially became a "physics & astronomy" department in 1984. The Center for Astrophysics was established within the department in 1985, and the hiring of numerous prestigious astronomy faculty ramped up throughout the 1980's and into the 1990's. And of course, the decision to build what became the Bloomberg Center for Physics and Astronomy on the JHU Homewood campus across the street from STScI was and is an important part of this legacy as well.

In the 1990's and 2000's, Hopkins professor Warren Moos led the successful development and operation of NASA's Far Ultraviolet Spectroscopic Explorer (FUSE) project, which was run from a control center in the Bloomberg building. FUSE is still the largest and most complex astrophysics project that NASA has handed off to a university to build and operate.  Hopkins professors and research staff have played major roles in the Hubble Space Telescope, its instruments, and servicing missions, sharing many staff jointly with STScI over the years.

One outstanding example of this synergy was Prof. Holland Ford, who was Co-investigator on the original Faint Object Spectrograph on Hubble (as was Arthur Davidsen).  Holland and Ball Aerospace engineer Jim Crocker were the architects of the COSTAR package that was flown on the first Hubble Servicing mission in 1993 to repair the spherical aberration problem for some of the instruments on Hubble. Crocker later joined JHU and was a key player in managing the development of the Sloan Digital Sky Survey project.  Ford went on to become the Principal Investigator of the team that build the Hubble Advanced Camera for Surveys that was installed on Hubble in 2002, Servicing Mission 3B.  And of course, the story continues with JHU involvement in the development, operations, and launch of the James Webb Space Telescope and further plans going forward.

## The Legacy of Science

In many ways, HUT opened the 900-1200 Angstrom spectral region to detailed scrutiny for the first time, certainly for everything other than the brightest nearby stars that had been observed earlier with the Orbiting Astronomical Observatory-3 (OAO-3 *Copernicus*) satellite.  Whether it be the plasma torus around Jupiter or an active galaxy nucleus hosting a massive black hole, or just about every kind of object in between, each HUT observation was ground-breaking at some level.
The most obvious scientific highlight for HUT was of course the detection and characterization of the so-called intergalactic medium using observations of the moderate redshift quasar HS1700+64, which was Arthur Davidsen's pet project.  The success of this observation would not have been possible without an Astro-2 mission and would not have been as successful as it was had the improvements to HUT between Astro-1 and Astro-2 not come to pass.  This project was highlighted in pre-mission releases and garnered a ton of press coverage after the Astro-2 mission.  (See for instance the article from June 1995 in the New York Times Science section, P84.)  It is largely due to this science result that the Smithsonian National Air & Space Museum contacted JHU and requested HUT to be placed on display, which occurred in November 2001(see P93); sadly, Arthur Davidsen passed away just a few months prior (see Obits directory in the archive).

The coverage of this single science program, however, does not do justice to the breadth of science accomplished by HUT on the Astro missions.  Each member of the HUT science team had one or more scientific programs of interest, with targets ranging from the solar system to nebulae, stars and star clusters in our Milky Way galaxy, active galaxies in the local universe and to many other topics as well. Again, the uniqueness of the far-UV (900 – 1850 Angstrom) spectral region makes many of these observations still relevant to this day.

But it wasn't just the 900-1200 Angstrom coverage that was important. An often-overlooked aspect of HUT was the simultaneous spectral coverage up to 1850 Angstroms, which was important for several reasons. It allowed the 900-1200 Angstrom spectral features to be put in context with more commonly observed features at longer wavelengths (for example, O VI 1032,1038 could be accurately compared to N V 1240, C IV 1550, etc., in stellar winds, supernova remnants, hot stars, and other objects). It provided for cross calibration of IUE and HST spectroscopic measurements, again mostly above 1150 Angstroms. For objects like cataclysmic variable stars, fitting the continuum across the entire HUT range was an important diagnostic for the accretion disks in those systems, along with observing time variability of the various components (continuum, emission line intensities and shapes). The list goes on.

## HUT Becomes History, Earns Spot in Smithsonian

by Alice Knox

The Hopkins Ultraviolet Telescope, which completed successful missions aboard the space shuttle Columbia in 1990 and Endeavour in 1995, has been permanently transferred to the Smithsonian Institution's National Air and Space Museum, where it will become part of a new exhibit.

The NASA-funded project was a collaboration between APL and the Hopkins Department of Physics and Astronomy. The telescope, which made astronomical observations in the ultraviolet spectrum during the two shuttle flights, was designed and built at the Homewood campus and APL.

"The data from the flights helped scientists better understand the ratio of hydrogen and helium during the formation of the universe," says **Glen Fountain**, HUT program manager for the first mission. "This information helped us to verify the current theory of the origin of the universe."

When the new Smithsonian exhibit, "Explore the Universe," premiers in September, the doors on HUT will be open so the public can see the mirrors and spectrograph inside, says **Steve Conard**, senior research engineer from the JHU Department of Physics and Astronomy. An electronic tour of the upcoming exhibit can be viewed at: nasm.edu/galleries/gal111/universe/univ_pre.htm. ❖

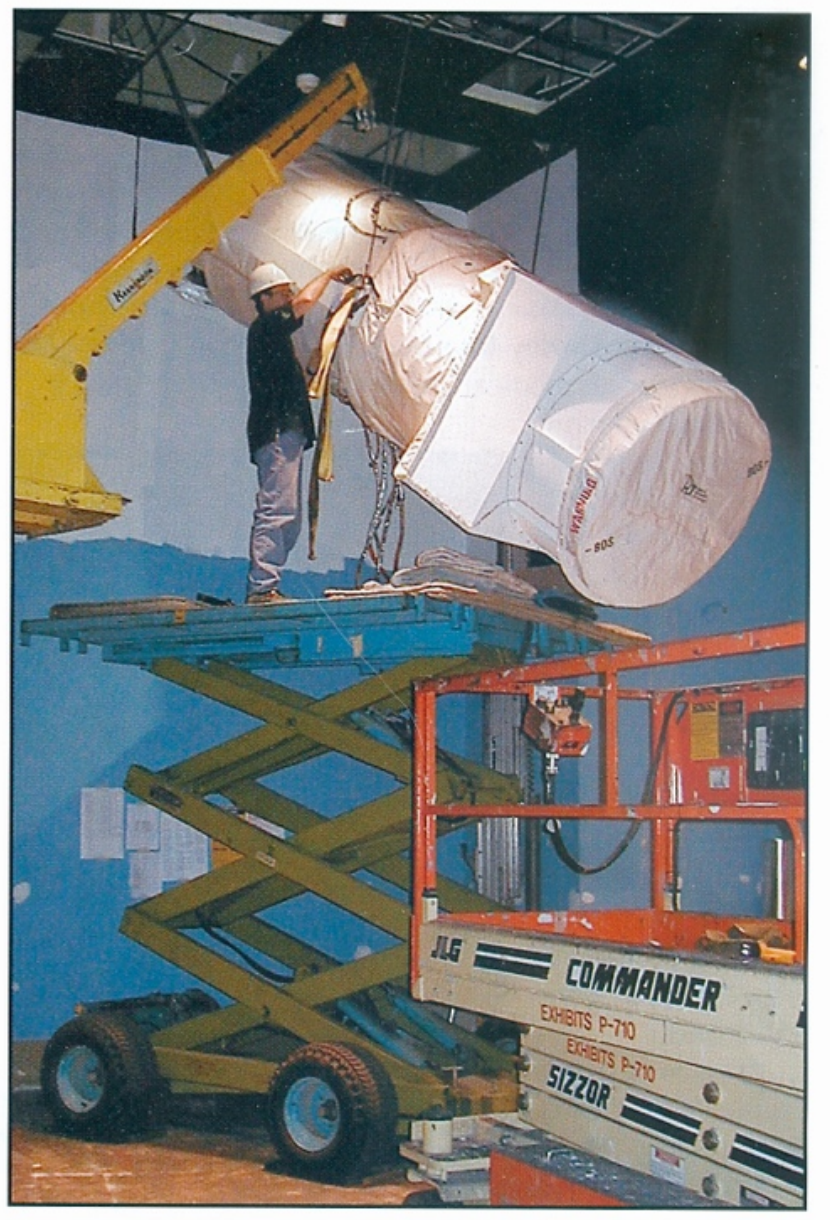

**Smithsonian employee Scott Neel suspends HUT from the ceiling of the National Air and Space Museum.**

A short article highlighting the installation of HUT at the NASM.

HUT observations of a range of white dwarf stars was interesting both because of the science on this class of object, but also because they were and are important calibration sources for UV missions. HUT was carefully calibrated in the lab by Jeff Kruk both before and after the Astro-2 mission, and the high signal-to-noise observations of numerous WDs throughout the mission produced an excellent calibration standard not only for the HUT data, but ultimately for HST and FUSE observations as well.

Another important but relatively minor aspect that should be mentioned is the handful of extreme ultraviolet (450 – 900 Angstrom) observations made during Astro-1. The second order of the HUT grating on Astro-1 allowed light from 450 – 900 Angstroms to be detected. While this spectral region is largely blocked by the interstellar medium of our Galaxy for most targets, very nearby white dwarf stars could be observed in a few cases, showing emission at the shortest wavelengths. NASA's Extreme Ultraviolet Explorer (EUVE) was not launched until June 1992, so these early observations were themselves groundbreaking.

Coincident with the preparation of this document in 2025, two gatherings were held to celebrate the success of the Astro missions. The first gathering was held at the U.S. Space and Rocket Center museum in Huntsville, Alabama, not far from NASA MSFC, attended primarily by operations staff and engineers from NASA KSC and MSFC. The second gathering was in early September 2025 at Johns Hopkins in Baltimore, attended mainly by science and engineering team members for the instrument teams (and HUT team in particular). As part of preparations for these gatherings, I spent some time devising searches in the Astrophysics Data System database trying to obtain a count of scientific articles based on data from one or more of the instruments on the Astro missions. This search may or may not be completely accurate, but it is sufficient to provide an indication of scientific impact of Astro. Here is what I compiled:

| Mission | Peer-reviewed Papers | Conference Proceeding and Other Papers |
|---|---|---|
| Astro-1 | 112 | 269 |
| Astro-2 | 187 | 287 |

Scientific publications summary for All Astro instrument teams.

These numbers included all of the telescope groups (including BBXRT on Astro-1) but the numbers were significantly dominated by HUT papers. This is an enormous number of papers to have resulted from two relatively short space shuttle missions, and while it is difficult to compare missions in different categories (like microgravity or other Spacelab missions), I dare say the Astro missions were easily the most successful science missions ever flown on the space shuttle.

## The Final Archive at MAST

As the HUT project ramped down in 1996, development for the FUSE mission (also headquartered at JHU) was ramping up. A number of key HUT personnel transferred to the FUSE project (myself included), and the local HUT archive of data (both raw and processed), processing programs, and supporting web pages continued to be maintained by the FUSE project. This was a symbiotic relationship as HUT observations that covered the 900-1200 Angstrom region were of significant use in planning FUSE observations at higher spectral resolution.

Data from both Astro missions were originally archived at the National Space Science Data Center (NSSDC), and later (in June 1999) they were transferred to the appropriately named Mikulski Archive for Space Telescopes (MAST) at STScI. However, at the time the data were originally delivered, there were no standardized criteria to follow. In retrospect, there was significant supporting science and engineering information for the observations that never made it into the file headers, especially various data that reflected the actual "as flown" conditions from real time operations. Hence, while reasonable "quick look" final extracted spectra were available in the archive, there were several aspects that made the archived data less useful than they could be to the general community for ongoing archival research.

As the FUSE mission itself came to the end of its operational life in October 2007, we realized that not only would the FUSE project need to be closed out, but also that the unique HUT materials that had maintained at JHU would also be going away (with a corresponding loss of significant background and supporting information for the HUT observations). In mid-2008, a Astrophysics Data Analysis Proposal (ADAP) was crafted to perform a new processing and archiving of the HUT data for future use by the community at large. This proposal (see D51) laid out a compelling plan for improving the archived HUT data with additional keywords to provide the missing context from real time operations, and in a time-tagged format that would be a much more flexible and useful for archival users.

The plan also included producing an archive of the HUT TV guide camera images for each observation which had never been archived previously. These images provided "ground truth" for field orientation relative to guide stars as well as providing information on pointing stability (or not) during the observation. Finally, the archiving of the HUT web site from JHU was also to be transferred to MAST for future reference.

The proposal was accepted and ultimately completed as planned. However, it was indeed a challenge to accomplish in the context of the FUSE mission close-out activities, which also included all of the key people taking on new jobs while the project was in process. Although it took longer than expected, special kudos are due to Mary Romelfanger, Jeff Kruk, and especially Van Dixon for their efforts above and beyond to make it happen. The details are captured in a PASP article by Dixon et al. (2013, PASP, 125, 431) (see D54).

## The Legacy of the People from the HUT Team

The HUT project grew out of the extant sounding rocket program at JHU, a program that continues to this day at JHU and several other research universities around the country. One of the hallmarks of the sounding rocket program is giving people (students, postdocs, etc.) access to hands on experience with the development of flight hardware. This was certainly the case with the HUT project as well, and placed many HUT team members in position to make major contributions to future NASA missions.

The team that Arthur Davidsen developed in the early 1980's to construct and operate HUT was largely made up of early career scientists and engineers, with a few key graduate students, administrative support people, and a project business manager thrown in. Of course, the team on the Homewood campus was paired with a strong and experienced team at the Applied Physics Laboratory, which also included many young engineers as well.

Most of the scientists and graduate students were expecting to be with the project for a few years, fly the HUT instrument multiple times, and then move on to faculty or other positions from there. The *Challenger* disaster and the ensuing delays stretched this out by many years. Some team members had left or were planning to leave the project by the time HUT finally flew the first time. Some team members stayed on through the Astro-2 mission and then moved to other projects and others joined the team for Astro-2. But ultimately everyone moved on to other projects or jobs by 1996 when the HUT/Astro funding ended. Here is a very brief summary of where some key players went and what they accomplished over their subsequent careers.

**Harold (Hal) Weaver** was mainly involved in the early years of the HUT project, first as a graduate student, and then returning to JHU after a postdoc at GSFC to help with the construction of the telescope. He left JHU after the *Challenger* disaster in 1986 and worked at STScI for nine years, working to establish the moving target capability for Hubble. In 1996, he joined the Senior Professional Staff at JHU/APL where he spent the remainder of his career. Hal was a very active observer of comets and solar system objects and participated in numerous missions over the years, including as a Co-Investigator and the Project Scientist on NASA's New Horizons Mission to Pluto and beyond.

**Knox Long** was HUT project scientist from 1982 through the Astro-1 timeframe. He then accepted a job at STScI in 1990 after the Astro-1 mission, initially as project scientist for the data archive, and then took on a variety of leadership positions over the years related to STScI, Hubble, and eventually JWST. He also remained very active in science endeavors independent of his functional work. Knox returned to work on real-time operations for Astro-2 as well.

**Sam Durrance** came to JHU in 1982 and was key to the construction and integration of HUT into the Astro payload, as well as flying on both Astro-1 and Astro-2 as payload specialist. After Astro-2, Sam left JHU and worked at several universities in Florida, teaching and becoming active in NASA's Space Grant educational activites.

**Randy Kimble** arrived at JHU in 1983, working on the HUT hardware development as Deputy Project Scientist. Randy transitioned to a civil service position at NASA GSFC in 1990 but returned to work real-time operations for the Astro-1 mission. In his career at GSFC, Randy was a key player in the successful development of several Hubble instruments (STIS, ACS, and WFC3) and working on Hubble servicing missions. After joining JWST in 2009, he had a major role in the complex cryo-vacuum test campaigns that prepared the instruments and telescope for flight and in the commissioning of JWST after launch.

**Chuck Bowers** was a graduate student at JHU working under Arthur Davidsen and Paul Feldman, building and flying the HUT prototype spectrograph on a sounding rocket to study UV airglow for his Ph.D. thesis (1985). He continued working on the HUT project through Astro-1, worked briefly on the FUSE project, then moved to GSFC in 1993, where he worked for the rest of his career. He became a Co-investigator and Optical Scientist for the STIS instrument on Hubble, leading a team to calibrate the numerous optical elements of the instrument and participating in the testing and commissioning on orbit. He joined the JWST project at the end of 2006 and became Deputy Observatory Project Scientist, assisting in various aspects of design, characterization, testing and finally commissioning of JWST.

**William (Bill) Blair** remained at JHU throughout his career after arriving in mid-1984 to work on HUT. After supporting both Astro-1 and Astro-2 mission planning, he accepted a job with the FUSE project at JHU, initially as the head of mission planning and later as Chief of Operations throughout most of FUSE's operational lifetime. After FUSE wound down, Bill contracted with STScI to work on the James Webb Space Telescope planning and scheduling system and in 2017 was named Project Scientist for User Support for JWST, retiring in early 2023. Bill remained engaged in research and astronomical observing throughout his career.

**Jeff Kruk** came to JHU prior to Astro-1 in the late 1980's and worked both missions. Jeff also accepted a job to work on both FUSE development and as Deputy Chief of Operations, overseeing the technical challenges faced by FUSE. After the end of FUSE operations, Jeff took a civil servant job at NASA GSFC, overseeing several major projects including the construction and testing of the Roman Space Telescope in preparation for flight (work that has just come to fruition with the launch of Roman on Aug. 30, 2026).

**Gerard (Jerry) Kriss** came to JHU in 1985 and worked both missions, becoming HUT Project Scientist for Astro-2. After Astro-2, Jerry worked on FUSE development activities for two years before accepting a job at STScI, initially as the Instrument Scientist for the STIS instrument on Hubble and then numerous other leadership roles over the years, including a stint as Interim Deputy Director of STScI and as Associate Director, first for Operations and later for Instrumentation. Jerry maintained a very active research career as well.

**Wei Zheng** joined the HUT project after Astro-1 and participated in the preparations and flight of Astro-2. Wei also worked on the early years of Hubble with Arthur Davidsen and Holland Ford and remained active in research at JHU for many years after that.

**Mary Romelfanger** came to JHU in 1988 and supported the HUT project throughout the remainder of its time, participating in both missions and providing critical technical support of both the computer systems and the data processing and archiving. She also helped the project usher in the new world of the internet, playing a key role in establishing the HUT web pages. After 1996, Mary played a similar role in the FUSE project by leading the data processing and archiving of all FUSE data. After FUSE, Mary transitioned to STScI and held key positions in the ongoing data processing and archiving activities of multiple missions.

**Steve Conard** began his engineering career with JHU in 1982, supporting the sounding rocket program as well as HUT. He continued with the department working as lead optical engineer with the FUSE mission before moving to APL in 2002 where he remained until retirement in 2024. At APL, Steve was a lead engineer for optical instruments on the CONTOUR, New Horizons, and Lucy planetary missions.

**Van Dixon l**eft JHU after Astro-2 after receiving his Ph.D. in 1996 based largely on HUT results. After a brief postdoc at STScI, he accepted a job at the Space Sciences Lab at UC Berkeley working on several projects including the ORFEUS instrument that also flew on the space shuttle. He then returned to JHU five years later to work for FUSE. In 2011, Van was hired by STScI where he was an Instrument Scientist for the JWST NIRISS instrument and managed the MESA branch of STScI, among other tasks.

**Harry Ferguson** obtained his Ph.D. in (1989) and left JHU prior to Astro-1 for a postdoc in Cambridge, England, but returned to support the mission. He was later hired by STScI where he has held numerous key positions over the years for Hubble, JWST, and currently Roman Space Telescope. He returned to participate in the Astro-2 mission in 1995. Harry has maintained a very active research career despite his heavy load of functional work at STScI over the years.

**Tom Brown** was a JHU grad student who supported the Astro-2 mission and received his Ph.D. with Arthur Davidsen in 1996. After a postdoc at GSFC, Tom was hired at STScI and worked in several Hubble instrument support positions before first becoming Mission Head for Hubble Space Telescope, and currently (2025 and forward) he is Mission Head for JWST.

**Bradford Greeley** was a graduate student at JHU from the early 1990's through Astro-2, receiving his PhD based on HUT cataclysmic variable data in 1996. He moved to a civil service position at NASA GSFC, working on the optical systems for several astronomy and earth observing missions for the remainder of his career. Sadly, Brad passed away in early 2016.

# Appendix A: A Detailed Timeline of the HUT Project and Related Events

**(Version dated August 20, 2025)**

1975/08: Arthur Davidsen arrived at JHU. (Ph.D. Berkeley 1975; arrived later in the year.)

1977/04-16: Launch of AFD/Hartig sounding rocket to observe 3C 273 in the UV (per RAK).

1977/09-15: AFD+ Nature article on 3C 273 UV spectrum published.

1978/01-26: Launch of the International Ultraviolet Explorer (IUE) satellite, increasing access to the UV universe.

1978/?? NASA Call for Proposals released for potential space shuttle payloads.

1978/11-15: Original HUT proposal, “A Far Ultraviolet Telescope/Spectrometer for Spacelab,” submitted in response to NASA AO-OSS-2-78. Est total cost: $2,059,000. Proposed starting date: Phase A: 1 July 1979; Phase B: 1 Jan. 1980; Phase C: 1 Jan. 1982. (See 1978-HUT_orig-prop_tech.pdf and 1978-HUT_orig-prop_cost.pdf.)

1979/01: Arthur F. Davidsen awarded the AAS Helen B. Warner prize for the UV spectrum of 3C 273.

1979/03: AFD appointed to the AURA Board for a 3-year term.

1979/??: NASA RFP for a space telescope institute released, including a selection for hosting the institute.

1979/08-30: Mailgram received with tentative acceptance of original HUT proposal, and NASA HQ release no. 79-112 confirms, along with 30 other investigations (including the other Astro telescopes).

1979/11-20: Updated HUT Definition phase (Phase A/B) proposal submitted to NASA, in response to RFP 5-63420/278.

1980/02-01: HUT Phase A/B study contract begins.

1980/07??: Sam Durrance arrives at JHU, but apparently didn’t focus on HUT work until early 1992. (See pg. 25 of 1991-10_JHMag_article.pdf.)

1981/??: AURA/JHU proposal for STScI on Homewood campus announced. From STScI online history: "After reviewing applications from prestigious institutions and universities, NASA selected the Association of Universities for Research in Astronomy (AURA) in 1981, which named the Johns Hopkins University's Homewood campus in Baltimore, Maryland, as its base of operations. In the same year, Dr. Riccardo Giacconi was selected as STScI's first director."

1981/02?: RFP-5-15177/278 for HUT construction received. (See 1981-RFP-5-15117-278.pdf.)

1981/04-12: The first space shuttle launch, STS-1 (*Columbia*) from NASA Kennedy Space Center.

1981/05: HUT Construction Proposal submitted. (See 1981-HUT_construction_prop.pdf.)

1981/09-21: NAS 5-27000 NASA Contract begins. (See 1981-HUT_contract_NAS5-27000.pdf.)

1981/12-10: JHU Gazette article announces $5.9M NASA contract for HUT. (See 1981-12-10_JHUGazette.pdf; note: first public use of HUT acronym?))

1982/01: Sam Durrance joins the HUT team.

1982/??: Knox Long joins JHU as research scientist and HUT project scientist.

1982/06-15: Ted Gull appointed NASA Astro project scientist.

1982/08-16: Aviation Week article discusses early concerns related to NASA and comet Halley missions. (See 1982-08-16_AvWeek_article.pdf.)

1982/12?: Hal Weaver completes PhD at JHU and leaves for NASA/GSFC postdoc.

1983/07: HUT Investigation Development Plan submitted. (See 1983-HUT_Inst_Dev_plan.pdf.)

1983/08: Randy Kimble joins JHU/HUT as Assistant Research Scientist.

1983/10-31: Baltimore Sun article discusses AFD, STD, and KSL in running for PS. Says the NASA vetting process would be "in a few weeks" and selection in spring 1983. (See 1983-10-31_BaltSun_article.pdf.)

1983/11: JHU Newsletter article 1983-12-02_JHUNewsletter_article.pdf confirms that NASA candidate PS testing occurred in mid-November at JSC.

Circa 1983-84: JHU Physics department decides to stay together as one department and become the Physics & Astronomy Department. (Don't have exact date yet.)

1984/??: Hal Weaver returns to JHU as assistant project scientist to work on HUT; remains until post-Challenger period.

1984/02-27: AFD writes to George Lucas requesting permission to use "Jabba the Hutt" as the HUT mascot. Letter in response received 1984/03-22 providing permission. An apology for the delayed reply because they have been, "...extremely busy with the production of INDIANA JONES AND THE TEMPLE OF DOOM."

1984/05-31: JHU press release announces that HUT construction has begun in Rowland Hall, in preparation for a flight on the space shuttle in March 1986 to view Halley's comet, followed by a JHU Gazette article dated June 7, 1984 . (See 1984-06-07_JHUGazette_article.pdf.)

1984/06-03,04: First HUT whitewater rafting trip to western MD.

1984/06-20: NASA announces selection of Durrance, Parise, and Nordsieck as payload specialists for the Astro missions. (See articles shortly thereafter, 1984-06-21_BaltSun-article.pdf and 1984-07-12_JHUGazette_article.pdf.)

1984/07-01: Bill Blair joins HUT project as Associate Research Scientist.

1984/09-??: HUT leaves Rowland Hall for APL. (See 1984-09_HUT_leaving_JHU.pdf.)

1985/03/??: Completed HUT shipped from APL to KSC for Astro-Halley mission.

1985/Spring: First Astro/Halley mission color brochure produced by NASA. Mission is designated STS-61E.

1985/05-01: High intensity light bulb bursts over vertical HUT at KSC, damaging HUT primary mirror. (See account on pg. 40-41 of Feb. 1982 JHU Magazine article 1986-02_JHUMag_article.pdf ).

1985/05-04: JHU Center for Astrophysical Sciences announced by Dean George Fisher in a Baltimore Sun article. A CAS charter dated Jan. 30, 1985, established CAS within the Physics and Astronomy department. AFD is the first director, announce in late November. The charter puts forth the idea of non-tenured positions equivalent to Assistant, Associate, and full professor, leading to today's research staff positions within the university.

1985/06: Davidsen & Fountain APL Technical Digest article on HUT published. (See 1985-06_AFD_APL_Tech Review_on_HUT.pdf.)

1985/08: Jerry Kriss joins JHU/HUT team as Associate Research Scientist.

1985-12-11: Chronical for Higher Education article published, noting importance of Astro mission for observing comet Halley, and notes the inclusion of a "wide field camera" to observe it. Also notes, "According to NASA's current plan, the Astro package will be lifted into space aboard a space shuttle three times, for periods of about nine days each, over the next year and a half. After this spring's initial launch, Astro will be readied to fly without the wide-field camera in the fall of 1986 and again in July 1987, at a total cost of close to $100-million." (See 1985-12-11_ChronHE_article.pdf.)

1986/01-21: JHU Gazette announces "Put your name in orbit" event for Jan. 28 at noon in the Glass Pavilion. (See 1986-01-21_JHUGazette_article.pdf.)

1986/01-28: 11:39 a.m. EST: STS-51-L Challenger disaster.

1986/02-15: APL memo describes standdown status of HUT. (See 1986-02-25 APL HUT status memo.pdf.)

1986/03-06: Nominal pre-Challenger launch date for Astro-Halley. (See a number of articles published shortly thereafter: ex: 1986-02_JHUMag_article.pdf, and especially the Baltimore Magazine article 1986-05_BaltMag_article.pdf.)

1986/03-07: Astro IWG at GSFC.

1986/05-24,26: Second HUT weekend and rafting trip in western MD.

1986/06-01: Report of the Presidential Commission on the Space Shuttle Challenger Accident" is published. (The Rogers Commission; see 1986-06_Challenger_Report.pdf.)

1986/07-31: SpEx proposal deadline.

1986/08-14,15: Astro IWG in Madison.

1986/09-01: Aviation Week & Space Technology article notes, "15-18 Spacelab shuttle missions" to be cancelled; Astro is one of only three left standing, nominally targeted for January 1989. Article says, " Instead of flying three times over 18 months starting this year, the Hopkins Telescope is likely to have one flight in 1989 and a second in 1992-93." (See 1986-09-01_AvWeek_article.pdf.)

1986/09: Van Dixon comes to JHU as grad student. Remains through both Astro-1 and Astro-2, receiving PhD in 1996.

1986/10-06: AFD memo announces the first post-Challenger manifest, with flights beginning in Feb. 1988. HST is scheduled for the fifth flight (Nov. 17, 1988) and Astro-1 as a "dedicated mission" on *Columbia* on Jan. 19, 1989. The bad news, no Astro-2 is to be seen.

1987/01-04,09: AAS meeting in Pasadena.

1987/02-24: SN 1987a in LMC discovered; ultimately used as justification for adding BBXRT to Astro-1 mission.

1987/05-30,31: HUT whitewater rafting trip #3 to western MD.

1987/06-15,18: AAS meeting in Vancouver.

1987/07-06: Aviation Week & Space Technology article, "NASA Nears Key Shuttle Tests, Changes Initial Payload Schedule" notes first mission scheduled for June 2, 1988, with HST launch moved to the sixth slot, the previous Astro-1 slot on June 1, 1989. Astro-1 not yet re-manifested at this point.

1987/07-27,31: HUT aliveness testing at KSC.

1987/09-29,30: Astro IWG meeting (location?)

1987/10-05: JHU Groundbreaking for new Bloomberg Center.

1988/03-21,22: Astro IWG meeting in Huntsville.

1988/04-12,15: "10 Years of IUE" conference held at GSFC.

1988/07-??: Mary Romelfanger hired by HUT to oversee computer systems.

1988/08-01 through 11: IAU General Assembly meeting in Baltimore.

1988/09-30: SUSI SmEx proposal submitted.

1989/01-26,27: Astro IWG mtg at GSFC.

1989/05-30 to 06-2: HUT functional test at KSC.

1989/07-25 to 28: KSC testing.

1989/08-3,4: Joint Ops test at KSC.

1989/08-10,11: Level IV Mission Sequence Test

1989/09-7,8: Astro IWG mtg at GSFC

1989/10-02,06: MSFC training, Sim, and IWG mtg.

1989/10-20: "Final" JSCIPLAN for Apr. 1990 launch date delivered.

1989/10: Harry Ferguson completes JHU PhD thesis; leaves for postdoc but returns for both Astro-1 and Astro-2 missions.

1989/12: Lennard Fisk announces that Astro-1 will be the only Astro mission.

1989/12-28: AFD sends a 3-pg letter to Fisk expressing disappointment at the decision to terminate additional Astro flights after Astro-1. (See 1989-12-18_AFD_memo.pdf.)
1990/01/08-09: STScI Science Writers Workshop for HUT/Astro and HST

1990/01-16,18: Mission planning trip to MSFC.

1990/02-13,16: Sim at MSFC.

1990/03-20,22: Joint Integrated Sim#1 at MSFC.

1990/04: Long JHU Magazine article in the April 1990 issue (no doubt actually published in late Feb. or March) notes the interplay between HST and HUT/Astro launches and capabilities. HST scheduled for Apr. 12, 1990, and Astro-1 for May 9, 1990, at this point. (See the HUT portion of this article here: 1990-04_JHUMag_partial.pdf.)

1990/04-02: Astro-1 Target Book delivered.

1990/04: JIS#2 cancelled; HST launch delayed; STS-35 moved to May 9, 11:50 p.m.; subsequently moved to May 15.

1990/04-10: Near launch of HST on Discovery (STS-31). This had been scheduled for Apr. 12, but was actually moved earlier by two days, only to be scrubbed at T-4 minutes when an APU failed. (An FOS team meeting had been scheduled at the Cape during this time to watch the launch.)

1990/04-24: Hubble launch on STS-125 (Atlantis)

1990/04-27: Scare about whether Sam will fly or not (due to kidney stone). Luckily, problem resolved (May 4 Evening Sun article); a May 16 launch was planned at this time.

1990/05-16: Launch scrubbed (*Columbia* cooling system valve problem); postponed to May 30.

1990-05: Big buildup in the press ahead of May 30 launch attempt.

1990/05-29: Launch scrubbed 7 hrs before launch (*Columbia* H leak problem). Initially sounded like it may only take a few days to a week to resolve the problem.

1990/06: June Sky & Telescope carries Blair & Gull article about the Astro Observatory. (No doubt was actually published back in the April timeframe.) (See 1990-06_A1-SkyTel_AstroObs.pdf.)

1990/06-01: Science "Research News" article by M. Waldrop (no doubt went to press prior to May 30 launch attempt) titled, "Astro: the First and Last of Its Kind," gives good discussion of the changed landscape for shuttle/Spacelab missions, the launch delays (so far), and the likelihood that Astro-1 will be the only flight of the payload.

1990/06-07: Confirmation that *Columbia* will be rolled back to VAB.

1990/06-13: Launch now "no earlier than August" announced.

1990/06-13 to 15 and 18-20: Replanning trips to MSFC for an Aug. 12 launch date.

1990/06-19: Pres. George H. W. Bush visits NASA/MSFC POCC and HUT SOA during Astro-1 preparations.

1990/06-26: JHU Gazette carries story about STScI building being named after JHU president Steven Muller.

1990/06-27: NASA officially announces the Hubble mirror problem (although the problem had been known internally since several weeks after launch).

1990/07-20: JHU P&A department moves from JHU Rowland Hall to new Bloomberg building.

1990/08: Launch slipped to Sept. 1. A JIS was inserted for Aug. 14-15 at MSFC.

1990/09-01: Launch delayed to Sept. 6. Note: ALL shuttles have been grounded until the leak problem is understood and fixed.

1990/09-06: Launch delayed again, due to H leak in *Columbia*'s aft engine compartment. Slipped to Sept. 18. (See insightful WaPo article describing NASA bashing due to Hubble flaw and Columbia leak issues 1990-09-06_WaPo_article.pdf .)

1990/09-17: Launch delayed once again by H leak. Launch delayed "indefinitely" according to Sept. 18 WaPo article. (See 1990-09-18_WashPo_article.pdf .) Time magazine article on the delay intones, "Tune in next leak."

1990/09-17: AFD memo to team describes his reasoning for being at KSC for launch. (See 1990-09-17 Pre-launch letter from Arthur Davidsen.pdf.)

1990/09-20: F. Roylance article in Evening Sun says launch is delayed at least 10 weeks. Discovery launch with Ulysses must happen by Oct. 23 or be delayed by 18 months. DoD mission in line for November on Atlantis, making Dec. 1 the earliest *Columbia*/Astro could go. Quotes AFD as saying we need access to test aliveness of instruments since they haven't been run since April. (See 1990-09-20_BaltSun_article.pdf.)

1990/10-01 to 05: Mission Replanning trip to MSFC; timeline planned to Jan. 1, 1991.

1990/10-06: Ulysses launches on *Discovery*.

1990/10-29: JHU Gazette article notes NASA is about to test *Columbia* for leaks; if all goes well and Atlantis launches as planned around Nov. 10, Columbia could launch as early as Dec. 1. The article notes that an aliveness test was run successfully the previous week at KSC. (See 1990-10-29_JHUGazette_article.pdf.)

(Note: an *AmericaSpace* web article from 2013 provides a detailed description of the Columbia leak problems that caused the many launch delays above. See 2013-03-10_In the Shadow of Challenger (Part 2).pdf.)

1990/11-27: Baltimore crew leaves for Huntsville and settles in for the long haul.

1990/12-02: STS-35 Astro-1 launch on *Columbia* at 12:49 a.m. CST (6:49 UTC). (See 1990-12-03_BaltSun_article.pdf.)

*(A separate document provides my detailed diary of events during the mission. See 1991-12_A1_memoirs_blair.pdf. A number of newspaper articles from during the mission and from various sources have been scanned to PDF.)*

1990/12-10: STS-35 landing, 11:54 p.m. CST (05:54:09 UTC), at Edwards AFB, CA.

1990/12-11: A long "Science Times" article in the NYT demonstrates how lobbying for a reflight started almost immediately. (See 1990-12-11_NYT_Science_article.pdf.)

1990/12-12: Letter from JHU president Richardson congratulating on successful Astro-1. (See 1990-12-12_Richardson_letter.pdf.)

1990/12-21: *Science* runs article "Astro-1: From the Jaws of Defeat." Ends by saying "There is virtually no hope that Astro will fly again…" due to NASA funding issues. (See 1990-12-21_Science_article.pdf.)

1991/01:  AAS meeting #177 in Philadelphia, first Astro-1 science presentations.

1991/01-??: Ted Gull steps down as Astro project scientist.

1991/01-14: Blair memo to Applegate at MSFC providing post-mission feedback. (See 1991-01_A1_feedback_to_NASA.pdf.)

1991/02/14: "Welcome Back Sam" program and reception hosted by JHU Physics & Astronomy department, Schafler auditorium.

1991-04: Astro-1 "Preliminary Science report" submitted to NASA. (See 1991-04_A1_HUT_prelim_scirep.pdf.)

1991/05-09: HUT "neutrino" paper published in *Nature*.

1991/05-20: Barbara Mikulski event at JHU to announce Astro-2 mission. (See 1991-05-20_PR_materials.pdf and 1991-05-27_SpNews_article.pdf.)

1991/03-26: NASA Group Achievement award for Astro-1 Investigators Group; a separate award with the same date was received for the HST Faint Object Spectrograph Development Team.

1991/05-27ff:  AAS #178 in Seattle WA.

1991/06: Steve Maran published Sky & Telescope article on Astro-1.  (No doubt, the June issue came out earlier, likely early April 1991.)  (See 1991-06_Maran_SkyTel_article.pdf. Note the box at the end by Stuart Goldman discussing Astro's future.)

1991/06: HUT meeting notes from early June confirm that Ted Gull has taken himself out of the running for Astro-2 Mission Scientist.  We are waiting for NASA management structure to be defined.

1991/06-24: Official letter from Fisk to AFD approving an Astro-2 mission, with initial manifest date of "late 1993."

1991/08-8,9: First Astro-2 IWG meeting at JHU; discussed "lessons learned" and GO program.

1991/??: Knox Long leaves JHU for STScI, but returns for Astro-2 mission support.

1991/10: Long JHU Magazine article about Sam, Astro-1, and looking ahead.

1991/10-07: HUT meeting minutes note we had heard that Charles "Chip" Meegan (MSFC) had been appointed as Astro-2 Mission Scientist. (Memo to this effect received mid-November -- exact date unreadable.)

1991/10,11: Work at KSC to extract HUT from the cruciform and get it into the ATM clean room. HUT meeting notes from Nov. 11 indicate that this has occurred and initial aliveness testing was done.

1991/11-21: AFD reports in HUT meeting that Chip Meegan is official as MSCI; work starting on justification for an extended mission concept.

1991/12-09: AFD, KSL, and WPB attend Astro-1 1-yr anniversary symposium at MSFC, and also talk with Chip Meegan re: GI program and extended mission concept.

1991/12-16: HUT meeting notes indicate arrival of Wei Zheng and Richard Buss to the team.

1991/12 and 1992/01: HUT at NIST for post-flight calibration (JWK).

1992/01-25: AFD memo announcing official position assignments for Astro-2 (GAK as Project Scientist, WPB and JWK as Deputies, etc.).

1992/03-11: AFD "Astro-2: Justification for a Long Duration Mission" white paper released.

1992/03-27(?): AFD receives NASA Group Achievement Award for Astro-1 on behalf of the HUT team.

1992/04-6: HUT meeting minutes report a nominal launch time of Sep. 14, 1993.

1992/04-23: IPSWG/IWG mtg at MSFC. A "Continuous Improvement Mission Planning Team" (CI/MPT) was established.

1992/06-15 to 17: First meeting of the CI/MPT at MSFC.

1992/06-30: Jayroe memo to JSC (primarily with input from the teams) justifying request for a night launch.

1992/07-20ff: CI/MPT mtg in Madison; concept of timeline slipping discussed.

1992/07-27: HUT mtg notes clearly indicate HUT spectrograph improvements are underway (new slit plate; grating to be coated with SiC). Looking at Jan. 1993 ship to KSC, Sep. 1993 turnover to NASA.

1992/08: Graduate student Tom Brown arrives at JHU; joins HUT project.

1992/09-07?: UV-MOWG defines and approves A-2 GI program structure.

1992/09-11: Astro IWG mtg at GSFC. Manifest shows Sep. 1994 launch on *Endeavour* for at least 10 days. (*Endeavour* can be fit with extended mission pkg.) Major discussions of CI/MPT progress and GO program.

1992/09-24: HUT meeting notes indicate need for new slit plate and CTE delays. Now looking at Mar. 1993 delivery to KSC and August 1993 turnover to NASA.

1992/10-19: HUT meeting notes indicate SiC grating showing 70% improvement over A-1 grating. GSFC expresses interest in trying to coat HUT back-up mirror.

1992/11-9: HUT meeting notes indicate addition of Mike Swam to HUT team. Coating of back-up primary with SiC sounding likely. Manifest update now shows Astro-2 on Jan. 10, 1995.

1992/12-7: HUT meeting notes indicate thrashing on launch manifest, with dates of June 1994, Oct. 1994, and Jan. 1995 all being discussed (with other "deployable" payloads). Impacts on our delivery schedule. We still have not received new slit plate.

1992/12-12: "Two years since Astro-1" party at AFD's.

1992/12-15 to 17: CI/MPT, IWG, and IPSWG mtgs at MSFC.

1993/01: ***Science*** article by AFD, "Far-Ultraviolet Astronomy on the Astro-1 Space Shuttle Mission" published. (1993-01-15_Science_article.pdf.)

1993/01-15F: QUEST (Quick Ultraviolet Exploratory Survey Telescope) SmEx proposal submitted as a HUT follow-on. (Ultimately not accepted for development.) (See 1993-01_Quest_prop_Inv.pdf.)

1993/01ff (through at least April): advertised and interviewed a number of candidates for HUT-related postdocs. (None eventually hired.)

1993/01-25: HUT meeting notes indicate a current manifest of Jan. 18, 1995, on *Columbia* with the Wake Shield experiment. (Lots of good notes from this mtg.)

1993/02-03: Fisk memo to AFD confirming appointment of GAK, WPB, JWK, and STD as HUT Co-Investigators.

1993/02-08: HUT Astro-2 Target List and Science Program Summaries submitted to NASA in support of NRA for Astro-2 GI program.

1993/02-17: AFD circulated memo from Jack Jones praising Astro-1 success. (See 1993-02-17_Jones_memo.pdf.)

1993/03-9,10: CI/MPT mtg at MSFC, finalize file formats. (Word has it that MSFC mgrs. want to shut the group down due to the many changes invoked by the team's efforts to streamline the MP process.)

1993/03: HUT Collected Scientific Papers, Vol. 1 1991-1992 (the yellow "HUT Book") published.

1993/03-03: HUT Memo #3-1 (M. Swam) announces major update to HUTSIM simulation program and integration into an IRAF task.

1993/03-22: HUT meeting notes: STD shares rumors of current manifest, showing Nov. 3, 1994, launch and an extended mission. This was confirmed at the 4/5/93 HUT meeting, where we are shown on *Columbia* for a 13 day mission, paired with the Capillary Loop Experiment. Pulling launch back earlier takes away planned contingency time, causing concern.

1993/04-13: Astro IWG meeting at JHU, plus break out meetings on Crew procedures, ground commanding, and mission planning.

1993/04-19: Spectrograph nearing completion, but delays in getting a good coating on the plates are causing pressure on SRF time for calibration over the next few weeks; APL has assessed what it would take to change out the primary mirror (time and risk); loss of contingency causing concern as to whether a mirror swap can be done.

1993/04-30: NASA NRA 93-OSSA-14 Astro-2 Guest Investigator RFP released. (Proposals due 930730.) (See 1993-04_A2GI_RFP.pdf.)

1993/05: The HUT Handbook, ver. 1.0 published. (See 1993_HUT_Hndbk.pdf.)

1993/06-14: HUT meeting notes make it clear that the issue of whether to coat the primary mirror with SiC is coming to a head. At KSC, the telescope has been disassembled and reflectometer measurements on the primary are about to begin.

1993:/06-24: With a delay in the delivery schedule, the idea of recoating and flight-qualifying the back up mirror is back into the discussion. MSFC requests a detailed plan prior to actually recoating the backup mirror.

1993/07: Exact date unclear, but a formal "Proposal to Upgrade the Hopkins Ultraviolet Telescope with a Primary Coated with Silicon Carbide" was prepared and submitted to the project. (See 1993-07_HUT_Mirror_SiC_prop.pdf.)

1993/07-19: HUT meeting minutes note, “Sam took the backup mirror to GSFC last Friday [7/16] to be coated with SiC.”

1993/07-26: HUT meeting minutes note “Primary is being coated even as we speak.” Discusses upcoming reassembly schedule.

1993/07-30: Memo 7#3 dated July 30, 1993, from GAK to KSC describes the re-integration plan and schedule and notes, “The backup mirror has been coated, and we are now making calibration measurements to qualify it for flight.”

1993/08-02: HUT meeting notes indicate the spectrograph is at SRF for calibration over the next two weeks. Backup mirror is back at JHU for assessment. Discussion of increasing the size of the 1 cm^2 aperture to 4 cm^2 to allow observation of more WUPPE sources.

1993/08-09: HUT meeting notes indicate 37 GI proposals received by NASA, many requesting multiple instruments. HUT is prime on 21 (8-16 HUT notes). STD reports excellent results from the mirror coating and that it will be shipped to KSC in a few days.

1993/08-11: Kriss et al. memo to MSFC describing the recoating of the backup mirror. (1993-08-11_HUT_Mirror_SiC_result.pdf )

1993/08-16: HUT meeting notes indicate SRF results confirming suspicion (from CTE) that the spectrograph performance is lower than expected. (Discussion of possible reasons.) Primary is in Florida and witness mirrors are clean.

1993/08-30: HUT meeting notes indicate mirror change out went smoothly and telescope reassembly is in progress. Spectrograph scare has gone away with further analysis (repeller grid problem): Aeff curve is essentially as expected.

1993/08-31 to 09-1: AFD and STD attend Payload OWG at Houston (first serious involvement of JSC in Astro-2 planning). Discussed launch window, SAA avoidance, IPS operations and checkout philosophy, and MSFC/JSC communications. Next launch manifest expected to show a slip of the mission into December 1994.

1993/09-13: HUT meeting notes indicate spectrograph and TV camera being packed and shipped this week. At KSC, the new mirror has been installed and the telescope reassembled; just waiting for the spectrograph and TV.

1993/09-30: WPB and WZ supported Astro-2 proposal review. Sequence file reviews start in earnest.

1993/10-11: HUT meeting notes indicate 10 GIs selected by NASA and assigned to the various teams; since so few WUPPE proposals, some who requested HUT + WUPPE were assigned to WUPPE. HUT GIs include Claus Leitherer, Nolan Walborn, David Findley, John

Raymond, and Brian Espey (but assigned to WUPPE). Focus and alignment activities at KSC.

1993/11-2: Astro IWG#6 mtg at MSFC. First meeting with GIs present.

1993/11-15: HUT meeting notes indicate Dec. 1, 1994, is new official launch date. Reassembly and alignment activities continue at KSC.

1993/11-18: HUT GIs meeting at JHU.

1993/11-22: Payload Crew Training meeting at JHU.

1993/12-02: STS-61, first servicing mission to HST launches. COSTAR (developed under the tutelage of Holland Ford at JHU) was successfully installed on Dec. 7, 1993.

1993/12-6: HUT meeting notes indicate final alignment last week at KSC. Final baffle and door install this week. Upcoming schedule for functional testing and delivery discussed.

1993/12-03,13: STS-61 on *Endeavour* repairs HST optical aberration and installs new instruments.

1993/12-12: HUT holiday party at AFD's.

1993/12-13: HUT meeting notes indicate reassembly of HUT complete. (Some MSFC concern about some of the bolts used.)

1994/01: Numerous interactions and negotiations to incorporate GI planning into the mission planning "block scheduling" concept.

1994/01-31: HUT meeting notes highlight that HUT has been turned over to KSC and mounted on the cruciform, with alignment activities underway. Changes to DEP s/w outlined.

1994/02-14: HUT meeting notes indicate POCC position assignments have been made, in preparation for training activities and Sims. Also, discussion of recent HST result on He II G-P observation of QSO 0302-003 and how it may impact HUT observations.

1994/03-14: HUT meeting notes indicate new 20 degree ram avoidance constraint has been added to mission planning to avoid impacts to SiC coating of primary mirror.

1994/03: Astro-2 MPHIRD (Mission Planning Handbook and Interface Requirements Document) published. (Was basically the culmination of the reworked mission planning process derived out of the CI/MPT.) (See 1994-03_MPHIRD.pdf.)

1994/03-15: Launch slip to Jan. 12, 1995, announced.

1994/04-25:  HUT meeting minutes note MSFC rumor about pulling launch back to early December as a cost-cutting move, with major impacts to mission planning as well.

1994/04-28: Kruk memo JWK 4-94b reports the completion of the Level IV Interface Verification  Test on 4/20-21/1994.

1994/05-19,20: Level IV Mission Sequence Test

1994/07: AFD goes to Cospar.  (See 1994-07_COSPAR_viewgraphs.pdf for compilation of viewgraph hard copies to support this talk.)

1994/07: HUT Handbook (ver. 1.1) published in support of GI program. (See 1994-07_HUT_Hndbk_ver_1.1.pdf.)

1994/08-03,05: Instrument Verification Test at KSC.

1994/08-10,11: Level III/II Mission Sequence Test at KSC.

1994/08-11: Short circuit on DEP board during last hour of Level III/II mission sequence test at the Cape.

1994/08-12-17: MSFC: POCC training, PI/Cadre Sim#1, & Astro IWG meeting.

1994/08-29:  HUT meeting minutes note rumor of Feb. 9, 1995, launch date.

1994/09-12:  HUT meeting minutes note rumor of Feb. 23, 1995, launch date; once again this wreaks havoc with mission planning.

1994/10-03:  HUT meeting minutes note a new timeline planning exercise is underway for delivery in December, supporting the Feb. 23, 1995, launch date.

1994/10-04: DEP repair completed; MST completed 241022. Cost recovery proposal after the fact. (See 1994-11_DEP_repair_cost_recovery_prop.pdf.)

1994/10-17:  HUT meeting minutes: Kruk notes that a KSC error in switching pumps for HUT may have exposed the spectrograph to air.

1994/10-24,27: PI/Cadre Sim#2 at MSFC; HUT attempted to go to three shifts, but debrief raised a number of issues.  Decision still TBD.

1994/11-07: Long JHU Gazette article by Emil Venere lays groundwork for Astro-2, and publicly announces availability of a HUT web page. (See 1994-11-07_JHUGazette_article.pdf.)

1994/12-01: Email from Chip Meegan notes March 2, 1995 as the new "official" launch date. The extension to 16 days is expected to be finalized the next day.

1994/12-05,08: PI/Cadre Sim#3 at MSFC

1994/12-18: New color HUT brochure for Astro-2 delivered to JHU!

1995/01-03,06: Astro-2 JIS#1 at MSFC

1995/01-23,26: Astro-2 JIS#2 at MSFC

*(A separate document provides my detailed description of events from June 1994 through week of Feb. 20, 1995. See 1995_08_A2_premission_memoirs_blair.pdf.)*

1995/02-16: L-14 press conferences were held.

1995/02-18: HUT team pre-launch party at AFD's (including Sam and Becky)

1995/02-27: Nice pre-mission Aviation Week article on Astro-2. (See 1995-02-27_AvWeek_A2Preview_article.pdf.)

1995/03-02: STS-67 Astro-2 launch on *Endeavour* at 06:38:13 UTC. (See 1995-03-02_AP_launch_article.pdf, and numerous other articles from during the mission.)

*(A separate document provides my detailed diary of events during the mission. See 1995-08_A2_memoirs_blair.pdf.)*

1995/03-18: STS-67 landing at Edwards AFB at 21:47:01 UTC.

1995/03-23: (See Space News summary of mission, 1995-03-19_SpaceNews_article.pdf, and others.)

1995/04-19: AFD gives Astro-2 overview talk at STScI.

1995/05-11: Astro-2 Science Debrief meeting at JHU with the astronauts, including ritzy reception afterwards at Evergreen House.

1995/06-11ff: Numerous presentations at summer AAS #186 meeting in Pittsburgh, PA, including the headliner, He II Gunn Peterson result. (See subsequent press coverage, cf.

coverage in the NYT "Science" page: 1995-06-13_NYT_article.pdf, and Science News 1995-06-17_SciNews_article.pdf.)

1995/07: Spaceflight magazine mission summary (1995-07_Spaceflight_STS67_MissReport.pdf) and Final Frontier magazine mission summary (1995-07_FinalFrontier_STS67_summary.pdf) published.

1995/07-10: Jack Jones memo to AFD congratulating on He II G-P result. (See 1995-07-10_Jones_letter.pdf.)

1995/09: A-2 HUT Engineering Report submitted. (See 1995-09_A2_HUT_Eng_Report.pdf.)

1995/11-08: AFD sends letter to Dan Goldin (NASA Administrator) requesting consideration of an "Astro-3" mission to view comet Hale-Bopp in the Mar-Apr 1997 timeframe. (Declined.) (See 1995-11-08_A3_letter.pdf.)

1995/11-20: Special HUT Astro-2 ApJ Letters volume (vol 454, Number 1, Part 2, including 17 articles) published. According to a short article in the JHU Gazette (1/2/96, "Only four times before has the prestigious publication dedicated an issue to [results from] one instrument."

1996/01-15: AAS meeting #187 in San Antonio, TX.

19996/03-07: HUT "IGM" paper published in Nature (1996-03-07_AFD_IGM_Nature.pdf)

1996/09-16: JHU Gazette article notes that Sam Durrance has left JHU for a private telecomm company after 16 years at JHU. (See 1996-09-16_JHUGazette.pdf.)

1996/??: Van Dixon receives PhD from JHU.

1996/12?? End of HUT contract at JHU.

1996/12: Tom Brown receives PhD from JHU.

1997/02-11: Second HST Servicing mission, STS-82, is launched. Space Telescope Imaging Spectrograph is installed. (Warren Moos was a co-I, and Mary Beth Kaiser had major involvement with STIS. Chuck Bowers worked on STIS at GSFC after leaving JHU.)

1997/06-10: HUTSpa SmEx proposal to put HUT on a Spartan free-flyer was submitted. (Ultimately declined.) 1997-06-10_HUTSPA.pdf

1998/09: Jerry Kriss leaves JHU for STScI.

1999/03-10 and 11: Spacelab Accomplishments Forum held at NASA-HQ; AFD presents on ASTRO missions. (NASA CP-2000-210332 is the Proceedings, published Sep. 2000). (See 2000-09_Spacelab_Accomplishments_AFD.pdf.)

2001/07-19: AFD passes away. Obituaries in NYT, WaPo, etc. (See Obits directory.)

2001/11??: HUT installed at National Air & Space Museum in Washington, D.C.

2005/??: Astro-1 15th anniversary celebration at JHU.

2008/05-09: Ron Parise passes away. (See Obits directory.)

2008/06-10: HUT "Final Archive" ADAP proposal submitted. (Accepted, and led to the final reprocessing and archiving of the HUT data. See 2013-04_HUT_Archive_PASP.pdf .

2009/03-11: Art Code passes away. (See Obits directory.)

2011/03: Belated "Astro-1 20 year" celebrations at JHU. Also, the "Davidsen Display System" in the Bloomberg lobby was dedicated to AFD's memory. (See 2011-03_Davidsen_display_dedication.ppt.)

2013: Historical web articles, "In the Shadow of Challenger" written by Ben Evans, on the America Space web site (https://www.americaspace.com ). Original publication dates were March 9 and 10 for the two articles, one on "The Lost Mission pf STS-61E" which was Astro-Halley, and one on "The Found Mission of STS-35." (See 2013-03-09_In the Shadow of Challenger (Part 1).pdf and 2013-03-10_In the Shadow of Challenger (Part 2).pdf.)

2013/04: The final HUT archive is announced via publication of Dixon et al., 2013, PASP, 125, 431. (See 2013-04_HUT_Archive_PASP.pdf, as well as the A-1 and A-2 summary plot packages 2013-HUT1_plots.pdf and 2013-HUT2_plots.pdf.)

2015/12: Astro-1 25th Anniversary: NASA GSFC Colloquium. Blair was a keynote speaker. (See 2015-12_Astro_25yr_retrospective_WPB.pptx.)

2017/10-29: Ted Stecher passes away. (See Obits directory.)

2020/11: Scott Vangen (A-2 Alternate PS) and associates found the non-profit Astro Restoration Project, formalizing more than two decades of previous effort to find and restore Astro and Spacelab hardware as an education and outreach project. (Ongoing.). (See article from Nov. 2024 AAS Historical Astronomy Division newsletter 2024-12_ARP_HADNews.pdf.)

2023/05-05: Sam Durrance passes away. (See Obits directory.)

# Appendix B: HUT PR Article Scans Reference Listing

(See directory Astro_PR_articles at https://archive.stsci.edu/hut/hut-historical-archive/ )

P1 1978-01_S&T_article_3C273.pdf
P2 1981-12-10_JHUGazette.pdf
P3 1982-08-16_AvWeek_article.pdf
P4 1983-10-31_BaltSun_article.pdf
P5 1983-12-02_JHUNewsletter_article.pdf
P6 1984-05-31_JHU_PressRel.pdf
P7 1984-06-07_JHUGazette_article.pdf
P8 1984-06-21_BaltEveSun_article.pdf
P9 1984-06-21_BaltSun-article.pdf
P10 1984-07-12_JHUGazette_article.pdf
P11 1984-10-19_JHUNewsletter_article.pdf
P12 1985-05-05_BaltSun_CAS_announce_article.pdf
P13 1985-06_AFD_APL_Tech Review_on_HUT.pdf
P14 1985-10_AFD_w_model.pdf
P15 1985-10-22_JHU_Gazette.pdf
P16 1985-11_APLNews_article.pdf
P17 1985-11-08_JHUNews_CAS_announce_article.pdf
P18 1985-12-11_ChronHE_article.pdf
P19 1985-12-28_WaPo_article.pdf
P20 1985-Spr_JHUZeniada_article.pdf
P21 1986-01-06_BaltSun_article.pdf
P22 1986-01-21_JHUGazette_article.pdf
P23 1986-01-31_JHU_Newsletter_article.pdf
P24 1986-02_JHUMag_article.pdf
P25 1986-02-02_BaltSun_article.pdf
P26 1986-03-04_JHUGazette_article.pdf
P27 1986-03-10_TampaNews_article.pdf
P28 1986-04_JHUMag_article.pdf
P29 1986-05_BaltMag_article.pdf
P30 1986-09-01_AvWeek_article.pdf
P31 1986-10-26_BaltSun_article.pdf
P32 1990-01-23_JHUGazette_article.pdf
P33 1990-04_JHUMag_partial.pdf
P34 1990-05-27_BaltSun_article.jpg
P35 1990-06_A1-SkyTel_AstroObs.pdf
P36 1990-09-06_WaPo_article.pdf
P37 1990-09-18_WashPo_article.pdf
P38 1990-09-20_BaltSun_article.pdf
P39 1990-09-21_WaPo_article.pdf
P40 1990-10-29_JHUGazette_article.pdf
P41 1990-12-03_BaltSun_article.pdf
P42 1990-12-06_NYT_article.pdf
P43 1990-12-07_BaltSun_article.pdf
P44 1990-12-07_WaPo_article.pdf

P45 1990-12-08_WaPo_article.pdf
P46 1990-12-08-15_SciNews.png
P47 1990-12-10_AvWeek_article.pdf
P48 1990-12-10_HunstvilleTimes_article.pdf
P49 1990-12-10_JHUGazette_article.pdf
P50 1990-12-11_BirmPost_article.jpg
P51 1990-12-11_NYT_Science_article.pdf
P52 1990-12-12_NYT_article.pdf
P53 1990-12-17_23_SpaceNews_article.pdf
P54 1990-12-17_AvWeek_article.pdf
P55 1990-12-17_JHUGazette_article.pdf
P56 1990-12-21_Science_article.pdf
P57 1991_APLTech_A1_Fountain.pdf
P58 1991-01-18_BaltSun_article.pdf
P59 1991-02_FinalFrontier_STS35_summary.pdf
P60 1991-04_S&T_article.pdf
P61 1991-04_SkyTel_cover.png
P62 1991-05-14_BaltSun_article.pdf
P63 1991-05-27_SpNews_article.pdf
P64 1991-05-28_JHUGazette_article.pdf
P65 1991-06_Maran_SkyTel_article.pdf
P66 1991-10_JHMag_article.pdf
P67 1993-01-15_Science_article.pdf
P68 1993-10-25_JHUGazette_article.pdf
P69 1993-11-29_COSTAR_JHUGazette.pdf
P70 1994-02-07_JHUGazette_HST_Fix.pdf
P71 1994-11-07_JHUGazette_article.pdf
P72 1995-02-13_JHUGazette_article.pdf
P73 1995-02-27_AvWeek_A2Preview_article.pdf
P74 1995-03-02_AP_launch_article.pdf
P75 1995-03-03_NYT_article.pdf
P76 1995-03-05_NYT_article.pdf
P77 1995-03-06_BaltSun_article.pdf
P78 1995-03-13_JHUGazette.pdf
P79 1995-03-18_BaltSun_article.pdf
P80 1995-03-19_SpaceNews_article.pdf
P81 1995-03-19_WaPo_article.pdf
P82 1995-03-20_JHUGazette_article.pdf
P83 1995-03-27_JHUGazette_article.pdf
P84 1995-06-13_NYT_article.pdf
P85 1995-06-15_BaltEveSun_article.pdf
P86 1995-06-17_SciNews_article.pdf
P87 1995-07_FinalFrontier_STS67_summary.pdf
P88 1995-07_Spaceflight_STS67_MissReport.pdf
P89 1996-09-16_JHUGazette.pdf
P90 1996-12-16_JHU_Gazette.pdf
P91 1997-11-17_Crocker_JHUGazette.pdf
P92 2001-03-26_JHUGazette_article.pdf
P93 2001-09-24_JHUGazette_HUT_NASM.pdf
P94 2010-11-29_JHUGazette_HUT20.pdf

P95 2011-03-11_JHUGazette_Display.pdf
P96 2013-03-09_In the Shadow of Challenger (Part 1).pdf
P97 2013-03-10_In the Shadow of Challenger (Part 2).pdf
P98 2025-12_HUT_A&S Mag.pdf

## Appendix C: HUT Document Scan Reference List

(See directory Document-scans at https://archive.stsci.edu/hut/hut-historical-archive/ )

D1 1978_HUT_orig_prop_cost.pdf
D2 1978-HUT_orig-prop_tech.pdf
D3 1981-10-12 APL HUT project kickoff memo.pdf
D4 1981-12-10 contract award JHU article.pdf
D5 1981-HUT_construction_prop.pdf
D6 1981-HUT_contract_NAS5-27000.pdf
D7 1981-RFP-5-15117-278.pdf
D8 1983-HUT_Inst_Dev_plan.pdf
D9 1984_02ff_Lucas_letters.pdf
D10 1984-07-12_Lucas_followup_letter.pdf
D11 1984-09_HUT_leaving_JHU.pdf
D12 1985_Hurricane-Gloria.png
D13 1985-01-29_WPB_HUT_Broshure_comments.pdf
D14 1985-06-05_KSL Mirror_assess.pdf
D15 1986-02-25 APL HUT status memo.pdf
D16 1986-06_Challenger_Report.pdf
D17 1988-12_IAUGA_report.pdf
D18 1989-12-18_AFD_memo.pdf
D19 1990-09-17 Pre-launch letter from Arthur Davidsen.pdf
D20 1990-12-12_Richardson_letter.pdf
D21 1991-01_A1_feedback_to_NASA.pdf
D22 1991-02-05_A2_Justification.pdf
D23 1991-04_A1_HUT_prelim_scirep_App.pdf
D24 1991-04_A1_HUT_prelim_scirep.pdf
D25 1991-05-20_PR_materials.pdf
D26 1991-06-24_A2_letter.pdf
D27 1992-01-25_A2_duties_memo.pdf
D28 1992-03-11_Long_Duration_memo.pdf
D29 1992-05-05_A2_science.pdf
D30 1993-01_Quest_prop_Inv.pdf
D31 1993-02-03_HUT_CoIs_memo.pdf
D32 1993-02-17_Jones_memo.pdf
D33 1993-04_A2GI_RFP.pdf
D34 1993-05_HUT_Hndbk.pdf
D35 1993-07_HUT_Mirror_SiC_prop.pdf
D36 1993-08-11_HUT_Mirror_SiC_result.pdf
D37 1994_ca-Mission-Planning-overview.pdf
D38 1994-03_MPHIRD.pdf
D39 1994-07_COSPAR_viewgraphs.pdf
D40 1994-07_HUT_Hndbk_ver_1.1.pdf
D41 1994-11_DEP_repair_cost_recovery_prop.pdf
D42 1994-11-01_SrReview.pdf

D43 1994-11-02_CIMPT_summary.pdf
D44 1994-11-23_SrRevResults.pdf
D45 1995-07-10_Jones_letter.pdf
D46 1995-09_A2_HUT_Eng_Report.pdf
D47 1995-11-08_A3_letter.pdf
D48 1996-03-07_AFD_IGM_Nature.pdf
D49 1997-06-10_HUTSPA.pdf
D50 2000-09_Spacelab_Accomplishments_AFD.pdf
D51 2008-06_HUT_ADAP.pdf
D52 2013-03-09_In the Shadow of Challenger (Part 1).pdf
D53 2013-03-10_In the Shadow of Challenger (Part 2).pdf
D54 2013-04_HUT_Archive_PASP.pdf
D55 2013-HUT1_plots.pdf
D56 2013-HUT2_plots.pdf